\documentclass[%
 reprint, %linenumbers,
superscriptaddress,
 amsmath,amssymb,
 aps, physrev,
floatfix, pi
]{revtex4-2}
\usepackage{booktabs}
\usepackage{array}
\usepackage{xcolor}
\usepackage{colortbl}
\usepackage{multirow}
\usepackage{amsmath}
\usepackage{xcolor}
\usepackage{braket}
\usepackage[nice]{nicefrac}
\usepackage{mathrsfs}
\usepackage{graphicx}% Include figure files
\usepackage{dcolumn}% Align table columns on decimal point
\usepackage{bm}% bold math
\usepackage{fancyhdr}
\usepackage{subfig}
\usepackage{float}
\usepackage{tabularx}
\usepackage{array}
\usepackage{booktabs}
\usepackage{caption} % For customizing captions
\usepackage[colorinlistoftodos]{todonotes} %allows for todo notes

\begin{document}
%\begin{bibunit}[plain]

%\preprint{Distribution D}

\title{\textbf{Telecommunications fiber-optic and free-space quantum local area networks at the Air Force Research Laboratory} 
}% 

\author{Erin Sheridan}
 \affiliation{U.S. Air Force Research Laboratory, Information Directorate, Rome, NY 13441}%Lines break automatically or can be forced with \\

\author{Nicholas J. Barton}
\affiliation{Murray Associates of Utica, Utica, NY 13501
}%

\author{Richard Birrittella}
\affiliation{Booz Allen Hamilton, Rome, NY 13441
}%

\author{Vedansh Nehra}
 \affiliation{Technergetics, Rome, NY 13441}%Lines break automatically or can be forced with \\

\author{Zachary Smith}
 \affiliation{U.S. Air Force Research Laboratory, Information Directorate, Rome, NY 13441}%Lines break automatically or can be forced with \\

\author{Christopher Tison}
 \affiliation{U.S. Air Force Research Laboratory, Information Directorate, Rome, NY 13441}%Lines break automatically or can be forced with \\
 
\author{\\Amos Matthew Smith}
 \affiliation{U.S. Air Force Research Laboratory, Information Directorate, Rome, NY 13441}%Lines break automatically or can be forced with 

\author{Shashank Dharanibalan}
 \affiliation{Technergetics, Rome, NY 13441}%Lines break automatically or can be forced with \\

\author{Vijit Bedi}
 \affiliation{U.S. Air Force Research Laboratory, Information Directorate, Rome, NY 13441}%Lines break automatically or can be forced with \\

\author{David Hucul}
 \affiliation{U.S. Air Force Research Laboratory, Information Directorate, Rome, NY 13441}%Lines break automatically or can be forced with \\

 \author{\\Benjamin Kyle}
 \affiliation{U.S. Air Force Research Laboratory, Information Directorate, Rome, NY 13441}%Lines break automatically or can be forced with \\

\author{Christopher Nadeau}
 \affiliation{Booz Allen Hamilton, Rome, NY 13441
}%

 \author{Mary Draper}
\affiliation{Booz Allen Hamilton, Rome, NY 13441
}%

\author{John Heinig}
 \affiliation{U.S. Air Force Research Laboratory, Information Directorate, Rome, NY 13441}%Lines break automatically or can be forced with \\

 \author{Scott Faulkner}
\affiliation{Booz Allen Hamilton, Rome, NY 13441
}%

  \author{\\Randal Scales}
\affiliation{Booz Allen Hamilton, Rome, NY 13441
}%

 \author{Andrew M. Brownell}
\affiliation{Murray Associates of Utica, Utica, NY 13501
}%

 \author{Stefan Preble}
\affiliation{Microsystems Engineering, Rochester Institute of Technology, Rochester, New York 14623, USA
}%

\author{James Schneeloch}
 \affiliation{U.S. Air Force Research Laboratory, Information Directorate, Rome, NY 13441}

 \author{\\Samuel Schwab}
 \affiliation{U.S. Air Force Research Laboratory, Information Directorate, Rome, NY 13441}

 \author{Daniel Campbell}
 \affiliation{U.S. Air Force Research Laboratory, Information Directorate, Rome, NY 13441}

\author{Derrick Sica}
\affiliation{Murray Associates of Utica, Utica, NY 13501
}%

\author{Peter Ricci}
 \affiliation{U.S. Air Force Research Laboratory, Information Directorate, Rome, NY 13441} 
 
\author{Vladimir Nikulin}
 \affiliation{Binghamton University, Binghamton, NY 13902} %Lines break automatically or can be forced with \\
 
\author{John Malowicki}
 \affiliation{U.S. Air Force Research Laboratory, Information Directorate, Rome, NY 13441}%Lines break automatically or can be forced with \\

 \author{Jacob Hall}
 \affiliation{U.S. Air Force Research Laboratory, Information Directorate, Rome, NY 13441}%Lines break automatically or can be forced with \\

\author{Michael Fanto}
 \affiliation{U.S. Air Force Research Laboratory, Information Directorate, Rome, NY 13441}%Lines break automatically or can be forced with \\

 \author{Matthew D. LaHaye}
 \affiliation{U.S. Air Force Research Laboratory, Information Directorate, Rome, NY 13441}%Lines break automatically or can be forced with \\

 \author{\\Laura Wessing}
 \affiliation{U.S. Air Force Research Laboratory, Information Directorate, Rome, NY 13441}%Lines break automatically or can be forced with \\

\author{Paul M. Alsing}
\affiliation{U.S. Air Force Research Laboratory, Information Directorate, Rome, NY 13441}%Lines break automatically or can be forced with \\

\author{Kathy-Anne Soderberg}
\affiliation{U.S. Air Force Research Laboratory, Information Directorate, Rome, NY 13441}%Lines break automatically or can be forced with \\

\author{Donald Telesca}
\affiliation{U.S. Air Force Research Laboratory, Information Directorate, Rome, NY 13441}%Lines break automatically or can be forced with \\

\date{\today}% It is always \today, today,

\begin{abstract}
As quantum computing, sensing, timing, and networking technologies mature, quantum network testbeds are being deployed across the United States and around the world. To support the Air Force Research Laboratory (AFRL)'s mission of building heterogeneous quantum networks, we report on the development of Quantum Local Area Networks (QLANs) operating at telecommunications-band frequencies. The multi-node, reconfigurable QLANs include deployed optical fiber and free-space links connected to pristine laboratory environments and rugged outdoor test facilities. Each QLAN is tailored to distinct operating conditions and use cases, with unique environmental characteristics and capabilities. We present network topologies and in-depth link characterization data for three such networks. Using photonic integrated circuit-based sources of entangled photons, we demonstrate entanglement distribution of time-energy Bell states across deployed fiber in a wooded environment. The high quality of the entanglement is confirmed by a Clauser-Horne-Shimony-Holt inequality violation of $S=2.717$, approaching the theoretical maximum of $S=2.828$. We conclude with a discussion of future work aimed at expanding QLAN functionality and enabling entanglement distribution between heterogeneous matter-based quantum systems, including superconducting qubits and trapped ions. These results underscore the practical viability of field-deployable, qubit-agnostic quantum network infrastructure. 

\end{abstract}

\maketitle
\thispagestyle{fancy}

%\tableofcontents

\section{\label{Intro}Introduction}

Scaling up quantum systems, such as quantum processing units, sensor arrays, and clocks, requires the implementation of quantum networks to reach the utility scale. Distributed quantum computing \cite{Wehner2018, Caleffi2024}, distributed quantum sensing \cite{zhang2021, malia2022, Kim2024}, quantum clock networks \cite{komar2014, Nichol2022}, secure communications \cite{Bennett1993, Cacciapuoti2020}, and additional applications of networked quantum information systems %, Caldwell2023}% 
rely on robust entanglement distribution between remote nodes. Quantum network testbeds are being implemented across the globe \cite{Wengerowsky2018, dynes2019, Alshowkan2021, Chung2021, Du2021, Earl2022, Pompili2022, Monga2023, rakonjac2023, Craddock2024, Bersin2024, Bersin2024a, Rahmouni2024, Krutyanskiy2024, Thomas2024, Strobel2024, Kucera2024, Sundaram2025} with these use cases in mind. These testbeds typically rely on telecommunications-band (telecom) optical fiber links, while some have also explored free-space links \cite{Namazi2017, Lanning2021, krzic2023}. All-optical quantum networks have demonstrated a variety of capabilities, and hold promise for many known applications \cite{Azuma2015, Kim2024}, but it is widely recognized that full-functionality quantum networks require matter-based qubits to act as quantum memories and quantum processors, and efficient, low-noise interconnects for these matter-based systems \cite{Wehner2018}. Furthermore, heterogeneous quantum networks capable of connecting diverse quantum systems can unlock even greater functionality. As such, the deployed network infrastructure must be compatible with matter-based quantum information systems while also supporting low-loss and high-fidelity signal transmission.

Here, we detail the establishment of telecom fiber and free-space quantum network testbeds at the Air Force Research Laboratory Information Directorate in Rome, New York and Stockbridge, New York. The three Quantum Local Area Networks (QLANs) presented here have different network topologies and operating environments, allowing for quantum demonstrations in a diverse set of conditions.  We detail link characterization experiments, initial timing synchronization demonstrations, and Bell State distribution across the QLANs using an on-chip telecom entanglement source developed at AFRL. Comprehensive data sets taken from these multi-purpose quantum network testbeds include monitoring of environmental conditions, link characteristics and corresponding quantum/classical signal qualities over long time periods. These data can be used to design necessary stabilization and mitigation solutions, to inform quantum network simulations, and to feed into centralized models which can be used to engineer better quantum networks and more accurately predict their behavior. Below we detail the progress we have made toward building the requisite resources and capabilities for such a testbed.
 
While this work focuses on infrastructure development, network engineering, and all-optical demonstrations, we see this as a first step toward building fielded, heterogeneous quantum networks, and our main goal remains the integration of a variety of matter-based quantum systems. In the future, these QLANs will provide a qubit-agnostic backbone to connect heterogeneous qubit species, including quantum integrated photonic devices, superconducting circuits and trapped ions, other matter-based qubits, sensors and other devices. We intend these networks to serve as a cross-domain (land- and air-based) quantum networking testbed for the Department of the Air Force.

\section{\label{Topologies}QLAN Topologies}
We begin by introducing the three QLANs that we have set up at AFRL locations, each of which has a unique setup and operating environment. The networks are the Griffiss QLAN, operating in and around the Innovare Advancement Center (IAC)\cite{IAC} in Rome, NY, the Rome Research Site (RRS) QLAN operating at the AFRL Information Directorate in Rome, and the Stockbridge QLAN located at an AFRL field test site located about 20 miles away in Stockbridge, NY \cite{Smith1991}.

\subsection{\label{Griffiss Topology} The Griffiss Quantum Local Area Network}
The Griffiss QLAN consists of four telecom network nodes situated within a quantum networking laboratory located inside of the IAC, adjacent to the Griffiss International Airport and the SkyDome, which is the largest indoor/outdoor anechoic chambered uncrewed aerial vehicle experimentation facility in the United States \cite{skydome}. The quantum networking laboratory includes trapped ion, integrated photonic circuit, superconducting, and quantum transduction systems. Quantum transduction\cite{Sheridan2024, Schneeloch2025princ, Schneeloch2025, Sheridan2025} and quantum frequency conversion capabilities are under development within the laboratory to facilitate the connection of these matter-based quantum systems to the QLAN \cite{Campbell2023, Craft2024, Li2025, Li2025a}, as well as quantum photonic devices \cite{Scott2019, Smith2024, Schwab2024} and associated packaging and interconnect solutions \cite{Smith2023}. The superconducting laboratory setup includes a dilution refrigerator (Bluefors XLD) outfitted with infrared- and visible-wavelength optical fibers, in addition to complete superconducting qubit control and measurement hardware, to facilitate a variety of quantum network connectivity scenarios. A detailed discussion of those efforts is outside the scope of this work. The laboratory is connected to deployed fiber loop-backs that run from the IAC to buried fiber at the Griffiss Technology Park. The loop-back nature of the deployed fiber connection and the central laboratory location allow for quick reconfigurations of the QLAN topology to accommodate a variety of experimental setups and use cases.

\begin{figure*}
    \includegraphics[width = 1\linewidth]{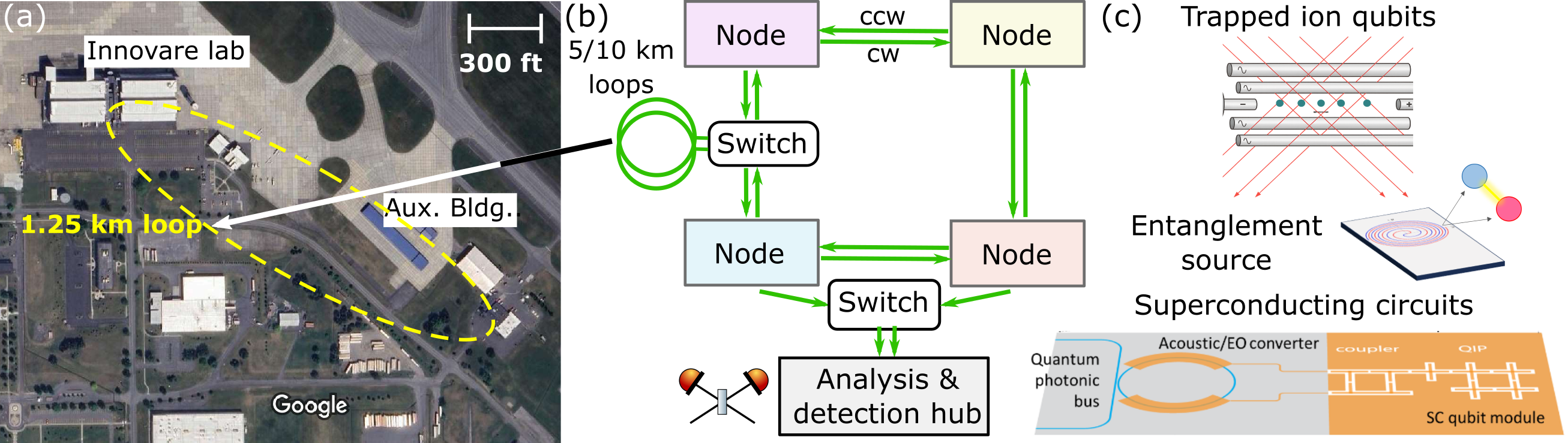}
    \caption{(a) Satellite image of Griffiss QLAN. Deployed fiber loops back between a quantum networking laboratory in the Innovare Advancement Center and an auxiliary building (Aux. Bldg.). Note that the exact fiber path is not shown. Image obtained from Google Maps, copyright Google 2025, Airbus 2025. (b) Simplified network topology diagram for Griffiss QLAN. Four network nodes are connected via duplex single-mode fiber, with clockwise (cw) and counter-clockwise (ccw) loops. Optical switches are used to route signals to deployed buried fiber of up to 15 km length, and to analysis and detection hardware. Example network nodes include (c) trapped ion systems (or networks of trapped ion systems) with quantum frequency conversion (QFC), hybrid superconducting systems with sub-components labeled, including a photonic bus, acoustic/electro-optic (EO) converter, tunable couplers, and qubits (QIP)), and photonic integrated circuit devices.  }
    %\vspace{128in}
    \label{Figure 1}
\end{figure*}

The Griffiss QLAN nodes are connected in a ring topology, shown in Figure \ref{Figure 1}(a). A ring topology lends itself well to the proposed ``Q-Fly" architecture for distributed quantum computation within a QLAN \cite{sakuma2024} and can serve as a platform for DLCZ-style \cite{Duan2001} entangling gates between remote network nodes. Coarse-wavelength division optical-add-drop multiplexers (CWDM OADMs) are the interface between each node and the ring. Each OADM has eight C-band wavelength channels, spanning $1470\; \mathrm{nm} - 1610\;\mathrm{nm}$ and spaced by $20\;\mathrm{nm}$,  and an O-band channel with a passband of  $1310\; \mathrm{nm} - 1410\;\mathrm{nm}$. The OADMs allow for bi-directional signal propagation in each of the duplex fiber links, so that the ring has separate clockwise and counter-clockwise signal propagation paths. The ring currently consists of four nodes, but can easily be expanded to include more nodes by adding more OADMs. The number of available wavelength channels at each node can also be expanded by dense wavelength division multiplexers (DWDMs) and O-band CWDMs. Each node can act as a Source node (all-photonic entanglement source), a Memory node (a matter-based qubit that emits an entangled photon) or an analysis node (Bell State analyzer, polarization state tomography, polarimeter, etc.) Of course, a node can have more than one of these capabilities at once. Detailed diagrams of the QLAN control plane, node components and analysis hardware are available in the Supplementary Information.

The Griffiss QLAN is equipped with a $4.3\;\mathrm{m}$-length free-space telecom link on an optical table in the laboratory. This link is used for the testing and development of free space quantum network capabilities in a controlled environment before they are moved to outdoor free space links at the Stockbridge QLAN.

Currently, three of the four ring network links consist of indoor duplex SMF-28 fiber spools of up to 1 kilometer (km) length, while the fourth link contains an optical switch that routes signals through up to 15 km of outdoor buried fiber (duplex, SMF-28), as shown in Figure \ref{Figure 1}(a). The 15 km loop consists of a 5 km loop and a 10 km loop connected in series. These loops themselves consist of multiple loop-backs between the IAC and a second building at the Griffiss Technology Park (Figure \ref{Figure 1}(c)), where each link is 1.23 km length. The QLAN can easily be reconfigured so that the 5 km and 10 km loops are assigned to two separate ring segments, if desired. As well, the loop-back nature of the QLAN allows for straightforward reconfiguration; a different topology, such as a switch-based star topology, can be implemented straightforwardly with existing hardware, and will be the subject of future work. Additional buried fiber is in place throughout the Griffiss Technology Park to accommodate future QLAN expansion.

\subsubsection{Detection and analysis capabilities}
The Griffiss QLAN Detection and Analysis manifold includes ten superconducting nanowire single photon detectors (SNSPDs, Single Quantum and PhotonSpot), three time-to-digital converters (TTX, Time Tagger X and Time Tagger Ultra, Swabian Instruments), fiber-based interferometers for number-state, polarization, or time-bin encoded photons, polarization tomography modules, WaveShaper filters for time-energy Bell state analysis (Finisar WaveShaper 1000A and 4000A), an OTDR (Jonard OTDR-1000) and polarimeters (Thorlabs PAX1000, Qunnect QU-APC). In general, for quantum networking demonstrations, we will be focused on interference-based heralding gates between remote nodes. For this, we have fiber and free-space telecom-wavelength interferometers that are configured to measure photonic Bell pairs in the number-state, polarization, time-bin, or energy-time bases. An optical switch routes signals from the ring network to the appropriate analysis module for a given experiment. In the near term, the interferometers will be used in all-optical entanglement swapping demonstrations. Diagrams of these modules are included in the Supplementary Information.

Additional equipment specific to each qubit technology is present in the laboratory, despite not being covered here, such as pump lasers, radio-frequency (RF) sources, and other control and readout hardware, and can be integrated into the QLAN as needed. This setup maximizes the reconfigurability of the network, so that nodes can be assigned as Source, Memory, Analysis, or a combination of these functionalities simultaneously while maintaining the relative autonomy of the separate experimental setups.

\subsection{\label{Stockbridge topology} The Stockbridge Quantum Local Area Network}
The Stockbridge Test and Experimentation Facility sits atop a 2,300 foot mountain in Stockbridge, NY, located about 20 miles (30 km) southwest of the Griffiss and RRS QLANs, with line-of-sight access to the RRS. The Stockbridge site is equipped with a controllable ambient radio frequency spectrum, a central command building and about 30 ``Pads" \textemdash outdoor test locations \textemdash spread throughout the woods, a $40\;\mathrm{m}$ walkup tower, an airstrip, and other facilities for electromagnetic characterization of sensor and communication technology. The Pads include military-grade shelters and adjacent outdoor concrete and gravel areas. Shelters have electrical connectivity and basic heating and cooling to maintain a relatively stable temperature. Pads contain sets of twelve duplex SMF-28 cables that connect them to the Command Center and the rest of the site, and some have duplex multi-mode cables as well. SMF-28 aerial fiber runs up the walkup tower, which is equipped with a weather monitoring station. All aerial fiber is fully conduited, which effectively shields the links from dark counts due to evanescent coupling of light pollution \cite{Kucera2024}.

Some of the Pads with line-of-sight access to each other are connected via infrared free-space optical (FSO) links \cite{Nikulin2024, Nikulin2024a}. The FSO links have lengths of $100 - 300 \; \mathrm{m}$, are typically operated in the telecom C-band, and currently connect three pads together. Optical heads are mounted on the exterior of the pads and connected to their interiors via SMF-28 fiber. Additional optical heads are currently being fabricated to increase the number of simultaneous FSO links available. Air turbulence and crosswind speed are monitored by a scintillometer installed in parallel to one of the FSO links. The test site also has a weather monitoring station which tracks temperature and wind speed at all times. 

The Stockbridge QLAN hub-and-spoke topology is anchored at the ``Quantum Pad" (QPad), a network hub located near the central building and walkup tower. An optical switch (Dicon) routes signals from the QPad to the central building, then to the walkup tower and elsewhere throughout the site. One specific configuration connects the QPad to Pads A, B, C, D and E. Pads A-C have line-of-sight access to each other, as do Pads D and E. Pads A-E constitute network ``spoke" nodes within which a variety of quantum systems can be housed and connected to the QLAN. This topology is distinct from the Griffiss ring topology; the multi-hop architecture creates a flexible, reconfigurable mesh \cite{alshowkan2024multihop} which allows for advanced network functionalities such as link recovery. Additional buildings and Pads are available for future integration as intermediate hubs and end nodes.

The QPad is equipped with two single photon avalanche diodes (SPADs, ID Quantiqe ID210), 12 SNSPDs in the telecom O- and C-bands, as well as a tunable telecom laser (Keysight Instruments), entanglement sources, modulators, and fiber-based analysis modules. The SPADs are portable and can be moved to alternative network nodes throughout the QLAN. The QPad also features a Rubidium timing standard and a White Rabbit (WR) WR-Z16 switch to distribute 1 pulse-per-second (PPS) and $10\; \mathrm{MHz}$ signals throughout the QLAN. WR-LEN receiver nodes are located within the QPad and at other pads. At the Stockbridge QLAN, as at Griffiss, we have the choice of sending WR signals over the same fiber as experimental signals or over dedicated fiber. As currently configured, the WR signals are sent over a dedicated LAN fiber that runs parallel to the experimental fiber link. 

One of the network nodes (Pad A) is equipped with a serial bit error ratio tester (BERT, Agilent N4901) to monitor the classical bit error ratio of the network links. The BERT runs at a $10\;\mathrm{Gbps}$ (gigabits per second) data rate. A configurable pseudorandom binary sequence (PRBS31) pattern is used to create a data transmission pattern that can accumulate errors and report bit error rates and ratios. The BERT electrical output is connected to an FPGA evaluation board with a clock recovery circuit and a small-form-pluggable (SFP) transceiver port. The SFP optical output wavelength and power can be controlled via choice of transceiver and is multiplexed into the deployed fiber. PAD A is also equipped with a digital communication analyzer (DCA, Keysight N1000A DCA-X with 86105C module) which can plot all BERT data into a data eye diagram to measure the data stream's quality in real time.

\begin{figure}
    \includegraphics[width = 1\linewidth]{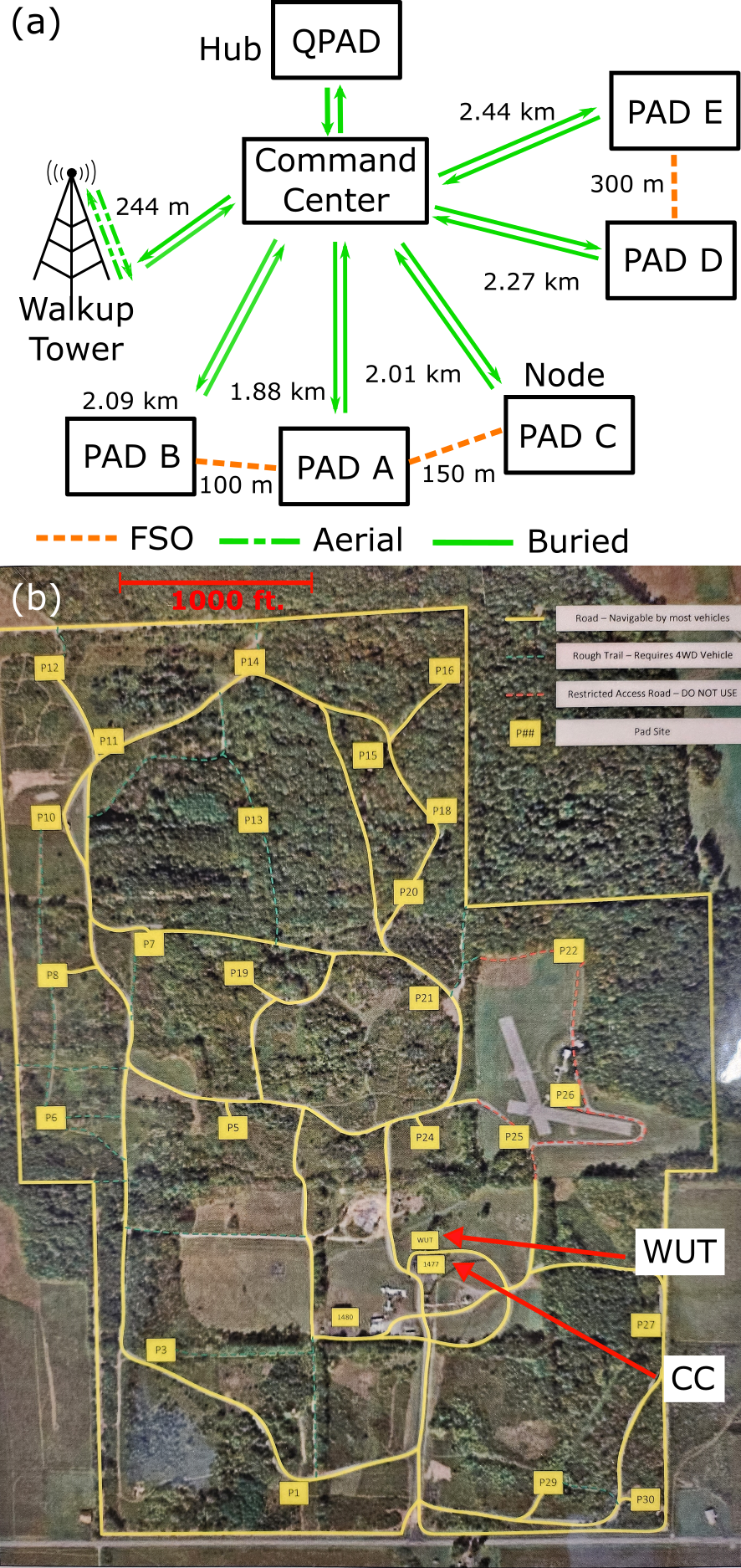}
    \caption{(a) Simplified network topology diagram for the Stockbridge QLAN. The Quantum Pad (QPad) serves as the hub in a multi-hop star configuration. The Command Center (CC) serves as an intermediate hub, where signals jump to/from the QPad to the CC then to the walkup tower or one of many possible network nodes. The link lengths as marked reflect the entire link length from the QPad to the end node, including hops through the CC. (b) Aerial image of the Stockbridge Test Site. Yellow lines denote navigable paths, and yellow squares denote pad locations. Note that the exact fiber paths and pads used in the QLAN are not marked. The CC and WUT are labeled.}
    %\vspace{128in}
    \label{Figure stock topology}
\end{figure}

\subsection{The Rome Research Site Quantum Local Area Network}
\begin{figure}
    \includegraphics[width = 1\linewidth]{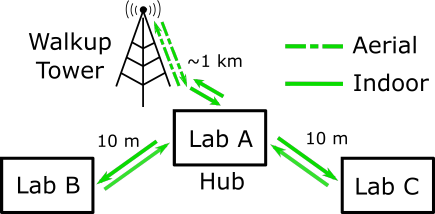}
    \caption{Simple diagram of the physical layout of the RRS QLAN. This setup lends itself well to either a ring or star-like network topology. Labs A, B and C are three adjacent labs located in the RRS Extreme Computing Facility.}
    %\vspace{128in}
    \label{Figure RRS network layout}
\end{figure}

Down the street from the Griffiss QLAN, we are implementing a third QLAN at the Rome Research Site. The RRS QLAN has a similar setup to the Griffiss QLAN, in that it has a network hub that is located in the central laboratory of a set of three quantum networking laboratories, which we refer to as Labs A, B and C, which are home to superconducting qubit and transducer experiments, quantum integrated photonics experiments, other aspects of quantum interconnect development including device packaging, robust and efficient routing of light and microwaves, and novel materials characterization. As in the IAC quantum networking laboratory, the superconducting setup includes a dilution refrigerator (Bluefors LD) outfitted with infrared- and visible-wavelength optical fibers and custom-installed SNSPDs. Fiber conduits connect the three laboratories together, which operate in adjacent rooms within one facility. The network hub also connects to four fiber circuits that run outdoors to aerial links, which run up a walkup tower, then loop back. This walkup tower serves as an end node for a line-of-sight free space optical link to the Stockbridge test site. The simple network layout diagram in Figure \ref{Figure RRS network layout} does not specify a network topology (e.g. ring or star-like), because with OADMs and optical switches available, we can implement either a Griffiss QLAN-like ring or a star-like topology and toggle between them quickly.

\section{\label{Fiber Characterization}Link Characterization}
Having introduced the three QLANs, we will dive into extensive network characterization measurements that have been carried out to benchmark and compare the performances of the QLANs with respect to each other and other quantum network testbeds.

\subsection{\label{OTDR Loss} Link losses}
We used an optical time-domain reflectometer (OTDR, Jonard) to characterize the loss of the deployed fiber loops in the three QLANs. At Griffiss, the measured total link loss is about $6-8\;\mathrm{dB}$ for both the $5 \; \mathrm{km}$ and $10 \; \mathrm{km}$ deployed buried links, which is unsurprising given that the links each consist of multiple loop-backs. The fibers themselves (SMF-28) are low-loss, with average loss values of $0.2 \;\mathrm{dB/km}$. Overall link loss can be mitigated in a straightforward manner by replacing lossy connectors with lower-loss connectors or splices. However, it is beneficial to have testbed links with realistic levels of loss due to imperfect connectors. 

The buried and aerial fiber links in the Stockbridge QLAN were characterized using the same OTDR. Measurements show that the fiber links are low-loss, with average loss values of about $0.2 \;\mathrm{dB/km}$, with link lengths ranging from about $1.5-3.2\;\mathrm{km}$ throughout the site. A $500\;\mathrm{m}$ aerial link was also measured and exhibits a loss of $0.233 \;\mathrm{dB/km}$. In general, these low-loss links allow for quantum signal transmission, potentially with multiple loop-backs to extend propagation distance if desired. 

Finally, the RRS QLAN outdoor aerial links were OTDR-tested. Each of the four circuits is about $1\;\mathrm{km}$ in length and has a loss value in the $6-10 \;\mathrm{dB}$ range, where the loss comes from adapters that connect the fiber loop through the laboratory and up and down the walkup tower. Loss will be mitigated in the future by decreasing the number of connectors used and implementing splices. Example OTDR plots can be found in the Supplementary Information.

The link losses measured at the Griffiss and RRS QLANs are similar to values measured in other quantum network testbeds, where readily usable deployed fiber links tend to carry $10\;\mathrm{dB}$ or more of loss. In this context, the Stockbridge QLAN has a noticeably lower loss value due to its uninterrupted fiber lengths of up to $3.2\;\mathrm{km}$ before any loopbacks. Together, the variety of loss environments allows us to vet quantum technologies in QLANs within a spectrum of different constraints. Technologies with differing loss tolerances can each be tested and pushed forward in a proper environment, and pushed to more challenging loss conditions when desired.

\subsection{Polarization drift and compensation}
The drift of the state of polarization (SOP) of an optical signal traversing a network link is a standard benchmark metric used to understand the stability of that link. Here, we report SOP monitoring results across the Griffiss, RRS and Stockbridge QLANs.

\subsubsection{Griffiss QLAN}
\begin{figure*}
    \includegraphics[width = 1\linewidth]{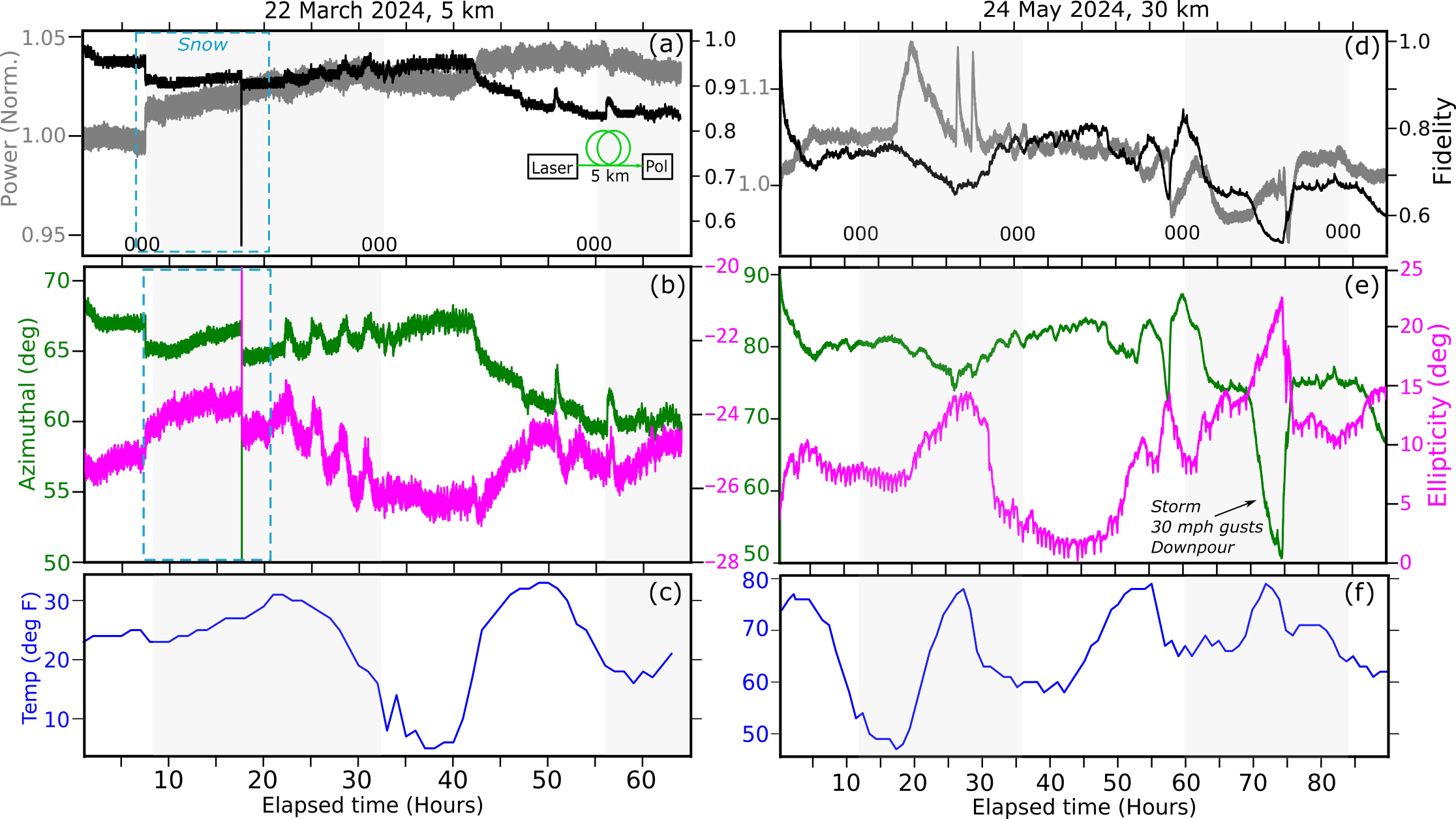}
    \caption{Example SOP drift measurements over Griffiss QLAN buried fiber in (a-c) March 2024, in wintry weather, and (d-f) May 2024, in summery weather. The top panels show the transmitted power (gray) and SOP fidelity (black) over time, while the center panel shows the SOP azimuthal (green) and elliptical (pink) angles, and the lower panels show the temperature at the Griffiss site during the experiments.}
    %\vspace{128in}
    \label{Figure griffiss sop drift}
\end{figure*}

To understand the effect of environmental perturbations on signals propagating over deployed fiber links, we monitored the transmitted power and SOP of a classically bright 1550 nm signal propagating over Griffiss buried fiber. A polarimeter is placed at the output of the experiment. Results are shown for two different experiments in Figure \ref{Figure griffiss sop drift}: for (a,b) 5 km and (c,d) 30 km deployed fiber length, without polarization compensation, at different times of the year. Data points are acquired at a rate of $1 /\mathrm{sec}$ during both experiments. Weather conditions are logged by a weather station located at the Griffiss International Airport \cite{weatherunderground2025}. The data show that the SOP is relatively stable against changes in environmental conditions. For example, snowfall occurred overnight March 22nd-23rd, during the experiment shown in (a,b). The sharp features occurring between 20-30 hours elapsed may correspond to snow plows clearing accumulated snow from the roadway. A second experiment is shown in Figure \ref{Figure griffiss sop drift} (c,d). Here, the SOP was monitored over 30 km of deployed fiber (15 km in either direction, with both paths looped back). The SOP drift is more significant in this experiment. A storm occurred on 27 May, and a downpour occurring between 70-75 hours elapsed appears to have a significant impact on the SOP of the link. Apart from the polarization, the transmitted power over the deployed fiber shown in Figure \ref{Figure griffiss sop drift} (a,c) holds steady, with a maximum variation of $<5\%$ for the 5 km link and a maximum variation of $<15\%$ for the 30 km link. The transmitted power through the links varies by up to $5\%$ (5 km, panel (a)) and up to $15\%$ (30 km, panel (c)) which indicates that there is a degree of polarization-dependent loss in the links. Results from Pearson linear correlation analyses are available in Tables \ref{tab: TOF correlations} and \ref{tab: SOP correlations} below.

\begin{figure}
    \includegraphics[width = 1\linewidth]{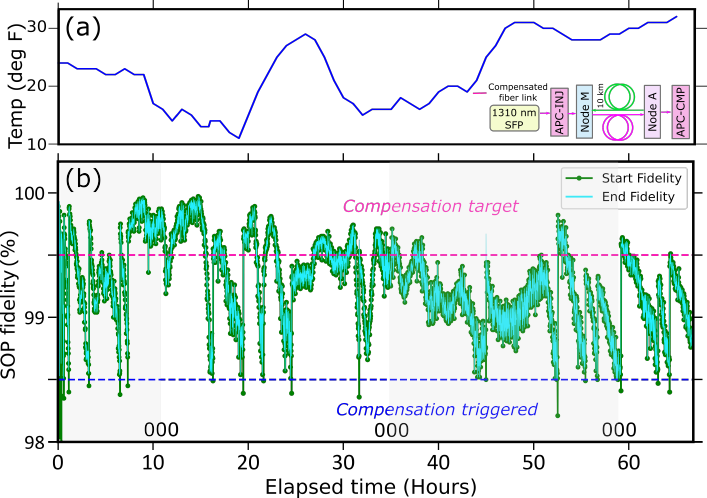}
    \caption{(a) Temperature over time over a period stretching from 13-16 December 2024. Inset: experimental setup. (b) Compensated fiber SOP fidelity. The red horizontal line denotes the target fidelity of 99.5\% and the blue line denotes the value, 98.5\%, at which a compensation is triggered. Over a nearly three-day period, compensation was triggered only 17 times.}
    %\vspace{128in}
    \label{Figure griffiss fiber comp}
\end{figure}

An automated polarization compensator (QU-APC, Qunnect\cite{Craddock2024}) is installed within the QLAN, and can be connected to either the clockwise or counter-clockwise ring loop, so that signals propagating through that ring undergo polarization compensation. See Figure \ref{Figure griffiss fiber comp} for an example experiment over 10 km deployed fiber in December 2024, where the QU-APC is used to stabilize the SOP while a classically bright O-band signal propagates over the link. Here, the SOP fidelity is maintained within $98.5\%$ over about three days. During this time period, compensation was triggered only 17 times, which indicates the relative SOP stability of the Griffiss QLAN buried fiber. Overall, the data shows that we can maintain an SOP fidelity of $\sim99\%$ or better for multiple days with ease, paving the way for quantum networking demonstrations with polarization-encoded photons. A second APC can be added, or time-multiplexing can be used, to allow for simultaneous compensation of both loops of the ring network.

\begin{figure}
    \includegraphics[width = 1\linewidth]{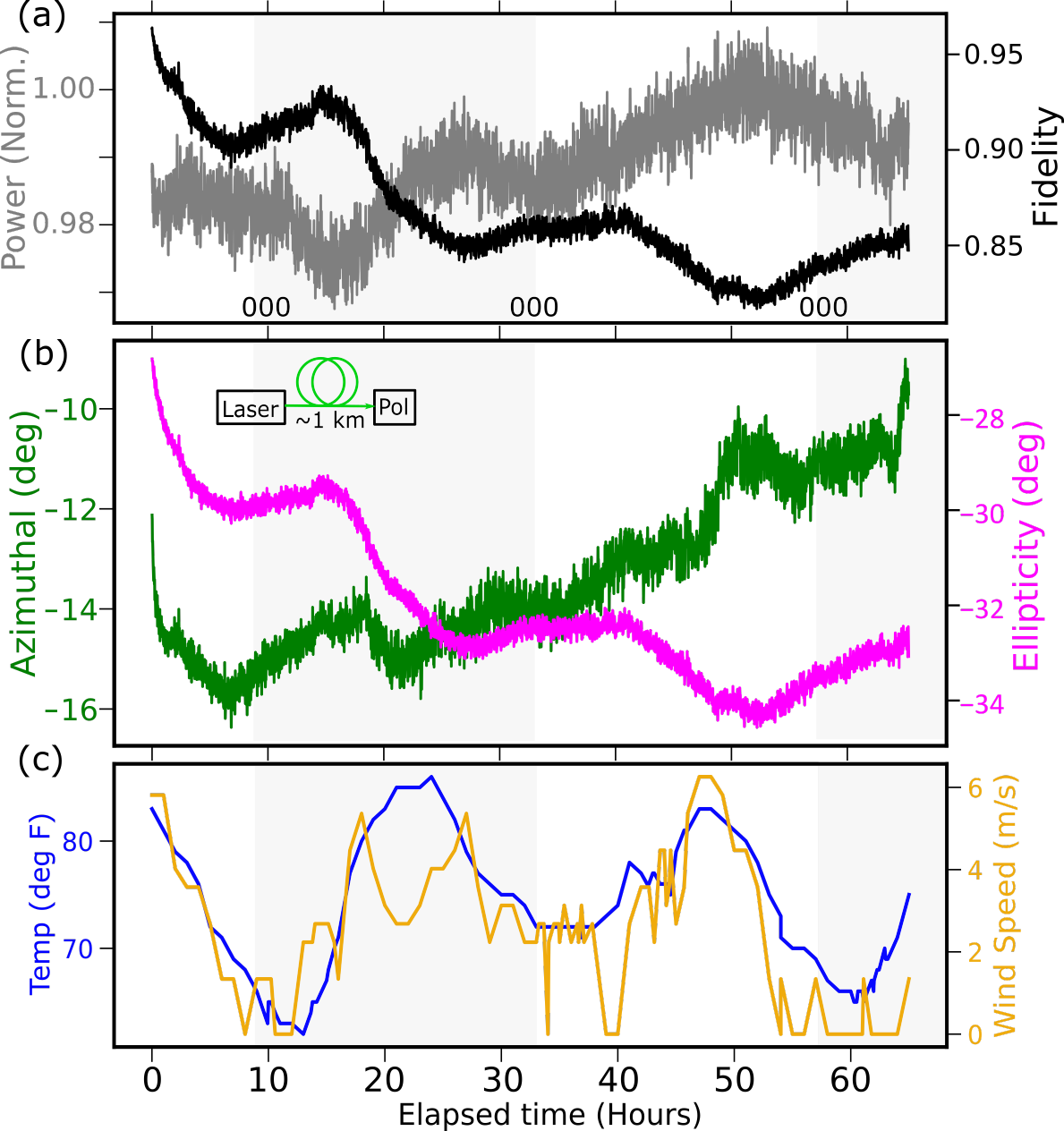}
    \caption{SOP drift measurement taken over ~1 km RRS link in July 2025. (a) SOP fidelity (black) and normalized power transmission (gray). (b) Azimuthal (green) and elliptical (pink) angle drift. (c) Temperature and wind speed data taken from the Griffiss International Airport.}
    %\vspace{128in}
    \label{Figure RRS SOP drift}
\end{figure}

\subsubsection{RRS QLAN}
In a similar manner, we monitored the SOP drift over the RRS indoor/aerial fiber link over multiple days. SOP/power drift results and associated weather data are shown in Figure \ref{Figure RRS SOP drift} for an experiment performed in July 2025. Despite the aerial nature of the link, the SOP and transmitted power are relatively stable, especially compared to aerial fiber drifts observed at the Stockbridge QLAN (below). The higher degree of stability at the RRS QLAN may be attributed to the placement of the walkup tower on which it is mounted: the tower is tucked behind a building in a settled area, unlike the walkup tower at Stockbridge that is standing on its own on a high hilltop.

\subsubsection{Stockbridge QLAN - fiber}
\begin{figure}[!htbp]
    \includegraphics[width = 1\linewidth]{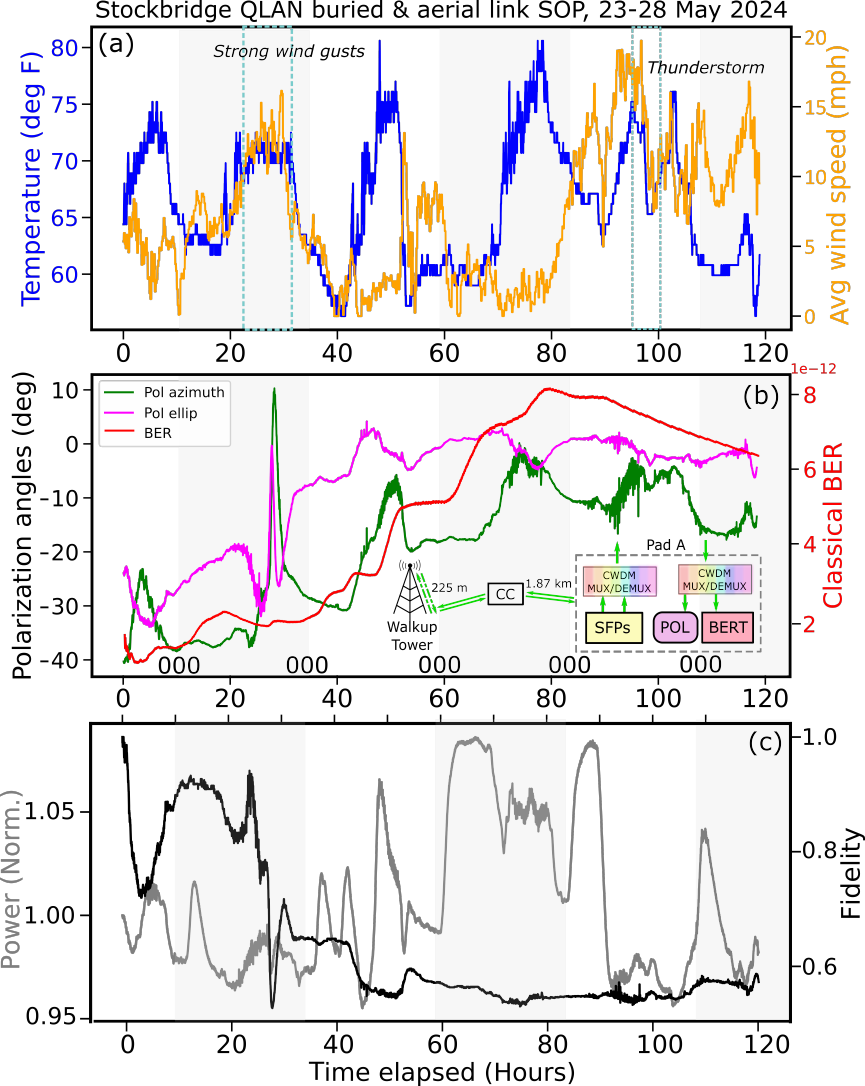}
    \caption{Fiber SOP and weather monitoring experiment at the Stockbridge QLAN for a combination buried-aerial link. (a) Weather data (b) polarization angle and bit error ratio data (c) transmitted power and calculated SOP fidelity over time.}
    %\vspace{128in}
    \label{Figure stock BA fiber pol}
\end{figure}

The Stockbridge QLAN fibers' SOP drift is characterized using a similar setup as used for the Griffiss QLAN. An example measurement incorporating a buried link and an aerial link is summarized in Figure \ref{Figure stock BA fiber pol}. $1550\;\mathrm{nm}$ and $1570\;\mathrm{nm}$ signals from SFP transceivers are multiplexed into the $1.87\;\mathrm{km}$ buried link connecting PAD A to the Command Center, then routed to the walk-up tower and looped back to PAD A and de-multiplexed. The $1550\; \mathrm{nm}$ signal output is sent to a polarimeter and the $1570\; \mathrm{nm}$ signal is sent to a serial BERT (bit error rate tester, Agilent N4901) for bit error ratio measurements. The (classical) bit error rate is a measure of signal integrity in telecommunication network links: it quantifies the number of transmitted bits that are received incorrectly. The bit error ratio, in contrast, measures the number of bit errors compared to the total number of bits received over a given time window. Bit errors in fiber links can arise due to attenuation, physical damage, chromatic/polarization dispersion, and noise from external signals, among other causes.

A weather station located on the walk-up tower records temperature and average wind speed. As can be seen in Figure \ref{Figure stock BA fiber pol}, the SOP exhibits a slow diurnal drift pattern that follows the regular increases and decreases in temperature that occur throughout the day - most notably the azimuthal angle. Fast changes in the SOP occur during a thunderstorm, along with a higher observed bit-error ratio. A sharp change in the SOP occurs at around 30 hours elapsed and can likely be attributed to a set of strong wind gusts in the same time period. The SOP of the same buried link was measured without the aerial link, as shown in the Supplementary Information, and exhibits a similar diurnal drift. Overall, the SOP behavior of the Stockbridge QLAN fibers differs significantly from that of the Griffiss QLAN. This difference can be attributed to the distinct environments in which the fiber is placed: the Griffiss buried fiber is located underneath paved roads, while the Stockbridge fiber is buried approximately $18$ inches underground in a conduit, with little insulation from environmental perturbations. Having access to these two different outdoor environments provides a diverse array of conditions for our testbeds, allowing us to better engineer, optimize and test robust quantum network devices, as demonstrated in the correlation analysis below. 

\subsubsection{Stockbridge QLAN - free space}
In addition to monitoring fiber links, SOP drift and BER were also measured over a stationary 100 meter free space optical (FSO) link. As shown in Figure \ref{Figure stock topology}, we currently have four FSO links in place, which can be used to, for example, connect pads A, B and C. The A-B link is 100 meters in length, while the A-C link is 150 meters in length. The A-C link is also equipped with a scintillometer (Scintec BLS Series) which continuously logs air turbulence and crosswind speed over the link. Light is transmitted and received by simple optical heads without adaptive optics or pointing-and-tracking functionalities, which are under development at AFRL\cite{Lanning2021} for daytime links, and will be explored and implemented in the near future. 

\begin{figure*}[!htbp]
    \includegraphics[width = 0.8\linewidth]{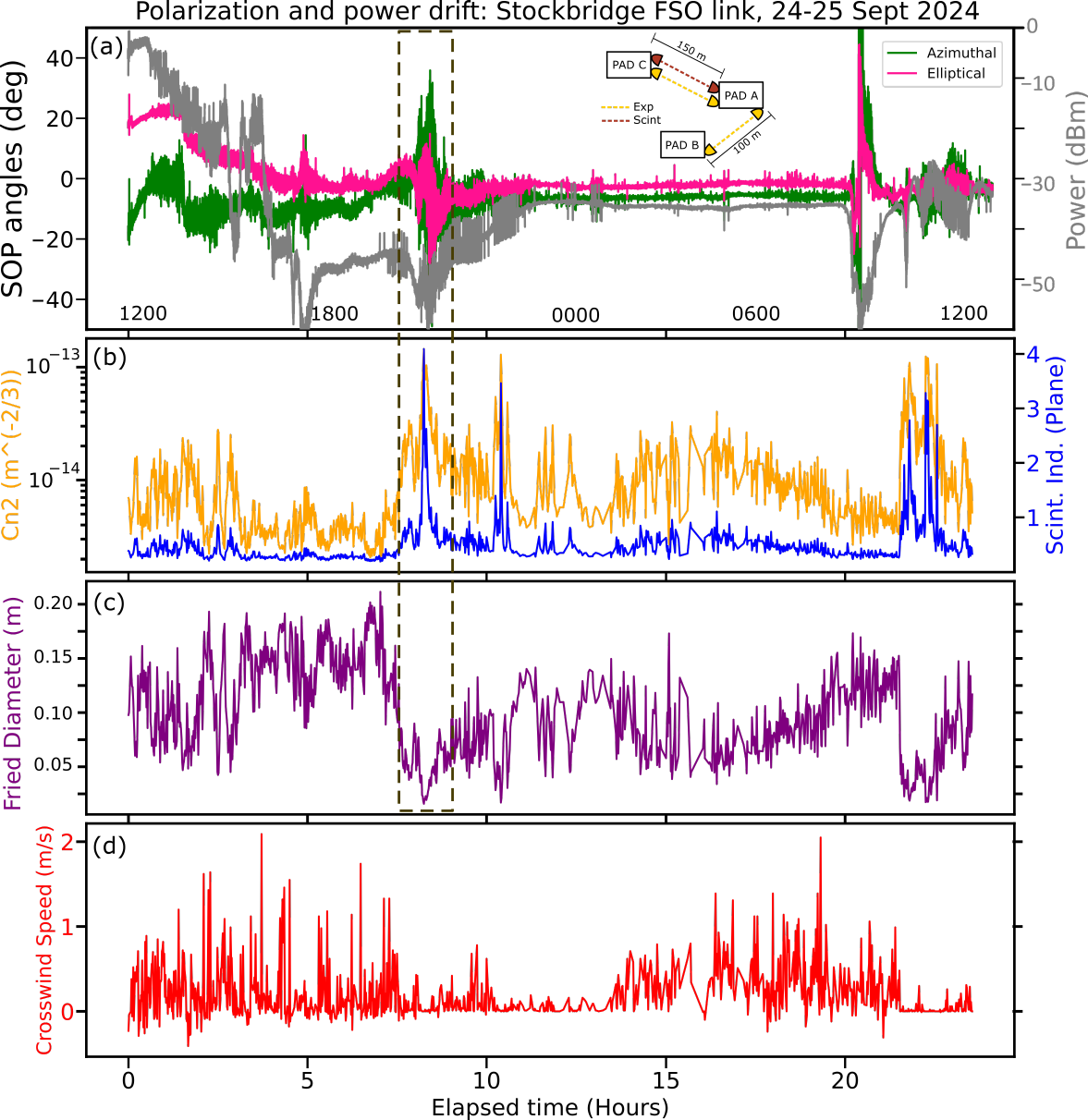}
    \caption{SOP and air turbulence measurements over a 100 meter FSO link at the Stockbridge QLAN over 24 hours. The polarization angles and transmitted power are shown in (a). Inset: simplified experimental diagram. Here, the scintillometer is measuring a path (red) nearly perpendicular to the experimental path (yellow, from Pad A to Pad B). Data measured by the scintillometer, including (b) the scintillation index and refractive index structure parameter $C_n^2$, (c) the Fried parameter, and (d) the crosswind speed are plotted for the same period.}
    %\vspace{128in}
    \label{Figure stock FSO}
\end{figure*}

Air turbulence can be characterized by the scintillation index ($\sigma_I$), which measures the normalized intensity variance caused by atmospheric turbulence:
\begin{equation}
    \sigma_I = \frac{\langle I^2 \rangle}{\langle I\rangle^2} -1
\end{equation} 
where $I$ is the received optical irradiance. A value of $10^{-3}-10^{-2}$ indicates weak scintillation, a value of $10^{-2}-10^{-1}$ indicates moderate scintillation, and a value of $10^{-1}-10^{0}$ indicates strong scintillation. More commonly, turbulence is measured via the structure function constant of refractive index fluctuations ($C_n^2$). The refractive index structure function  $D_n(\rho)$ measures the refractive index variations of the atmospheric channel as a function of radial distance $\rho$:
\begin{equation}
   D_n(\rho) = \langle n[\vec{r},t] - n[\vec{r}+ \rho, t] \rangle = C_n^2 |\rho|^{2/3} 
\end{equation}
for refractive index $n$. For terrestrial links, typical $C_n^2$ values fall in the $10^{-16}$ to $10^{-12} m^{-2/3}$ range, where values of $10^{-15}m^{-2/3}$ and below indicate stable or calm atmospheric conditions, and values of $10^{-13}m^{-2/3}$ and above indicate a highly turbulent atmosphere and the potential for considerable signal distortions \cite{tunick2005}. In addition to $\sigma_I$ and $C_n^2$, the Fried diameter $r_0$ measures the quality of optical transmission of a specific wavelength over a free space channel: it is the diameter of a circular area over which the root-mean-squared wavefront aberration due to passage through the atmosphere is equal to 1 radian. Typical $r_0$ values range from $5\;\mathrm{cm}$ (poor) to $20\;\mathrm{cm}$ (excellent). An ideal low-turbulence link will have $\sigma_I$ and $C_n^2$ as low as possible, and $r_0$ as high as possible. 

Results from a 24-hour FSO logging experiment are shown in Figure \ref{Figure stock FSO}. Link SOP and transmitted power are shown in panel (a), atmospheric turbulence data is shown in panels (b,c), and crosswind speed and site temperature are shown in (d). The transmitted power drops off steeply during the day due to temperature-induced misalignment of the optical heads, even for minor temperature changes, demonstrating that active alignment is needed to maintain low-loss transmission. Since these data were taken, the optical heads have been ruggedized to withstand temperature drift, and now can passively maintain alignment over multiple days within $\sim 10\;\mathrm{dB}$ of the original transmission value. Furthermore, we ace actively developing pointing and tracking solutions for active link alignment over long time periods. Even with high link loss, we can gather insightful SOP and turbulence data. 

Surprisingly, the data reveal exceptionally strong turbulence over the short FSO link. Indeed at certain times, the scintillation index value peaks at $3$ or $4$, and the corresponding $C_n^2$ sits in the high $10^{-13}$ range. Possible causes for this strong turbulence, given that weather monitoring rules out any storms, could have to do with ground effects and the densely wooded environment of the testbed. For example, as the temperature starts to decline between 5-10 hours elapsed, in the evening, a strong spike in turbulence is observed, along with an increase in fast SOP variations and an associated drop in transmission power, perhaps due to turbulence-induced beam wandering. 

Pads A, B and C, and the line-of-sight links between them, are located in small clearings within a dense temperate forest environment. The measurements in Figure \ref{Figure stock FSO} took place in September, before trees' autumn leaf shedding. Forest canopies are known to host complex air flow phenomena such as Kelvin–Helmholtz instability, among other effects \cite{Smyth2023}, and forest clearings add additional complexity to air and heat flows \cite{Lee2000}, which can contribute to locally strong turbulence. As such, the Stockbridge QLAN's free space optical links provide a unique platform for fascinating studies of quantum and classical signal propagation in a temperate forest, including the potential for extreme air turbulence that is natural to this environment. In the 5-10 hours elapsed window,  a quick change in site temperature may induce locally strong turbulence via differential cooling rates between the clearing and the canopy, thermal instability, and enhancement of these effects from the geometric confinement of the clearing that concentrates and accelerates air flows.

As compared to optical fiber SOP, the FSO link SOP remains quite stable in the overnight hours, with the elliptical and azimuthal angles holding steady within $\pm 5^{\circ}$. Additionally, in a time window of about 7-9 hours elapsed, in the evening, the air turbulence spikes, with a corresponding decrease in the Fried diameter. At this time, the SOP varies significantly, and the transmitted power drops below the measurement threshold. The SOP measurement is not reliable when the power drops below the noise floor. This feature may be indicative of significant turbulence spatially shifting the beam, so that it doesn't hit the receiving aperture, while also possibly inducing short-timescale SOP jitter. Comparing the FSO SOP to fiber SOP results shows that both links are susceptible to polarization instability, albeit with different causes. If polarization qubits are propagating through the QLAN, polarization stabilization will be necessary for high-fidelity transmission over both types of links.

\subsection{Polarization stabilization of free space link}
\begin{figure}[!htbp]
    \includegraphics[width = 1\linewidth]{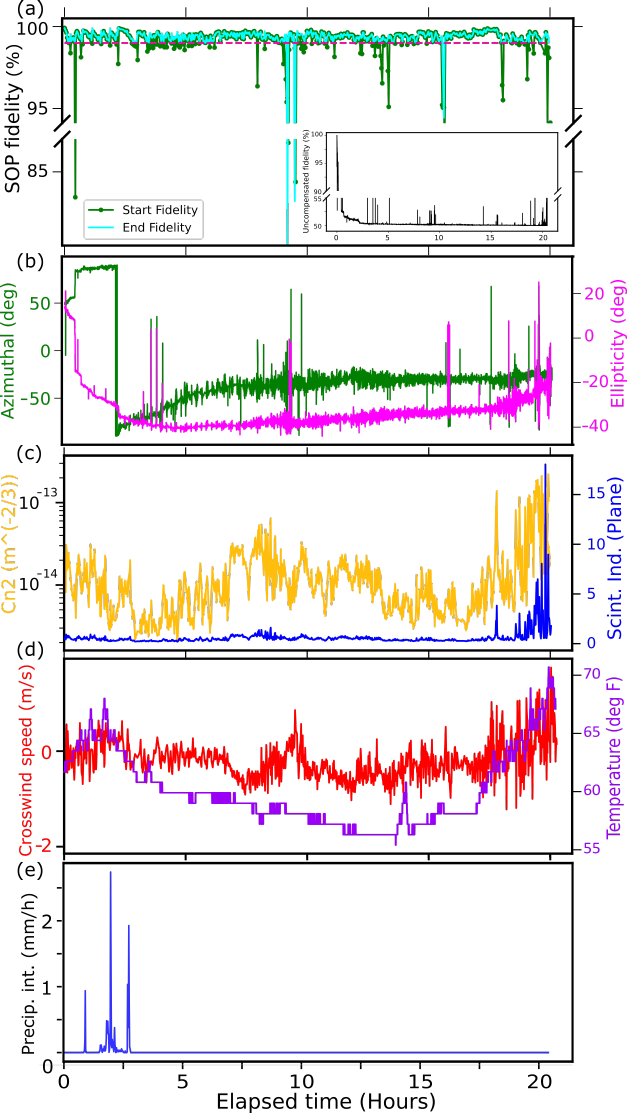}
    \caption{(a) Compensated SOP fidelity of network link with $4\;\mathrm{km}$ buried fiber and $100\;\mathrm{m}$ free space propagation at the Stockbridge QLAN from 29-30 May 2025. Inset: uncompensated SOP fidelity (black) and normalized transmitted power (gray, log scale). (b) Drift of elliptical and azimuthal angles. (c) Local $c_n^2$ (yellow) and scintillation index (blue) and (d) local crosswind speed (red) and site temperature (purple) give some hints as to how environmental perturbations affect SOP. (e) Disdrometer-captured precipitation intensity data reveals a rain event occurring simultaneously with a disturbance in the uncompensated SOP.}
    %\vspace{128in}
    \label{Figure stock FSO comp}
\end{figure}

To this end, we implemented polarization compensation over our free space link. As shown in Figure \ref{Figure stock FSO comp}, we use a QU-APC located in the QPad to run the compensation. A small percentage of the experimental signal is tapped off before compensation and sent to a separate polarimeter to log uncompensated SOP drift. Weather and air turbulence are monitored simultaneously. For the experiment shown here, the SOP and compensation are logged for 20 hours. The QU-APC is set to trigger a compensation if the SOP fidelity falls beneath $99\%$, it is compensated back to $99\%$. Over this time period, compensation is triggered frequently. The SOP fidelity falls below 99\% 102 times and is successfully compensated back to 99\% or higher 101 times on the first attempt. Note that a large disturbance, likely due to a rain event at about 2 hours elapsed, as measured by our disdrometer, causes the uncompensated SOP fidelity to plummet, but does not affect our ability to compensate within 99\% fidelity. The total link uptime for this measurement is 99.9\% (total downtime of about 67 seconds over approximately 20 hours). Polarization compensation succeeds even in the presence of rain and strong local air turbulence, which spikes in the last couple hours of the experiment and causes the amount of transmitted power to drop significantly. 

\subsection{Time of flight drift}
Time-of-flight (TOF) drift, e.g. the change in the time it takes for a photon to traverse a path induced by environmental perturbations, is another important figure of merit for quantum network links \cite{Bersin2024}. We measure the TOF drift at the Griffiss and Stockbridge QLANs by using an acousto-optic modulator (AOM) to generate 10-nanosecond duration pulses at 1550 nm. The pulsed signal is split by a fiber beamsplitter. One output is sent directly to a single photon detector inside the laboratory, and the second output is sent through the deployed fiber before reaching a second single photon detector. A time-to-digital converter (TTX) connected to the single photon detectors generates coincidence histograms. The coincidence peaks are fit with Gaussian functions to extract the central time delay $\tau$ between the two pulses. The TOF drift $\Delta \tau$ is defined as the change in the central time delay from its original value $\tau_o$: $\Delta \tau = \tau - \tau_o$. TOF drift measurements have not yet been performed at the RRS QLAN, but are planned for the near future.

\subsubsection{Griffiss QLAN}
\begin{figure}[!htbp]
    \includegraphics[width = 1\linewidth]{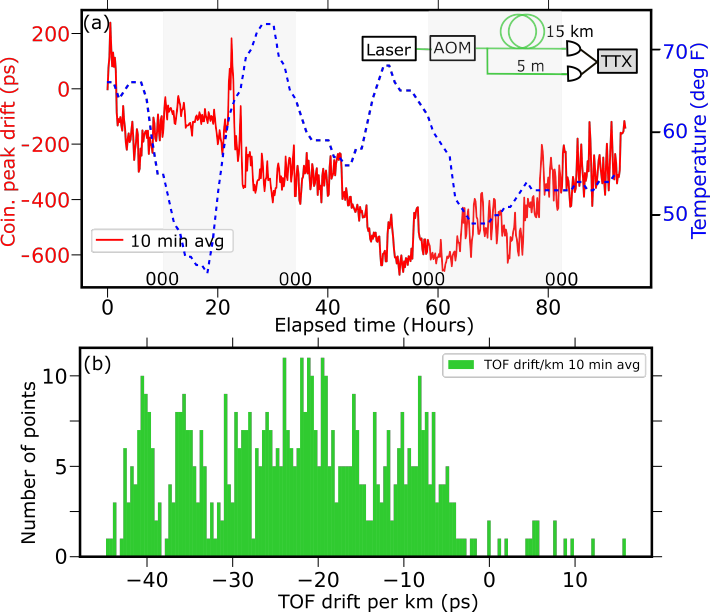}
    \caption{Results from a four-day TOF drift measurement on 15 km length buried fiber at the Griffiss QLAN on 2-6 May 2024. (a) Local temperature and TOF drift in ps, averaged over 10-minute intervals. (b) Histogram of averaged drift values in ps/km over the duration of the experiment, showing a very small amount of drift per km.}
    %\vspace{128in}
    \label{Figure griffiss TOF}
\end{figure}

Results from a four-day TOF drift experiment at the Griffiss QLAN are shown in Figure \ref{Figure griffiss TOF}. TOF drift is logged along with local temperature for 10-ns pulses traveling through 15 km of deployed fiber. Over four days, the total drift $\Delta \tau$ remains below $700\;\mathrm{ps}$, and the drift per km is below $50\;\mathrm{ps/km}$ for a maximum temperature change of $-23^{\circ }\mathrm{F}$ during the measurement period. The sub-nanosecond drift over such a long distance confirms the time-domain stability of the Griffiss fiber link over a long period, which holds promise for high-fidelity transmission of time-bin qubits in this QLAN without the need for active TOF compensation. The coincidence peak has a width of about $7.3\;\mathrm{ns}$ for the duration of the experiment. If we use this data to extract a $\Delta  \tau$ value normalized to path length and temperature changes, we get a relationship of $\Delta \tau = -0.27\mathrm{\frac{ps}{km \cdot K}}$ ($ R = -0.09, R^2 = 8.1\times 10^{-3}$), which is a significantly smaller than the value of $\Delta \tau = 37.4\mathrm{\frac{ps}{km \cdot K}}$ reported in \cite{Kucera2024} over a $14.4\;\mathrm{km}$ deployed link with $2.56\;\mathrm{km}$ aerial fiber.  This comparison shows the extent to which buried fiber is more insulated from temperature changes, and hence more stable in the time domain as compared to aerial fiber.  However, the $R$ and $R^2$ values for this fit are very low, showing that the temperature correlation is very weak, and the correlation explains less than $1\%$ of the data variance. To gain a more complete picture, we perform detailed linear correlation analysis below.  Also see below for an additional TOF drift measurement at the Griffiss QLAN that is implemented using two networked time taggers and White Rabbit timing synchronization.

\subsubsection{Stockbridge QLAN}
\begin{figure*}[!htbp]
    \includegraphics[width = 1\linewidth]{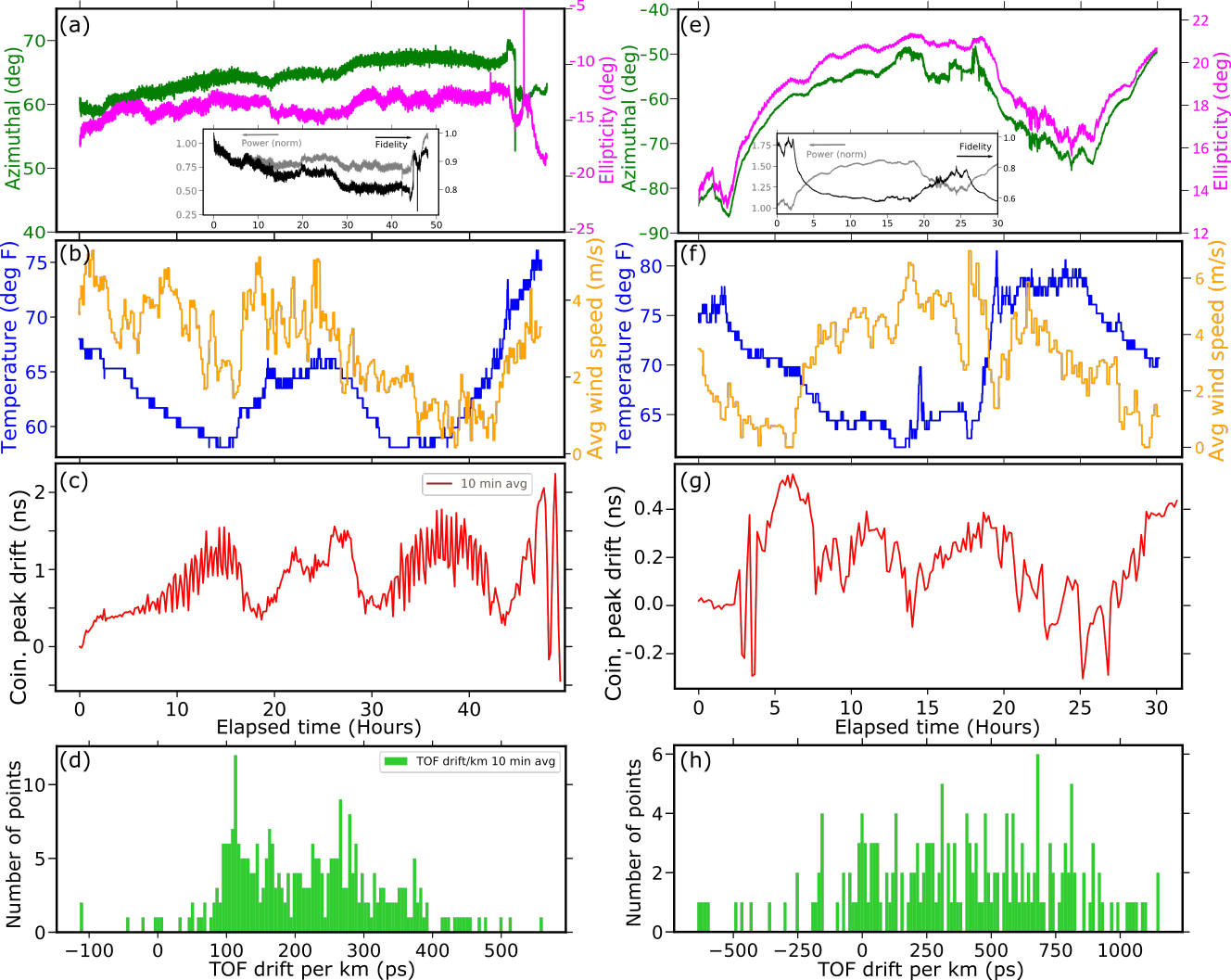}
    \caption{Full data set of simultaneous time of flight (TOF) and SOP drift measurements of (a-d) a buried 4 km fiber loop and (e-h) an aerial 450 meter aerial fiber loop at the Stockbridge QLAN. }
    %\vspace{128in}
    \label{Figure stock TOF}
\end{figure*}

TOF drift results for the Stockbridge QLAN, for buried and aerial fiber links, are shown in Figure \ref{Figure stock TOF}. Unsurprisingly, the TOF drift for the Stockbridge fibers is significantly higher than that measured at the Griffiss QLAN. A 4 km buried fiber link exhibits TOF changes that closely mimic changes in local temperature. The TOF drift sits mostly between values of $-100-500\;\mathrm{ps/km}$. On the other hand, the aerial fiber link exhibits large TOF drift, sitting in the range of about $-500-1000\;\mathrm{ps/km}$, that does not track temperature changes as closely.  

If we again extract relationships of $\Delta \tau$ to temperature changes with a simple correlation fit as in \cite{Kucera2024}, we obtain $\Delta \tau = -1.78\mathrm{\frac{ps}{km \cdot K}}$ ($R = -0.048, R^2 = 0.002$) for the buried link, much smaller than the sensitivity observed in the buried-aerial link in \cite{Kucera2024}, and $\Delta \tau = -0.05\mathrm{\frac{ps}{km \cdot K}}$ ($R = -0.3084, R^2 = 0.0951$)  for our aerial link, surprisingly the lowest sensitivity we measured in any of our links, and significantly lower than \cite{Kucera2024}. There are two potential explanations for this discrepancy: first, the average wind speed at our walkup tower impacts the TOF drift, which isn't accounted for here, and second, our walkup tower aerial fiber is constrained to a larger extent than typical hanging aerial fibers, which may reduce the amount of stress variations due to temperature changes. Nevertheless, such significant TOF drifts, regardless of their environmental cause, will require active path length stabilization for network functions that rely on high-speed and high-fidelity propagation of time-bin encoded photons, or on long-term TOF stability for phase-sensitive applications such as distributed quantum sensing. To supplement this initial analysis and understand how well our TOF drift data fits to a linear correlation with temperature and wind speed, we include more detailed linear correlation results below. 

\section{White Rabbit-based timing synchronization}

White Rabbit precision time protocol (WR) is an ethernet-based time-transfer protocol that can achieve timing synchronization between remote network nodes with sub-nanosecond accuracy and few-picosecond jitter\cite{Lipinski2011, Rizzi2018}. As a commercially available solution, it is commonly used within quantum network testbeds \cite{Alshowkan2022, McKenzie2023}. 

Both the Griffiss and Stockbridge QLANs are equipped with WR switches (WR Z-16, Z-16 LJ) and WR standalone/receiver nodes (WR-LEN, Safran Trusted 4D Inc.) for fiber and free space network timing synchronization. At both QLANs, a timing distribution node (WR Z-16 or WR-LEN) distributes 1 pulse-per-second (PPS) and $10\;\mathrm{MHz}$ signals from a Rubidium timing standard (FS725, Stanford Research Systems) to WR-LENs. The receiver nodes are then connected to equipment at each network node, such as digital delay generators, modulators, time-to-digital converters, and more. 

A variety of WR fiber arrangements have been tested,including propagating WR signals over (a) separate, dedicated fiber links and (b) shared, wavelength-multiplexed fiber links with experimental signals. A WR link can be established through up to $10\;\mathrm{km}$ distance in the Griffiss deployed fiber loop using $80\;\mathrm{km}$ range transceivers before link loss becomes prohibitive. A WR link has been established using the $1310/1490\;\mathrm{nm}$,  $1490/1570\;\mathrm{nm}$ and $1490/1550\;\mathrm{nm}$ wavelength pairs. In the future, we can achieve WR synchronization over longer links by replacing lossy connectors. Details about these experiments at the Griffiss QLAN can be found in the Supplementary Information. We have also implemented WR over an indoor free space link of $4.3\;\mathrm{m}$ at the Griffiss QLAN. A detailed discussion of these experiments is reserved for future work. 

\begin{figure}[!htbp]
    \includegraphics[width = 1\linewidth]{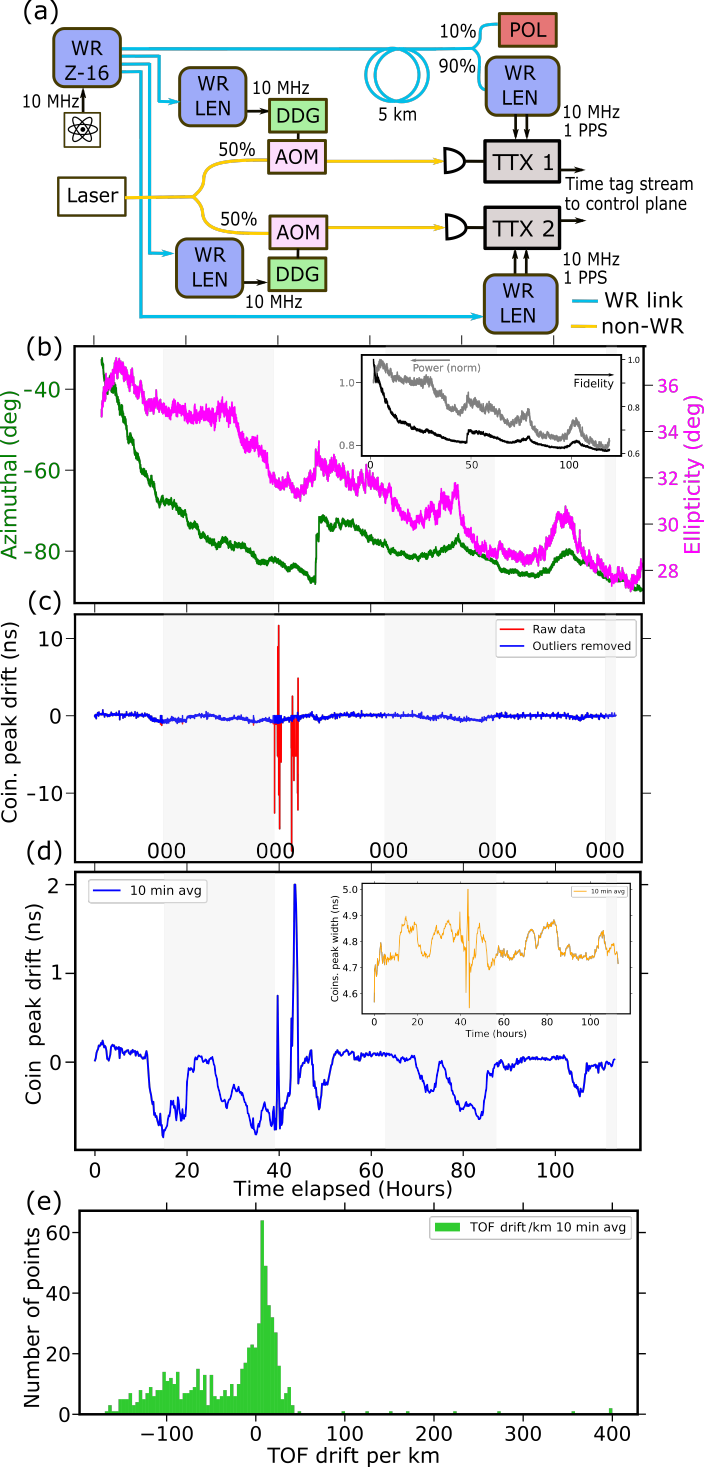}
    \caption{Distributed coincidences with White Rabbit. (a) Experimental setup. A WR-Z-16 distributes a $10\; \mathrm{MHz}$ signal to WR-LEN nodes which synchronize two digital delay generators (DDGs) that drive acousto-optic modulators (AOMs), producing 10 ns duration 1550 nm pulses, which are incident upon SNSPDs connected to two time taggers (TTXs). A fiber splitter taps 10\% of the WR signal for SOP monitoring. (b) SOP drift. Inset shows the normalized power and SOP fidelity. The drift of the center of the coincidence peak is shown, unaveraged, with and without outliers in (c). Outliers are removed then the data is averaged over 10-minute windows in (d). Inset: width of the peak (averaged, outliers removed). (e) Histogram of the TOF drift values per km.}
    %\vspace{128in}
    \label{Figure WR sync}
\end{figure}

In Figure \ref{Figure WR sync} we detail an experiment in which two remote TTX's are synchronized via WR. A clock signal is distributed from a WR-Z16 to WR-LEN receiver nodes using a $5\;\mathrm{km}$ dedicated fiber link. The experiment employs four WR-LEN nodes to synchronize two remote sources (here classical laser pulses generated by acousto-optic modulators (AOMs) and two remote TTXs. The WR-LEN nodes output $10 \; \mathrm{MHz}$ to digital delay generators (DDGs, Stanford Research Systems DG645), which control fiber AOMs, and both $10\;\mathrm{MHz} + 1\; \mathrm{pps}$ signals to two input channels at each of the TTXs. We developed custom software that records coincident detection events between two remote TTXs, each of which streams its time tags to the network's control plane. 

Here, coincident detection events registered from single clicks on both TTX 1 and TTX 2 are reconstructed via time tag streaming over a network server. The resulting coincidence peak is fitted with a Gaussian $f(x) = ae^{\frac{-(x-c)^2}{w^2}}$ to extract its center time delay $c$ and width $w$. The center (Figure \ref{Figure WR sync}(c,d)) and width (Figure \ref{Figure WR sync}(d), inset) are plotted over time to test how well WR can maintain synchronization over a deployed link. This experimental setup is similar to the TOF characterization experiment, except instead of characterizing the TOF drift over a deployed link itself (test pulses are kept indoors), this experiment characterizes the amount of timing drift/error associated with the WR-based synchronization of remote AOMs and time taggers. Results show that our network can generate distributed coincidence events over a period of five days with less than a nanosecond of timing drift overall, and an average coincidence peak width of $4.5\;\mathrm{ns}$ (limited by the $10\;\mathrm{ns}$ pulse duration). Note that at about 40 hours of elapsed time, an unidentified environmental perturbation causes the SOP to change abruptly. The transmitted power (panel (b), inset) increases slightly. At this point, the coincidence center time shifts and its width increases, showing that the cause of the abrupt SOP change also has a deleterious effect on the WR synchronization between the remote TTXs. An interesting feature of this data set is that the histogram of TOF drift values in Figure \ref{Figure WR sync}(e) sees its maximum centered exactly at  $0\;\mathrm{ps/km}$, with tails extending from $-150-$ to $50\;\mathrm{ps/km}$. This is in contrast with TOF drift measurements taken without WR synchronization, whose TOF drift histograms are not centered at $0\;\mathrm{ps/km}$.

\section{Comprehensive modeling of environmental impacts}
Figures \ref{Figure griffiss TOF}, \ref{Figure stock TOF}, \ref{Figure WR sync} show histograms of the TOF drift in units of $\mathrm{ps/km}$ for fiber links at Griffiss and Stockbridge. The buried fiber at the Griffiss QLAN clearly has much more time-domain stability; TOF drifts below $50\;\mathrm{ps/km}$ over four days. When we expanded the experiment to include WR-based timing synchronization of remote time taggers, the TOF remained below $200\;\mathrm{ps/km}$, even including outlier points where WR failed.

Using the data obtained across the Griffiss and Stockbridge QLANs, we analyzed the correlations between TOF drift and polarization angle drift and environmental parameters, focusing on temperature and average wind speed. We performed linear fits using Pearson Correlations. Fits of linear relationships between two variables output a ``Pearson's $r$", which is a metric that quantifies the strength and direction of the relationship. $r=1 (-1)$ indicates a perfect positive (negative) linear correlation, while $r=0$ indicates no linear correlation. For our fits we extract $r$ and coefficient of determination ($R^2$) values. $R^2 \in [0,1]$ quantifies how well a regression fits its associated data points, with $R^2=0$ indicating no fit and $R^2=1$ indicating a perfect fit. 

Tables \ref{tab: TOF correlations} and \ref{tab: SOP correlations} summarize results for TOF drift and SOP drift measurements on buried and aerial fiber links. Unsurprisingly, the Griffiss QLAN buried link exhibits weak dependencies on environmental parameters, consistently showing correlation coefficients below $|0.3|$. Since the fiber is better insulated than the buried fiber at Stockbridge, this is intuitive. The indoor/aerial link at RRS also exhibits relatively high SOP stability. The Stockbridge QLAN fibers, both buried and aerial, exhibit a stronger TOF drift correlation with temperature, particularly when asymmetric heating about a threshold temperature is taken into account.

Overall, TOF drift correlations with temperature are in the weak-moderate range at Griffiss, and in the moderate-strong range at Stockbridge. The wind speed dependence is mostly weak, with some moderate dependence exhibited by the Stockbridge buried fiber link.  Across both QLANs tested, the SOP drift shows moderate-to-strong dependence on temperature and wind speed changes, indicating that polarization stabilization of these network links is required for high-fidelity transmission of polarization-encoded qubits. Note that significantly smaller correlations are observed at Stockbridge when significant relative lengths of buried and aerial fiber are included in the measurement. This decrease in strong linear correlations indicates that the combination of link types leads to more complex, possibly nonlinear correlations that require more in-depth analysis.

A counter-intuitive result from this analysis is that the SOP angle drift for buried fiber depends more strongly on ambient wind speed, both at Griffiss and Stockbridge, than even the Stockbridge aerial fiber. Further investigation is warranted to explain this result, which may indicate that wind loading on surface structures is transmitting mechanical stress into the fibers. Wind speed may also be indicative of ambient barometric pressure, and such pressure changes may lead to additional mechanical stress on fiber. Unlike correlations observed elsewhere \cite{Bersin2024}, we do not observe any non-negligible correlation between SOP drift \emph{rate} and the square of wind speed. This difference may be explained by the significantly shorter length of aerial fiber as compared to the BARQNET \cite{Bersin2024} testbed.

% Define colors for shading
\definecolor{headergray}{gray}{0.8}
\definecolor{veryweak}{gray}{0.9}
\definecolor{weak}{RGB}{144,238,144}
\definecolor{moderate}{RGB}{255,255,153}
\definecolor{strong}{RGB}{255,182,193}

\begin{table*}[htbp]

\centering
\caption{TOF Drift Environmental Correlations. G: Griffiss/ S: Stockbridge. B: buried/ A: aerial. Shading indicates the strength of the correlation as determined by $r$. Very weak (grey): $|r|< 0.2$. Weak (green): $0.2<|r|< 0.4$. Moderate (yellow): $0.4<|r|< 0.6$ (red): $|r|> 0.6$. Linear fit equations provided in Supplementary Table S1.}
\resizebox{\textwidth}{!}{%
\begin{tabular}{|c|c|c|c|c|c|c|c|c|c|c|}
\hline
\rowcolor{headergray}
\textbf{Figure} & \textbf{QLAN} & \textbf{Link type} & \multicolumn{2}{c|}{\textbf{TOF drift vs wind speed}} & \multicolumn{2}{c|}{\textbf{TOF drift vs temp overall}} & \multicolumn{4}{c|}{\textbf{TOF drift vs temp split}} \\
\hline
\rowcolor{headergray}
 &  &  & $\mathbf{r}$ & $\mathbf{R^2}$ & $\mathbf{r}$ & $\mathbf{R^2}$ & \multicolumn{2}{c|}{\textbf{Below threshold}} & \multicolumn{2}{c|}{\textbf{Above threshold}} \\
\hline
\rowcolor{headergray}
 &  &  &  &  &  &  & $\mathbf{r}$ & $\mathbf{R^2}$ & $\mathbf{r}$ & $\mathbf{R^2}$ \\
\hline
\ref{Figure griffiss TOF} & G & B & \cellcolor{weak}$-0.227$ & \cellcolor{weak}$0.051$ & \cellcolor{veryweak}$-0.130$ & \cellcolor{veryweak}$0.017$ & \cellcolor{veryweak}$-0.060$ & \cellcolor{veryweak}$0.004$ & \cellcolor{veryweak}$0.010$ & \cellcolor{veryweak}$0.000$ \\
\hline
\ref{Figure WR sync} & G & B-WR & \cellcolor{weak}$0.222$ & \cellcolor{weak}$0.049$ & \cellcolor{weak}$0.275$ & \cellcolor{weak}$0.076$ & \cellcolor{moderate}$0.421$ & \cellcolor{moderate}$0.177$ & \cellcolor{veryweak}$-0.081$ & \cellcolor{veryweak}$0.007$ \\
\hline
\ref{Figure stock TOF}(c) & S & B & \cellcolor{weak}$-0.365$ & \cellcolor{weak}$0.133$ & \cellcolor{veryweak}$-0.048$ & \cellcolor{veryweak}$0.002$ & \cellcolor{moderate}$-0.518$ & \cellcolor{moderate}$0.269$ & \cellcolor{weak}$0.278$ & \cellcolor{weak}$0.077$ \\
\hline
\ref{Figure stock TOF}(g) & S & A & \cellcolor{veryweak}$-0.162$ & \cellcolor{veryweak}$0.026$ & \cellcolor{weak}$-0.308$ & \cellcolor{weak}$0.095$ & \cellcolor{strong}$0.659$ & \cellcolor{strong}$0.434$ & \cellcolor{moderate}$-0.511$ & \cellcolor{moderate}$0.600$ \\
\hline
\end{tabular}%
}
\label{tab: TOF correlations}
\end{table*}

\begin{table*}[htbp]

\centering
\caption{Ellipticity and Azimuthal Drift Environmental Correlations. Same color legend as Table I-A. As opposed to TOF drift, SOP angular drifts show stronger environmental correlations across the three QLANs tested. R:RRS. A: aerial B: buried B-wr: buried with White Rabbit Linear fit equations are provided in Supplementary Table S2.}
\resizebox{\textwidth}{!}{%
\begin{tabular}{|c|c|c|c|c|c|c|c|c|c|c|}
\hline
\rowcolor{headergray}
\textbf{Figure} & \textbf{QLAN} & \textbf{Link type} & \multicolumn{2}{c|}{\textbf{Ellipticity drift vs wind speed}} & \multicolumn{2}{c|}{\textbf{Ellipticity drift vs temp}} & \multicolumn{2}{c|}{\textbf{Azimuthal drift vs wind speed}} & \multicolumn{2}{c|}{\textbf{Azimuthal drift vs temp}} \\
\hline
\rowcolor{headergray}
 &  &  & $\mathbf{r}$ & $\mathbf{R^2}$ & $\mathbf{r}$ & $\mathbf{R^2}$ & $\mathbf{r}$ & $\mathbf{R^2}$ & $\mathbf{r}$ & $\mathbf{R^2}$ \\
\hline
\ref{Figure griffiss sop drift}(a) & G & B & \cellcolor{moderate}$0.4376$ & \cellcolor{moderate}$0.192$ & \cellcolor{moderate}$0.588$ & \cellcolor{moderate}$0.345$ & \cellcolor{moderate}$-0.447$ & \cellcolor{moderate}$0.199$ & \cellcolor{moderate}$-0.541$ & \cellcolor{moderate}$0.293$ \\
\hline
\ref{Figure griffiss sop drift}(b) & G & B & \cellcolor{moderate}$0.594$ & \cellcolor{moderate}$0.353$ & \cellcolor{moderate}$0.530$ & \cellcolor{moderate}$0.281$ & \cellcolor{weak}$0.282$ & \cellcolor{weak}$0.080$ & \cellcolor{moderate}$0.462$ & \cellcolor{moderate}$0.213$ \\
\hline
\ref{Figure RRS SOP drift} & R & A & \cellcolor{veryweak}$-0.070$ & \cellcolor{veryweak}$0.005$ & \cellcolor{veryweak}$-0.122$ & \cellcolor{veryweak}$0.015$ & \cellcolor{moderate}$-0.349$ & \cellcolor{moderate}$0.122$ & \cellcolor{veryweak}$-0.138$ & \cellcolor{veryweak}$0.019$ \\
\hline
\ref{Figure WR sync} & G & B-WR & \cellcolor{weak}$0.392$ & \cellcolor{weak}$0.154$ & \cellcolor{weak}$0.381$ & \cellcolor{weak}$0.145$ & \cellcolor{strong}$0.646$ & \cellcolor{strong}$0.418$ & \cellcolor{weak}$0.277$ & \cellcolor{weak}$0.077$ \\
\hline
\ref{Figure stock TOF}(a) & S & B & \cellcolor{moderate}$-0.559$ & \cellcolor{moderate}$0.312$ & \cellcolor{strong}$-0.714$ & \cellcolor{strong}$0.509$ & \cellcolor{strong}$-0.738$ & \cellcolor{strong}$0.544$ & \cellcolor{moderate}$-0.465$ & \cellcolor{moderate}$0.216$ \\
\hline
\ref{Figure stock TOF}(e) & S & A & \cellcolor{moderate}$0.426$ & \cellcolor{moderate}$0.181$ & \cellcolor{strong}$-0.771$ & \cellcolor{strong}$0.594$ & \cellcolor{weak}$0.324$ & \cellcolor{weak}$0.105$ & \cellcolor{strong}$-0.756$ & \cellcolor{strong}$0.572$ \\
\hline
\ref{Figure stock BA fiber pol} & S & B \& A & \cellcolor{veryweak}$-0.044$ & \cellcolor{veryweak}$0.002$ & \cellcolor{veryweak}$-0.096$ & \cellcolor{veryweak}$0.009$ & \cellcolor{veryweak}$0.114$ & \cellcolor{veryweak}$0.013$ & \cellcolor{weak}$0.338$ & \cellcolor{weak}$0.114$ \\
\hline
SM & S & B & \cellcolor{strong}$0.637$ & \cellcolor{strong}$0.406$ & \cellcolor{moderate}$0.456$ & \cellcolor{moderate}$0.208$ & \cellcolor{strong}$0.577$ & \cellcolor{strong}$0.333$ & \cellcolor{strong}$0.770$ & \cellcolor{strong}$0.593$ \\
\hline
\end{tabular}%
}
\label{tab: SOP correlations}
\end{table*}

One notable result is the asymmetric response of the TOF drift of fiber links, at Stockbridge in particular, to temperature changes. Specifically, for  at the Stockbridge aerial fiber link, when the temperature is below a threshold value of $70.5^{\circ }F$, the linear relationship is positive, moderate strength, with $r = 0.659$ and $R^2 = 0.434$, whereas for temperatures above the threshold, the linear relationship is negative, with $r = -0.511$ and $R^2 = 0.600$. (Correlation plots for this experiment are included in the Supplementary Information.)

Our experiments and analysis also allow for the study of correlations between TOF drift and SOP drift, correlation lag analysis, relationships between rates of change of each variable, as well as nonlinear relationships, though a detailed analysis of these other relationships is reserved for elsewhere. Detailed results from the full correlation analysis can be found in the Supplementary Information.

\section{Quantum-classical signal coexistence}
Measurements exploring wavelength crosstalk-induced noise from co- and counter-propagating quantum and classical signals in fiber and free space, including WR signals, were performed. We showed that we can co-propagate CWDM-multiplexed quantum and classical signals over a $5\;\mathrm{km}$ buried fiber link at the Griffiss QLAN and maintain our ability to measure entangled coincidences, even though the presence of the classically bright light degrades the achievable signal-to-noise ratio (SNR).  For WR specifically, we showed that polarization multiplexing effectively mitigates noise in the quantum channels due to the presence of WR signals. Finally, we explored the use of double-clad fiber couplers\cite{Wroblewski2023} to multiplex single-mode quantum and multi-mode classical signals over free space and demonstrated our ability to do so in the laboratory with no significant SNR degradation.  More details can be found in the Supplementary Information.

\section{Bell's Inequality violation with chip-based entanglement sources}
\subsection{\label{PIC source}Photonic integrated circuit-based entangled photon source}
\begin{figure}[!htbp]
    \includegraphics[width = 1\linewidth]{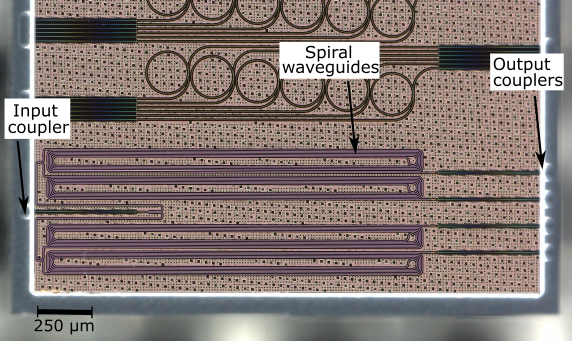}
    \caption{Microscope image of a PIC spiral source designed by our team and fabricated at the AIM Photonics foundry \cite{Fahrenkopf2019}. The image shows a set of four sources. One input waveguide splits into four separate spirals, which are connected to four separate output couplers. As such, one chip can act as four separate entanglement sources. The overall chip dimensions are $1.866\;\mathrm{mm} \times 4\;\mathrm{mm}$. The devices located above the spiral sources are unrelated to the present work.}
    %\vspace{128in}
    \label{Figure PIC source chip}
\end{figure}

The Griffiss QLAN has multiple commercial entangled photon sources available, including a broadband spontaneous parametric down-conversion (SPDC) C-band source (ADVR) and a narrow-band bichromatic $795\;\mathrm{nm}-1364\;\mathrm{nm}$ source using warm Rubidium vapor (Qunnect)\cite{Craddock2024source}. Here, we focus on networking demonstrations performed with a photonic integrated circuit (PIC) source designed by our team at AFRL and the Rochester Institute of Technology and fabricated at the AIM Photonics foundry \cite{Fahrenkopf2019}. 

The PIC source, shown in Figure \ref{Figure PIC source chip} consists of a a CMOS-compatible Si waveguide clad in SiO\textsubscript{2}. The waveguide has a spiral shape that maximizes propagation length within the photonic chip dimensions ($1.9\;\mathrm{mm} \times 4.1\;\mathrm{mm}$). Four-wave mixing (FWM) in the waveguide produces broadband frequency-and polarization entangled photon pairs. Typically, we pump the chip with a $1550 \;\mathrm{nm}$ tone for spontaneous FWM, and use a CWDM with $20\;\mathrm{nm}$ channel spacing and $6.5\;\mathrm{nm}$ pass band to select the $1530\;\mathrm{nm}$ and $1570\;\mathrm{nm}$ signal-idler pair. The source divides input light into four independent identical (up to fabrication defects) spiral waveguides to produce four separate entanglement sources per pump laser. The source carries the advantage of requiring no temperature stabilization to achieve phase matching. Also, with chip dimensions of $1.9\mathrm{mm} \times 4\mathrm{mm}$, it is highly compact, and can be packaged straightforwardly with optical fiber connections and a 3D-printed case. All that is required for operation is a pump laser in the telecom C-band, and a set of polarization paddles to optimize the input polarization to the chip.

We pump the PIC source with a $1550\;\mathrm{nm}$ continuous wave (CW) laser to produce entangled photon pairs (biphotons) via spontaneous FWM. The biphotons have a non-separable joint-spectral intensity (JSI) defined in terms of the full biphoton state 
\begin{equation}
\label{eq:PIC source JSI}
\ket{\psi} = \int d \omega_s d \omega_i \psi(\omega_s, \omega_i) a^{\dagger}(\omega_s) b^{\dagger}(\omega_i)\ket{vac},
\end{equation}
\noindent where $\psi(\omega_s, \omega_i)$ is the biphoton wavefunction and $\text{JSI}(\omega_s,\omega_i)=|\psi(\omega_s, \omega_i)|^{2}$.  Conservation of energy guarantees that for pump photons of angular frequency $\omega_p$, the source produces signal-idler pairs at angular frequencies $\omega_s + \omega_i = 2\omega_p$. The phase matching condition for pair generation is $\Delta k = k_s + k_i - 2k_p $. Photons generated via spontaneous FWM are entangled through both their time of creation and the conservation of momentum. The Heisenberg uncertainty principle for time and energy  $\Delta E \Delta t \geq \hbar /2$ ensures that the energy of a pair of photons and their relative time of arrival are coupled. See the Supplementary Information for a detailed theoretical description of the device.

\begin{figure}[!htbp]
    \centering \includegraphics[width=0.9\linewidth ]{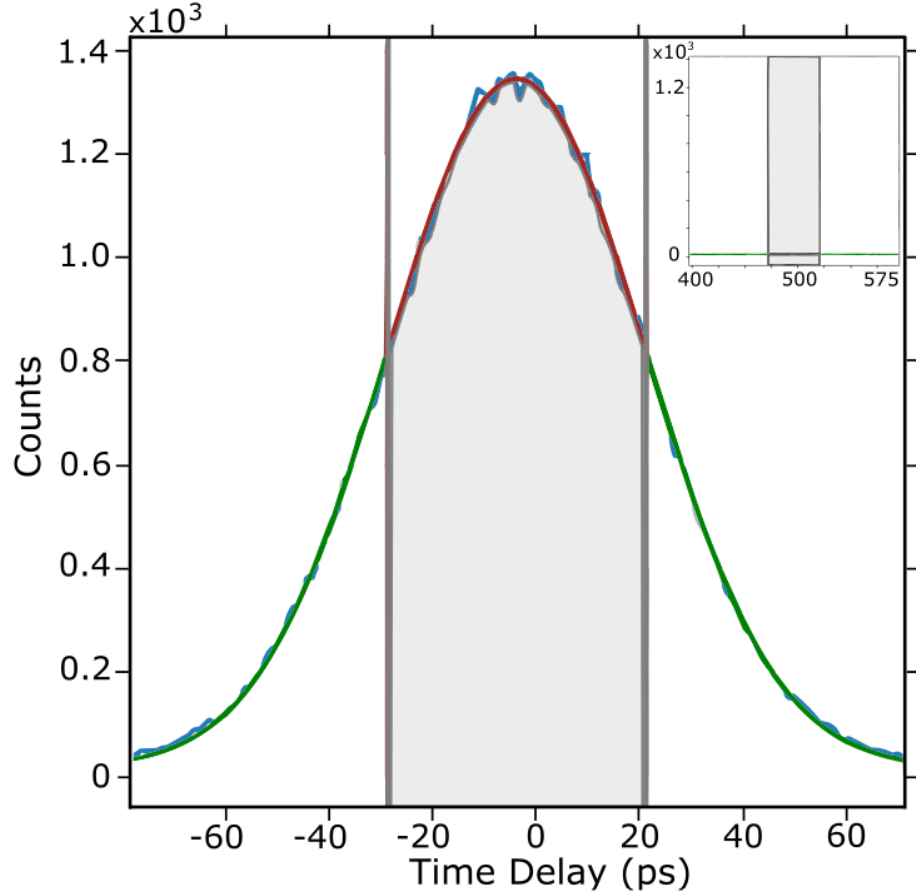}
    \caption{ Example coincidence histogram from PIC source, pumped with  $1550\;\mathrm{nm}$ CW at $15\;\mathrm{mW}$. Gray: integrated area. Green/red: experimental data. Blue: Gaussian fit. Inset: integrated accidental counts. The scan is taken with 1 picosecond bin widths, 5 second integration time, and no averaging. An integral over the shaded area contains $57.5\times10^3$ coincidence counts. The extracted coincidence to accidental ratio is $75$, with a visibility of $97.4\%$.}
    %\vspace{128in}
    \label{Figure pic source}
\end{figure}

With a CW pump (Keysight 8160 family laser) power of $10\;\mathrm{mW}$, the source outputs $3-9 \times 10^6 \; counts/sec$ in both the $1530 \;\mathrm{nm}$ and $1570\;\mathrm{nm}$ CWDM channels, after filtering (with 2 dB insertion loss at the CWDM). The source produces $10-20\times10^3\;pairs/sec$. The coincidence plot in Figure \ref{Figure pic source} reveals a coincidence-to-accidental ratio of $75$ and a visibility of $97.4\%$ ($1\;\mathrm{ps}$ time bin, $5\;\mathrm{s}$ integration time). The source's brightness can be increased by integrating low-loss edge couplers on-chip, and by removing the $1\times 4$ splitters, but is limited by higher-order effects. Higher-brightness devices have been fabricated and are undergoing initial characterization. The source is pumped with horizontally polarized light. Spontaneous FWM annihilates two horizontally-polarized photons, and, preserving polarization (unlike Type-I and Type-II spontaneous parametric down-conversion), creates a signal-idler photon pair of horizontal polarization \cite{Lin2007}. 

Below, we detail how the PIC source can be used to generate energy-time Bell states. It is in principle also possible to create polarization Bell pairs with the source, though the spectral correlation of the signal-idler pairs will limit the spectral purity and therefore the quality of polarization entanglement. The spectral purity of the biphotons can be tuned via the pump laser bandwidth, as discussed in the Supplementary Information. We are also developing new PIC-based entanglement sources that have higher spectral purity, though a detailed discussion of those devices is reserved for future work. 

To quantify the quality of entanglement of the PIC source's biphotons, we perform quantum correlation measurements on the signal-idler photon pairs produced by the source. We use a custom WaveShaper filter profile to effectively simulate the frequency response of a Franson interferometer to generate and analyze time-energy Bell states and experimentally verify the non-classical correlations between the generated signal/idler pairs. These experiments demonstrate that we can generate and distribute entanglement through our QLANs.

Bell's theorem \cite{Bell_Aspect_2004} states that local hidden variable theories, based on local realism, cannot reproduce the predictions of quantum mechanics. A system of two supposed entangled particles is shown to be non-classical in nature when the correlations between the two particles violate Bell's inequality, e.g. when the correlations surpass a maximum achievable value for classical systems. One popular formulation of Bell's Inequality is the Clauser-Horne-Shimony-Holt (CHSH) inequality \cite{CHSH1969}, a specific inequality that lends itself well to experimental measurements. We demonstrate a violation of Bell's inequality through a CHSH inequality measurement at the Griffiss QLAN. As discussed in detail below, classical systems have a maximum CHSH score $S$ of $2$, whereas quantum entangled systems have a maximum score of $2\sqrt{2}$. An experimental measurement of $S \geq 2$ serves as a violation of Bell's inequality and verification of quantum correlations.

\subsection{\label{Franson}Franson interferometry for time-energy Bell states}
For polarization-entangled photons, CHSH measurements are performed by rotating polarizers to induce variable constructive and destructive interference between the photons. Corresponding changes in the output coincidence histogram allow for the measurement correlations and a means to confirm a violation of Bell's Inequality. We have developed a new experimental method to CHSH measurements on entangled photon pairs from our PIC entanglement source in the energy-time basis, using a Franson interferometer comprised of programmable liquid crystal optical modulators.

Franson interferometers are used to generate and measure time-energy Bell pairs \cite{Franson1989, Sun2017}: these consist of two identical, asymmetric Mach-Zehnder Interferometers (MZIs), as shown in Figure \ref{Figure CHSH indoor}(a). Two-photon interference measurements are taken by measuring coincidence events between detectors placed at the interferometer's outputs. Just as the orientation of a polarizer is rotated to either pass or block a photon and achieve modulation in the polarization basis, phase shifters (or variable time delays) within the two MZIs can be used for modulation in the energy-time basis. 

Let us revisit the wavefunction for the PIC biphoton source output. As described above, we pump our PIC source with a $1550\;\mathrm{nm}$ CW laser tone to produce entangled photon pairs (biphotons) with the joint spectral intensity (JSI) from Equation \ref{eq:PIC source JSI}. For a sinusoidal phase modulation applied by a Franson interferometer, the JSI becomes
\begin{multline}
\ket{\psi} = \int d \omega_s d \omega_i \psi(\omega_s ,\omega_i) \cos(a_1 \omega_s + b_1) \\  \cos(a_2\omega_i + b_2) a^{\dagger}(\omega_s) b^{\dagger}(\omega_i)\ket{vac}.
\end{multline}
the phase modulation term can be simplified to 
\begin{multline}
   \phi(\omega_{s},\omega_i) = \cos(a_1 \omega_s + b_1) \cos(a_2\omega_i + b_2) = \\
    \frac{1}{2}\left( \cos(a) + \cos(b \omega_i) + c\right)
\end{multline}
where $a,b,c$ are newly defined constants. To understand how this frequency modulation will manifest in the time domain, we take the Inverse Fourier transform and get:
\begin{multline}
   \mathscr{F}^{-1} \{\phi(\omega_{s},\omega_{i})\} = \\
   \frac{1}{2}\sqrt{\frac{\pi}{2}}e^{-ic}\delta(t-a) + \frac{1}{2}\sqrt{\frac{\pi}{2}}e^{ic}\delta(t+a) + \sqrt{\frac{\pi}{2}}\cos(b)\delta(t).
   \label{eqn:fft}
\end{multline}
This expression tells us that three peaks will appear in the coincidence histograms at the Franson interferometer's output.

For a relative time delay $\Delta t$ between the two MZIs, measuring coincidence events at detectors 1 and 2 gives a JSI of
\begin{multline}
    \ket{\psi} = \int d \omega_s d \omega_i \psi(\omega_s ,\omega_i) (1+e^{i\Delta t \omega_s})\\(1+e^{i\Delta t \omega_i}) a^{\dagger}(\omega_s) b^{\dagger}(\omega_i)\ket{vac}\\
    = \int d \omega_s d \omega_i \psi(\omega_s, \omega_i) (1+e^{i\Delta t 2 \omega_p} + e^{i\Delta t \omega_s}+ e^{i\Delta t \omega_s})\\ a^{\dagger}(\omega_s) b^{\dagger}(\omega_i)\ket{vac}.
    \label{eqn:biphoton_state}
\end{multline}

The $1+e^{i\Delta t 2 \omega_p}$ term corresponds to a central coincidence peak at time delay $t=0$. Coincidences at $t=0$ correspond to either short-short (SS) paths or long-long (LL) paths traveled by each photon in the entangled pair. It is impossible to distinguish between these two scenarios, so we are left with a Bell pair
\begin{equation}
    \ket{\psi} = \frac{1}{\sqrt{2}}(\ket{SS}+ \ket{LL}).
\end{equation}

By choosing coincident events at $t=0$ and discarding others, we post-select to generate and measure this Bell state. Setting the phase $\theta = \Delta t \omega_p$, the coincidence detection probability $P(\theta)$ is
\begin{equation}
    P(\theta) = \frac{1}{4}\Big|\left(1+e^{i2\theta}\right)\Big|^2 = \frac{1}{2}\Big(1+\cos(2\theta)\Big)
    \label{eq: cos phase}
\end{equation}
which motivates the use of a sinusoidal phase modulation term in our WaveShapers, as detailed below. Simulated Franson interference plots are shown in the Supplementary Information. 

\subsection{Time-Energy CHSH Violation Measurements}
To perform a CHSH measurement for two photons entangled in the polarization degree of freedom, two polarizers (A and B) are placed in the paths of the photons. The polarizers are rotated to specific angles, and coincidence histograms are obtained at various combinations of angles to measure correlations. The CHSH score is calculated via the expression
\begin{multline}
    S = |E(\alpha, \beta) - E(\alpha, \beta') + E(\alpha', \beta) + E(\alpha', \beta')|\\
    = |E0-E1+E2+E3|
\end{multline}
where the correlation values $E$ are defined, for angles $\alpha$ and $\beta$, as e.g.,  
\begin{equation}
    E(\alpha, \beta) = \frac{C(\alpha, \beta) - C(\alpha, \beta_{\perp}) - C(\alpha_{\perp}, \beta) + C(\alpha_{\perp}, \beta_{\perp})}{C(\alpha, \beta) + C(\alpha, \beta_{\perp}) + C(\alpha_{\perp}, \beta) + C(\alpha_{\perp}, \beta_{\perp})}.
\end{equation}
Here $C$ denotes the measured number of coincidence counts at a given pair of angles, and $\alpha_{\perp} = \alpha + 90^\circ$ and $\beta_{\perp} = \beta + 90^\circ$. Values of $S \geq 2$ demonstrate a violation of the CHSH inequality, confirming non-classical correlations exist between the two photons. The maximum achievable value is $S=2\sqrt{2}\approx2.828$, known as the Tsirelson's bound.

To perform a CHSH measurement in the energy-time basis, the phase applied by the MZIs in a Franson interferometer acts as a polarizer. We apply sinusoidal phase offsets to entangled photon pairs using our programmable optical modulators, which act as MZIs (see Equation \ref{eq: cos phase}). Specifically, each MZI is instantiated as a Finisar WaveShaper 1000A (WaveShapers A and B), a programmable optical filter which can precisely shape the amplitude and phase of optical signals. The WaveShapers are used to apply both bandpass amplitude filters (either square or Gaussian shaped) and sinusoidal phase offsets to emulate phase shifters within bulk-optics interferometers. The applied WaveShaper filter function is 
\begin{equation}
    f(\nu;\lambda_c) = \cos\left(\left(\frac{\nu - c}{\lambda_c}\right)(\pi/\tau)+B\right)
\end{equation}
versus frequency $\nu$, centered at $\lambda_c$ wavelength, with a period of $\tau$ and phase offset $B$. $B$ is understood as analogous to a polarizer angle, which is ``rotated." Cycles of constructive and destructive interference in the coincidence counts are generated by gradually adjusting the relative phase offset between the two WaveShapers in this way. The visibility of the interferometer, as composed, is
\begin{equation}
    V = \frac{C_{\theta=0}-C_{\theta=\pi/2}}{C_{\theta=0}+C_{\theta=\pi/2}}
\end{equation}
for a maximum coincidence count number $C_{\theta=0}$ for constructive interference and a minimum coincidence count number $C_{\theta=\pi/2}$ for destructive interference, where $\theta$ is the relative phase different between waveshapers. We routinely obtain visibilities exceeding $90\%$ with this setup. Note that these visibilities are for a given  time integration of 20 ps; this informs the experimenter that if one wanted to perform channel gating for heralded operations with high visibility, one would need to herald within a 20ps window of the gating photon detection.  More details are available in the Supplementary Information.

\begin{figure*}[!htbp]
    \includegraphics[width = 1\linewidth]{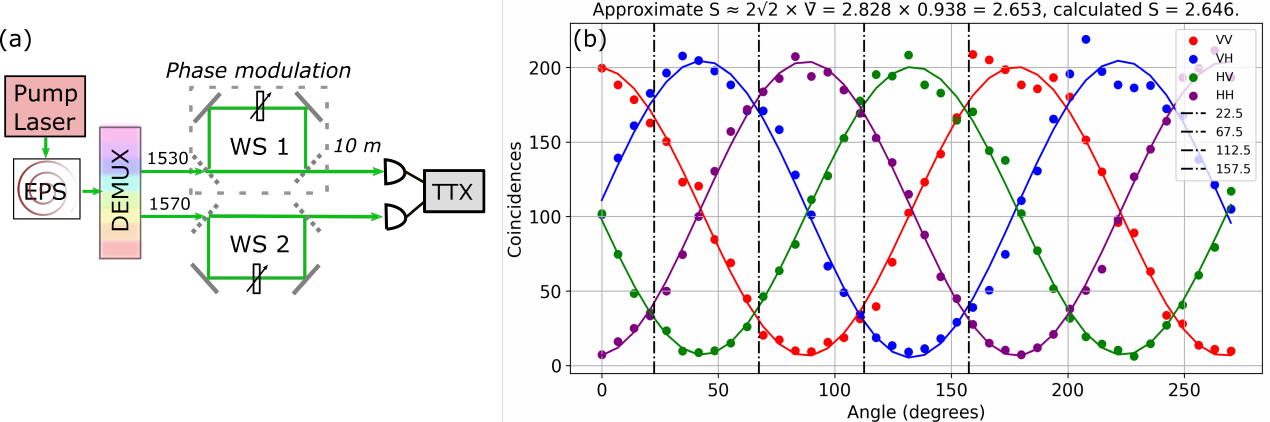}
    \caption{WaveShaper-based Franson interferometer for CHSH measurements. (a) Diagram of experimental setup. EPS: entangled photon soure, DEMUX: de-multiplexer, WS A, WS B = WaveShapers imparting Franson interference frequency modulation, TTX = time tagger. (b) Experimental data from a CHSH measurement using our EPS and a WaveShaper-based Fransen interferometer. The phase offset (or angle) of WaveShaper B is swept from $0^\circ$ to $275 ^\circ$. Each curve on the plot corresponds to a different fixed phase offset for WaveShaper A. Data points correspond to experimental results, which closely follow the $\cos^2$ solid curve fits. Using indoor fiber links of 10-meter length, we achieved modulation visibilities of 93.75\% and a corresponding CHSH score of $2.646\pm0.0732$. }
    %\vspace{128in}
    \label{Figure CHSH indoor}
\end{figure*}

Coincidence counts are measured when WaveShaper A is kept at a constant phase value and the WaveShaper B phase offset is rotated from $0^{\circ}$ to $275^{\circ}$ in steps of $7^{\circ}$. The coincidences vs. phase curve follows a $C \sim \frac{1}{2}\cos^2(\alpha - \beta)$ relationship. Output curves are fit with this function to extract the $C(\alpha, \beta)$ terms needed to calculate the correlation values. We report potential violations from coincidence curve fits: measurements are not taken simultaneously, they are taken consecutively.  The fidelity of the potential violation is ultimately determined by the fidelity of the coincidence curve fits.  For our experiments, the CHSH measurement angles are set to
\begin{multline}
    \alpha = 0 ^{\circ} \; , \;\beta = 22.5 ^{\circ} \; , \;\alpha' = 45 ^{\circ} \; , \;
    \beta' = 67.5 ^{\circ} 
\end{multline} 
so that the correlation values are defined as
\begin{alignat}{2}
   E_0 &= E(0^{\circ}, 22.5^{\circ}), \quad &\quad
   E_1 &= E(0^{\circ}, 67.5^{\circ}), \nonumber\\
   E_2 &= E(45^{\circ}, 22.5^{\circ}), \quad &\quad
   E_3 &= E(45^{\circ}, 67.5^{\circ}).
\end{alignat}
For an indoor fiber link of 10 meters, (see Figure \ref{Figure CHSH indoor}) using a Gaussian shape amplitude filter, we obtain a CHSH score of
\begin{multline}
    S = |E(\alpha_0, \beta_{22.5}) - E(\alpha_0, \beta_{67.5}) +\\ E(\alpha_{45}, \beta_{22.5}) + E(\alpha_{45}, \beta_{67.5})| \\
     = 0.60602 + 0.71185 + 0.70779 + 0.61990 = 2.646\pm0.0732,
\end{multline}

\noindent or a significance factor of roughly nine standard deviations, $9\sigma$. We provide derivations for the uncertainty and corresponding significance factor in the Supplementary Information.  As secondary verification, the coincidence plots can be rescaled in terms of the measurement visibility such that $E_{i}(\phi)\sim V_{i}\cos(2\phi+\delta_i)$ where $\delta_i$ represents the measurement phase.  Defining $\phi=(\alpha-\beta)/2$ and shifting the curves to match the canonical model we can write $E_{i}(\alpha,\beta)\sim V_{i}\cos(\alpha-\beta)$, leading to

\begin{align}
    S_{\text{approx.}}&\approx \left|V_{1}\cos(\alpha-\beta)+V_{2}\cos(\alpha-\beta^\prime)+\right. \nonumber\\
    &\;\;\;\;\;\;\;\;\;\;\;\left.+V_{3}\cos(\alpha^\prime-\beta)-V_{4}\cos(\alpha^\prime-\beta^\prime)\right|.
    \label{eqn:tsire_approx1} 
\end{align}

\noindent Plugging in the canonical angles $\alpha=0,\;\alpha^\prime=\tfrac{\pi}{2},\;\beta=\tfrac{\pi}{4},\;\beta^\prime=-\tfrac{\pi}{4}$ immediately yields

\begin{align}
    S_{\text{approx.}}&\approx \left[\frac{V_{1}\sqrt{2}}{2}+\frac{V_{2}\sqrt{2}}{2}+\frac{V_{3}\sqrt{2}}{2}-\left(-\frac{V_{4}\sqrt{2}}{2}\right)\right] \nonumber\\
    &= \frac{1}{4}\sum_{i}V_{i}\times2\sqrt{2},
    \label{eqn:tsire_approx2}
\end{align}

\noindent or simply $S_{\text{approx.}}=2\sqrt{2}\;\bar{V}$, a fair standard to our experimental data.  

\begin{figure}[!htbp]
    \includegraphics[width = 1\linewidth]{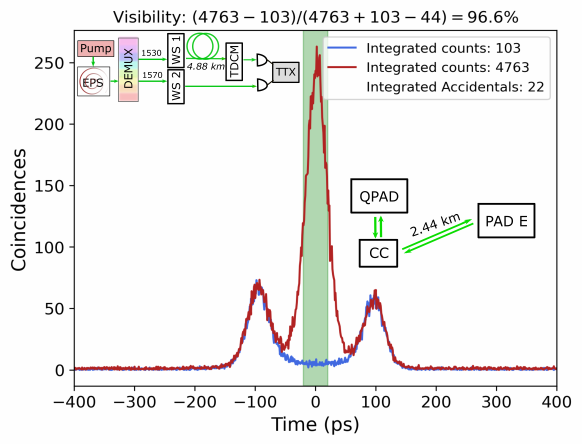}
    \caption{Measurement visibility across the 4.88 km loop at the Stockbridge QLAN.  Counts are integrated over a 20 ps window around the SS+LL peak and the visibility is computed via $V=\tfrac{C_{\theta=0}-C_{\theta=\pi/2}}{C_{\theta=0}+C_{\theta=\pi/2}-2C_{\text{acc.}}}\approx96.6\%$, where $\theta$ is the relative phase difference between waveshaper profiles. Note that the FS TDCM corrected for $\sim5$ps of dispersion-related jitter. Insets: experimental setup diagram (top left) and network layout diagram (bottom right).}
    %\vspace{128in}
    \label{fig:5km_vis}
\end{figure}

\subsection{Spectral Dispersion Compensation}
Beyond laboratory demonstrations, we demonstrate Bell's Inequality Violations within a deployed, long-distance fiber-optic network. In this scenario, one photon from the entangled pair travels through $~5\;\mathrm{km}$ of deployed fiber at either the  Griffiss or the Stockbridge QLAN, while the other photon stays within the laboratory/Pad. The addition of long fiber links presents a challenge in the form of chromatic dispersion. A cumulative effect, dispersion causes light at different frequencies to propagate through the fiber at different speeds. FOr normal dispersion, shorter wavelengths travel faster than longer wavelengths. The chromatic dispersion introduces a wavelength-dependent phase shift 
\begin{equation}
    \Psi_{CD}=\beta_{2} (\lambda)L
\end{equation}
for a fiber of length $L$, where 
 \begin{equation}
     \beta_{2} (\lambda)= -D\lambda^2/2 \pi c 
 \end{equation} 
is the propagation constant, in units of $ps^2/km$, for a group velocity dispersion (GVD) parameter $D$, in units of $ps/(nm\cdot km)$. 
%
%The dispersion phase can be written as 
%\begin{equation}
%    H_{disp} = e^{\frac{i\beta_2 L(\omega - \omega_0^2)}{2}} = e^{\frac{-i \pi %c DL (\lambda - \lambda_0^2)}{2\lambda_0^2}}
%\end{equation}
%Where
In the time domain, the dispersion manifests as a broadening of coincidence peaks. For our experiments, the broadening will cause the three coincidence peaks to merge into each other, limiting our ability to post-select time-energy Bell States.  The coincidence peak broadens significantly for $L = 5\;\mathrm{km}$ as opposed to $L = 10\;\mathrm{m}$ for the indoor measurement. As a result, the visibility of the Franson interferometer degrades. As the visibility $V$ decreases, the maximum achievable CHSH score $S_{\text{max}}\leq 2\sqrt{2}V$ falls, as per Eq.~(\ref{eqn:tsire_approx2}). For long-distance CHSH demonstrations, we use either a tunable dispersion compensation module (TDCM, FS) or WaveShapers for dispersion compensation; details on the WaveShaper-based setup, with before and after coincidence histograms, are available in the Supplementary Information. 

\subsection{CHSH inequality violation over 5 km deployed fiber}
\begin{figure}[!htbp]
\centering
    \includegraphics[width = 1\linewidth]{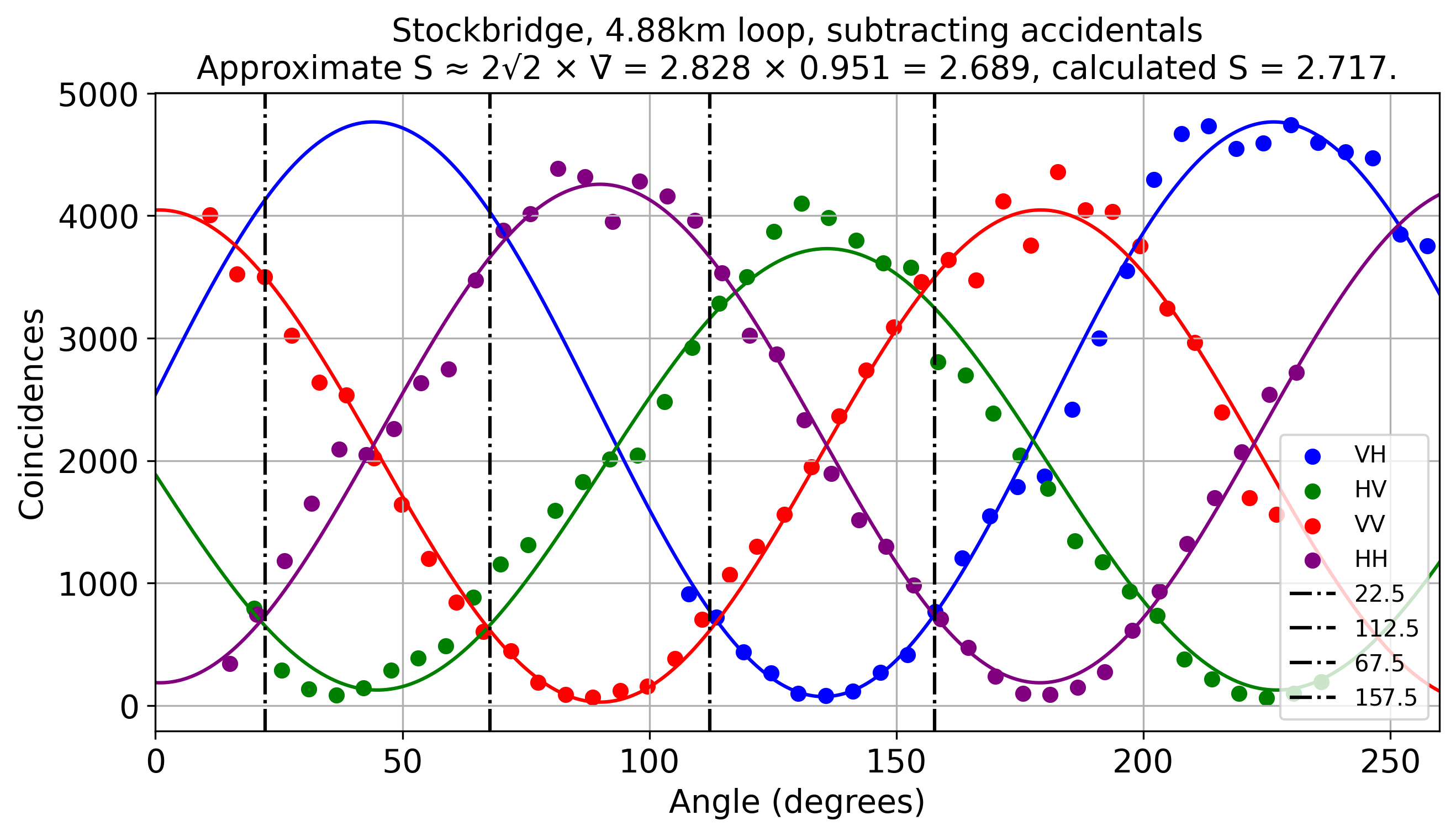}
    \caption{CHSH measurement results over $4.88\;\mathrm{km}$ deployed fiber at the Stockbridge QLAN. Alice's WaveShaper is set to a fixed phase while Bob's WaveShaper performs phase sweeps around a fixed phase denoting orthogonal measurement schemes. Accidentals were subtracted to yield an average visibility of $\bar{V}\sim95\%$ yielding $S=2.717\pm0.016$, in line with the canonical result derived using Eq.~(\ref{eqn:tsire_approx2}), $S_{\text{approx.}}\sim2.689$.}
    %\vspace{128in}
    \label{fig:5km_CHSH_stock}
\end{figure}

\begin{figure}[!htbp]
\centering
    \includegraphics[width = 1\linewidth]{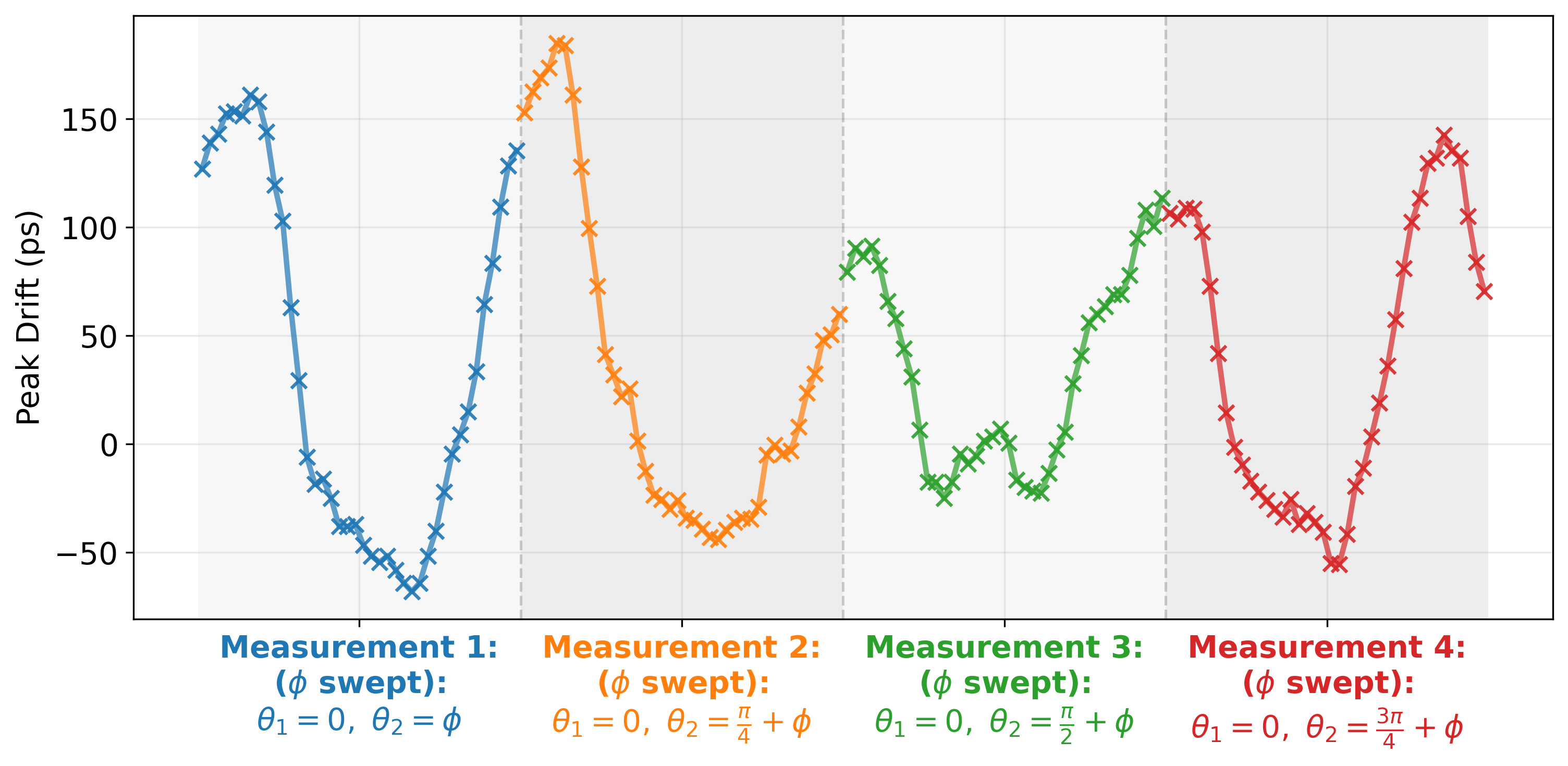}
    \caption{TOF drift for the LL+SS coincidence peak spanning the duration of the experiment in Figure \ref{fig:5km_CHSH_stock}.  For each orthogonal measurement, marked by a different plot color, the phase $\phi$ is swept from $0$ to $2\pi$. Each measurement was recorded over a roughly 6 hour period.}
    \label{fig:5km_CHSH_1ttx_peakdrift}
\end{figure}

Once we measured a CHSH Inequality violation in the laboratory, we moved our setup to the Stockbridge QLAN to repeat the experiment on a fielded network. An experimental diagram is shown in Figure \ref{fig:5km_vis}(a). A spiral source located in the QPad is pumped with a $1550\;\mathrm{nm}$  CW laser tone. The $1530\;\mathrm{nm}$ signal is sent to a WaveShaper in the QPad, then routed directly to an SNSPD, while the $1570\;\mathrm{nm}$ signal is sent through a second WaveShaper, then out to a buried fiber link with a total length of $4.88\;\mathrm{km}$ and looped back to the QPad for dispersion compensation and detection. A CHSH measurement is performed in the same manner as above. Results are shown in Figure \ref{fig:5km_CHSH_stock}. For an achieved average modulation visibility of $95.8\%$ after dispersion compensation, a CHSH score of $S = 2.717\pm0.026$ is achieved ($S=2.691\pm0.026$ before background subtraction), which approaches the Tsirelson bound. Note this is a significance factor of nearly 27.5 standard deviations, $27.5\sigma$ (roughly $26\sigma$ before subtracting accidentals). Included in Fig.~(\ref{fig:5km_CHSH_1ttx_peakdrift}) are time-of-flight measurements in the biphoton coincidence domain.  This is done by tracking the LL+SS peak time (or more accurately, we track the midpoint between SL and LS peaks since the LL+SS  coincidences destructively interfere for certain relative phases).  We also performed this experiment over 5 km deployed fiber at the Griffiss QLAN. For this experiment, detailed in the Supplementary Information, we did not achieve a CHSH violation. We attribute this poorer result to diminished performance our our SNSPDs at Innovare; at the time of this measurement, the SNSPD cryostat was at a higher base temperature, significantly impacting the efficiency of the detectors. A combination of lower efficiency and higher dark counts from the lab environment, and a slightly higher timing jitter, led to a diminished achievable visibility. We plan to repeat this experiment at the Griffiss QLAN at a later date.

It is notable that the achievable visibility at the fielded Stockbridge QLAN surpasses that achieved in our laboratory environment; we attribute the improvement to a decrease in detector dark counts, since the QPad is an isolated environment, unlike a high-traffic laboratory, and has minimal stray light impinging upon optical fibers. The lower loss in a single fiber loopback at Stockbridge, as compared to the Griffiss QLAN, allows for shorter integration times during measurements, which also decreases noise. The achievement of such a strong CHSH violation over a fielded network with significant environment perturbations indicates the promising potential for our QLANs to achieve robust, high-quality entanglement distribution in real-world operating conditions, even before active link stabilization is implemented. Compare this to previously reported results in laboratories and deployed networks: our value of $S=2.700$ exceeds scores achieved in other laboratory-based demonstrations with all-optical networks \cite{Carvacho2017, Poderini2020}, and is also higher than that achieved in deployed networks, such as Ref. \cite{Hensen2015}, where a value of $S=2.42$ was obtained over a much shorter deployed link of 1.3 km, albeit using emissive matter-based qubits. Our results are on par with CHSH values obtained recently over Deutsche Telekom's deployed fiber network for polarization-entangled photons \cite{Sena2025}. While our link distances are shorter than theirs, we may be able to achieve longer distances by repeating our experiments over the longer Griffiss QLAN links, or looping back links at the Stockbridge QLAN. Unlike some other lab-based demonstrations that achieve, for example $S\approx 2.5$ \cite{Lima2010} without the post-selection loophole, our measurements utilize post-selection and hence are not loophole free \cite{Aerts1999}.

Figure \ref{fig:5km_CHSH_1ttx_peakdrift} shows the TOF drift of the coincidence peak (SS + LL) used for CHSH measurements over the duration of the measurement. The center of the peak exhibits an overall drift of $\sim250\;\mathrm{ps}$ over the $\sim24\;\mathrm{hr}$ duration. This TOF drift over $4.88\;\mathrm{km} $ is significantly less than that observed during TOF drift measurements using $10\;\mathrm{ns}$ coherent classical pulses over a different buried link at Stockbridge; the difference in TOF drifts may be attributed to different environmental conditions and different network links, and is still being explored. Moving from the time-energy basis to the polarization or time-bin basis will remove the need for spectral dispersion compensation, but will require stabilization in those other domains. The relevant stabilization solutions are either in place or are under active development at our QLANs.

% \todo[inline]{add remaining data points}

 \subsection{Two time tagger CHSH measurement}
\begin{figure}[!htbp]
\centering
    \includegraphics[width = 1\linewidth]{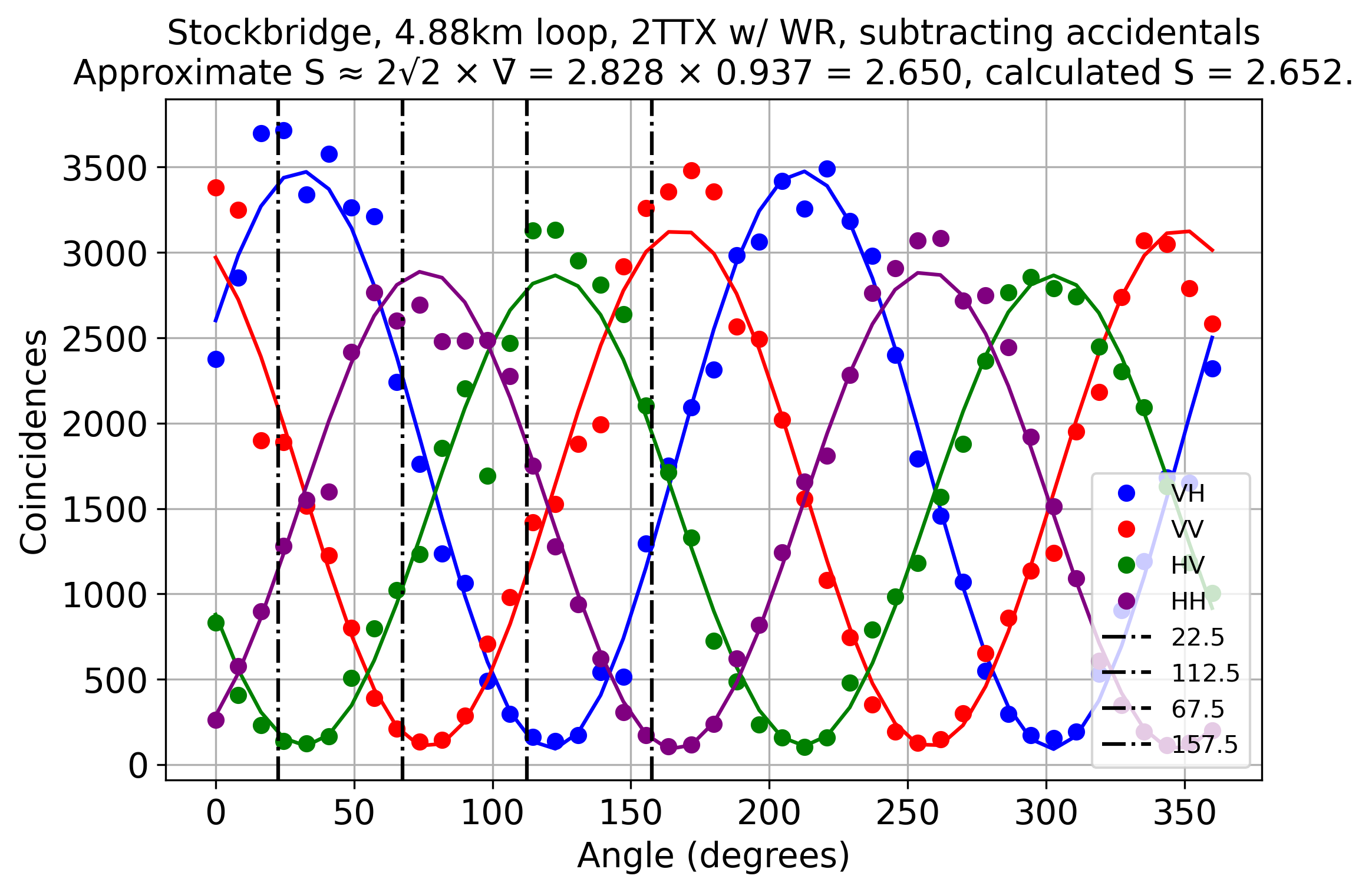}
    \caption{CHSH measurement results over $4.88\;\mathrm{km}$ deployed fiber at the Stockbridge QLAN using tagger channels from separate TTXs. The TTX clocks are synced via Z16 WR Grandmaster and distributed via WR LEN modules. Channel phases are synced through first occurrence of a distributed PPS signal. Accidentals were subtracted to yield an average visibility of $\bar{V}\sim93.7\%$ yielding $S=2.652\pm0.0301$, in line with the canonical result derived using Eq.~(\ref{eqn:tsire_approx2}), $S_{\text{approx.}}\sim2.650$.}
    %\vspace{128in}
    \label{fig:5km_CHSH_2ttx_stock}
\end{figure}

\begin{figure}[!htbp]
\centering
    \includegraphics[width = 1\linewidth]{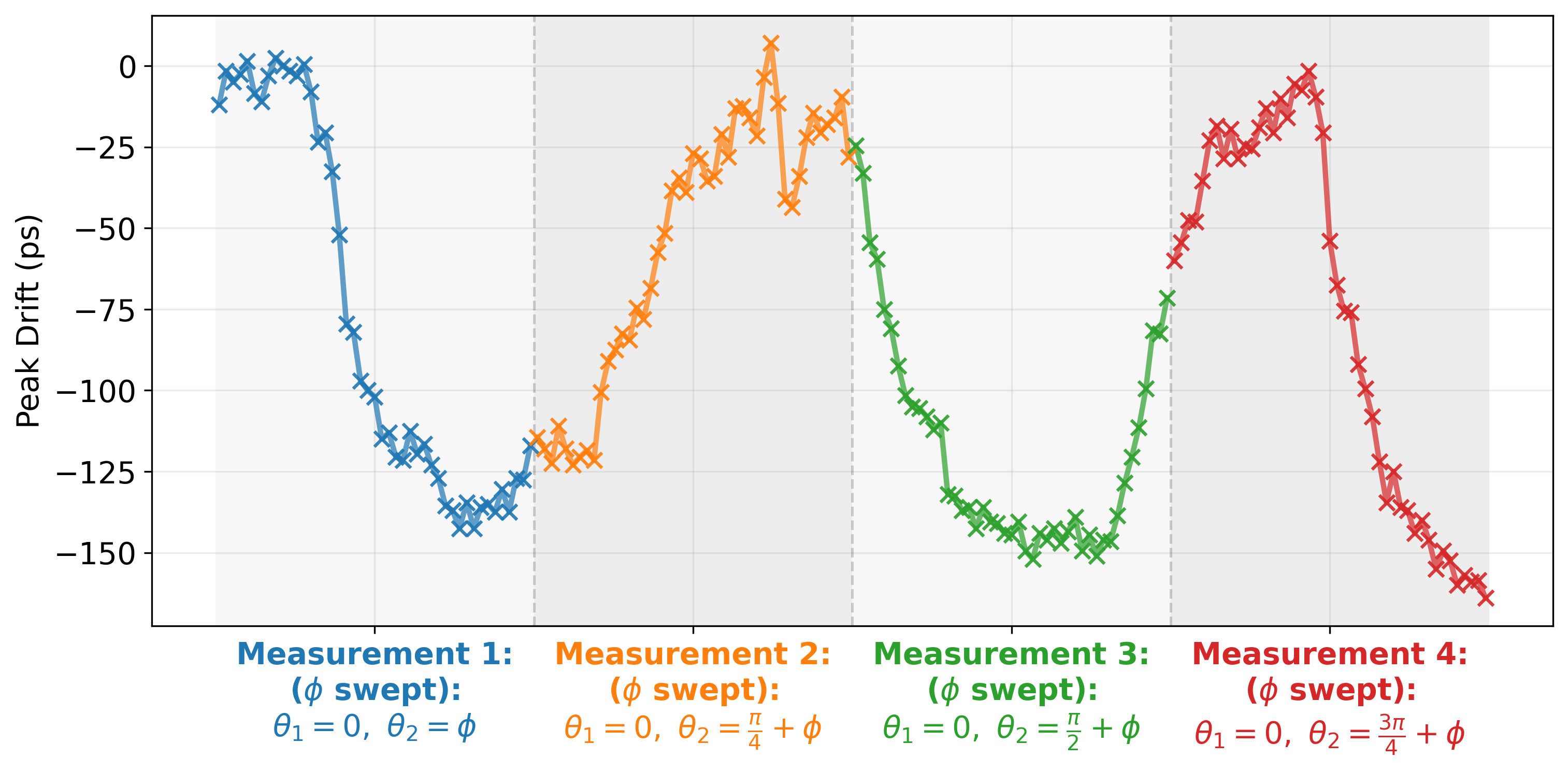}
    \caption{TOF drift for the LL+SS coincidence peak spanning the duration of the experiment in Figure \ref{fig:5km_CHSH_2ttx_stock}.  For each orthogonal measurement, marked by a different plot color, the phase $\phi$ is swept from $0$ to $2\pi$. Each measurement was recorded over a roughly 6 hour period.}
    \label{fig:5km_CHSH_2ttx_peakdrift}
\end{figure}

Finally, we can use two remote and networked time taggers to repeat this demonstration over the same 4.88 km deployed fiber link at Stockbridge, but with distributed coincidence measurements synchronized via WR. As with the experiment shown in Figure \ref{Figure WR sync}, we use the 10 MHz and 1 PPS outputs from WR-LEN nodes to synchronize the TTXs to the distributed clock signal. Time tags are streamed to a network server, and coincident detection events are reconstructed within our software. This capability allows us to perform synchronized networking protocols such as Remote State Preparation. Results are shown in Figure \ref{fig:5km_CHSH_2ttx_stock}. We achieved a visibility of $93.7\%$, about a two percentage points lower than achieved with a single time tagger, and a CHSH score of $S=2.652\pm0.0301$ with a significance factor of roughly $22\sigma$, constituting a slightly weaker violation than what was obtained using two channels from the same TTX. We note that this value obtains after accidentals are subtracted; with accidentals, we achieve $S=2.639\pm0.03$ with a significance factor of roughly $21\sigma$ and an average visibility of $\bar{V}=93.2\%$.  The average accidentals rate was roughly $6$ counts per run. Switching from a single TTX setup to a two-TTX remote coincidence setup typically increases overall timing jitter (per coincidence ) by $7\%$ and decreases the signal-to-noise ratio by $9\%$, which likely causes the slight decrease in the achievable visibility. As with the single TTX CHSH measurement, we plot the TOF drift of the SS+LL coincidence peak in Figure\ref{fig:5km_CHSH_2ttx_peakdrift}.  Here, the drift resembles that observed for the White-Rabbit synchronized TOF drift measurement at the Griffiss QLAN (Figure \ref{Figure WR sync}): TOF drift has periods of hovering around $0\;\mathrm{ps}$, then drifts only in the negative direction, with an overall maximum drift reaching about $150\;\mathrm{ps}$. Overall, this measurement confirms our ability to perform quantum networking protocols over deployed fiber that rely on WR-based synchronization, and to obtain high performance matching that achieved in a laboratory.

\section{\label{Discussion}Discussion and Outlook}

\begin{figure}[!htbp]
    \includegraphics[width = 0.9\linewidth]{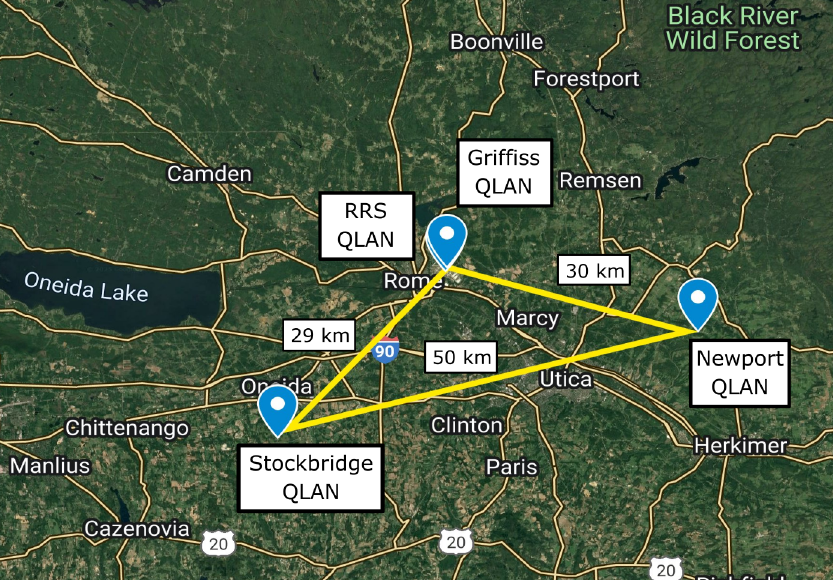}
    \caption{Future QMAN vision. Satellite image with the Stockbridge, Griffiss and RRS QLANs labeled, plus the site of a future QLAN at AFRL's Newport Site. These QLANs can be connected via fiber and/or free space to form a QMAN in Central New York. Satellite image obtained via Google MyMaps, copyright Google 2025, NASA 2025.}
    %\vspace{128in}
    \label{Figure QMAN map}
\end{figure}

Here, we have detailed the implementation of three QLANs at the Air Force Research Laboratory (AFRL) Information Directorate in Central New York. The QLANs consist of indoor and outdoor buried and aerial fiber links, as well as telecom free-space optical links. Characterization of the deployed infrastructure in a variety of conditions shows the diversity of environments available at our quantum networking testbeds. Unlike other quantum network testbeds that have been established, the AFRL testbeds allow for demonstrations both in a settled, high-traffic city environment, and in a heavily wooded, rugged outdoor environment that offers control of the local RF spectrum. As far aw we aware, there is only one other quantum fiber network testbed operating at least partially in a wooded environment \cite{Kucera2024}, though there, the fiber simple runs through a forest to connect two urban areas. Furthermore, the Griffiss and RRS QLANs offer proximity to trapped ion qubits and superconducting circuit/transduction systems, as well as quantum photonic integrated devices, which opens the door to heterogeneous quantum networking demonstrations that have not yet been achieved. We are actively developing these quantum network nodes and interconnects toward telecom network integration. A key aspect in pushing these technologies forward is the use of photonic integrated circuits wherever feasible.

The rapidly expanding ecosystem of deployed quantum networking testbeds provides increasingly diverse insights into practical deployment challenges. While metropolitan networks \cite{dynes2019, Craddock2024, Bersin2024} have validated quantum networking in urban environments, and research laboratory-based testbeds \cite{Alshowkan2021, Chung2021, Monga2023} have pioneered flexible architectures and topologies, our multi-environment QLAN approach specifically addresses the systematic characterization of environmental effects on quantum network performance in fiber and free space. This work complements recent advances in matter-qubit integration \cite{Bersin2024a, Krutyanskiy2024} and quantum-classical coexistence \cite{Rahmouni2024, Thomas2024} by providing the environmental stability analysis needed to deploy such capabilities in diverse real-world conditions.

Going forward, we will pursue QLAN fiber and free-space inter-connectivity toward the establishment of a Quantum Metropolitan Area Network (QMAN), as shown in Figure \ref{Figure QMAN map}. With the two QLANs in Rome (RRS and Griffiss) separated from each other by less than a kilometer, and from Stockbridge by about 30 kilometers, the three QLANs can eventually be connected to each other, and to a future QLAN to be set up at AFRL's Newport test site. The Newport test site is equipped with an elevated outdoor antenna test range and a buried optical fiber network similar to the one at the Stockbridge test site. Together, these inter-connected QLANs can form a QMAN for the Air Force in Central New York. This QMAN can be connected to other quantum testbeds in New York State \cite{Craddock2024, Sundaram2025, Du2021}.

In the near term, we will refine and expand our QLAN infrastructure, with a continued focus on cryogenic quantum nodes, atom-based quantum nodes, mobile quantum nodes, and link stabilization and control. Specifically, we anticipate that in the next one to five years, with proper quantum transduction and frequency conversion, trapped-ion and superconducting-circuit based qubits, and quantum/classical sensors will be connected to our QLANs for advanced quantum networking demonstrations. For example, our ability to control the RF spectrum at the Stockbridge QLAN makes it a perfect location for quantum RF sensing networks. Going forward, AFRL's QLANs will serve as qubit-agnostic telecom platforms for heterogeneous quantum networking, with applications in distributed quantum computing, distributed quantum sensing, long range interferometry, quantum communications, and other quantum-enabled capabilities.

\definecolor{yesgreen}{RGB}{144,238,144}
\definecolor{devyellow}{RGB}{255,255,153}
\definecolor{notetorange}{RGB}{255,204,153}
\definecolor{nogrey}{RGB}{211,211,211}
\definecolor{subheadergray}{RGB}{192,192,192}

\begin{table*}[htbp]
\centering
\caption{Quantum Local Area Network (QLAN) Capabilities and Resources Comparison}
\resizebox{\textwidth}{!}{%
\begin{tabular}{|l|c|c|c|}
\hline
\rowcolor{headergray}
\textbf{Capability/Resource} & \textbf{Griffiss QLAN} & \textbf{Stockbridge QLAN} & \textbf{RRS QLAN} \\
\hline
\hline
\multicolumn{4}{|c|}{\cellcolor{subheadergray}\textbf{PHYSICAL INFRASTRUCTURE}} \\
\hline
Buried outdoor fiber & \cellcolor{yesgreen}Yes & \cellcolor{yesgreen}Yes & \cellcolor{nogrey}No \\
\hline
Aerial outdoor fiber & \cellcolor{nogrey}No & \cellcolor{yesgreen}Yes & \cellcolor{yesgreen}Yes \\
\hline
Free space link & \cellcolor{yesgreen}Yes, indoor & \cellcolor{yesgreen}Yes, outdoor & \cellcolor{nogrey}No \\
\hline
RF spectrum control & \cellcolor{nogrey}No & \cellcolor{yesgreen}Yes & \cellcolor{nogrey}No \\
\hline
Walkup tower connectivity and access & \cellcolor{nogrey}No & \cellcolor{yesgreen}Yes & \cellcolor{yesgreen}Yes \\
\hline
\hline
\multicolumn{4}{|c|}{\cellcolor{subheadergray}\textbf{NETWORK INFRASTRUCTURE}} \\
\hline
Parallel fiber links for dedicated classical LAN & \cellcolor{yesgreen}Yes & \cellcolor{yesgreen}Yes & \cellcolor{yesgreen}Yes \\
\hline
Computer/ethernet LAN control plane & \cellcolor{yesgreen}Yes & \cellcolor{yesgreen}Yes & \cellcolor{devyellow}Under development \\
\hline
SFP transceivers and switches & \cellcolor{yesgreen}Yes & \cellcolor{yesgreen}Yes & \cellcolor{yesgreen}Yes \\
\hline
White Rabbit-based timing synchronization & \cellcolor{yesgreen}Yes & \cellcolor{yesgreen}Yes & \cellcolor{yesgreen}Yes \\
\hline
GPS Network Time Protocol Server & \cellcolor{notetorange}Not yet & \cellcolor{yesgreen}Yes & \cellcolor{notetorange}Not yet \\
\hline
\hline
\multicolumn{4}{|c|}{\cellcolor{subheadergray}\textbf{QUANTUM SOURCES \& DETECTION}} \\
\hline
PIC entanglement sources & \cellcolor{yesgreen}Yes & \cellcolor{yesgreen}Yes & \cellcolor{yesgreen}Yes \\
\hline
SNSPDs and TTXs & \cellcolor{yesgreen}Yes & \cellcolor{yesgreen}Yes & \cellcolor{yesgreen}Yes \\
\hline
Quantum state tomography and CHSH & \cellcolor{yesgreen}Yes & \cellcolor{yesgreen}Yes & \cellcolor{devyellow}Under development \\
\hline
Photon basis interconversion modules\cite{Nehra2025} & \cellcolor{yesgreen}Yes & \cellcolor{notetorange}Not yet & \cellcolor{notetorange}Not yet \\
\hline
\hline
\multicolumn{4}{|c|}{\cellcolor{subheadergray}\textbf{ENVIRONMENTAL MONITORING}} \\
\hline
Weather monitoring (e.g. temp, wind) & \cellcolor{yesgreen}Yes & \cellcolor{yesgreen}Yes & \cellcolor{yesgreen}Yes \\
\hline
Air turbulence monitoring & \cellcolor{nogrey}No & \cellcolor{yesgreen}Yes & \cellcolor{nogrey}No \\
\hline
Disdrometer monitoring & \cellcolor{nogrey}No & \cellcolor{yesgreen}Yes & \cellcolor{nogrey}No \\
\hline
Air visibility monitoring & \cellcolor{nogrey}No & \cellcolor{yesgreen}Yes & \cellcolor{nogrey}No \\
\hline
\hline
\multicolumn{4}{|c|}{\cellcolor{subheadergray}\textbf{MATTER QUBIT INTEGRATION}} \\
\hline
Proximity to trapped ion nodes & \cellcolor{yesgreen}Yes & \cellcolor{notetorange}Not yet & \cellcolor{notetorange}Not yet \\
\hline
Proximity to superconducting nodes & \cellcolor{yesgreen}Yes & \cellcolor{notetorange}Not yet & \cellcolor{yesgreen}Yes \\
\hline
Classical/quantum sensor nodes & \cellcolor{notetorange}Not yet & \cellcolor{devyellow}Under development & \cellcolor{notetorange}Not yet \\
\hline
\hline
\multicolumn{4}{|c|}{\cellcolor{subheadergray}\textbf{LINK STABILIZATION \& CONTROL}} \\
\hline
Link polarization stabilization & \cellcolor{yesgreen}Yes & \cellcolor{yesgreen}Yes & \cellcolor{notetorange}Not yet \\
\hline
Link TOF stabilization & \cellcolor{devyellow}Under development & \cellcolor{notetorange}Not yet & \cellcolor{notetorange}Not yet \\
\hline
Link phase stabilization & \cellcolor{devyellow}Under development & \cellcolor{notetorange}Not yet & \cellcolor{notetorange}Not yet \\
\hline
Pointing and tracking for free space optical links & \cellcolor{devyellow}Under development & \cellcolor{devyellow}Under development & \cellcolor{notetorange}Not yet\\
\hline
Phase-stabilized fiber and free space interferometers & \cellcolor{devyellow}Under development & \cellcolor{notetorange}Not yet & \cellcolor{notetorange}Not yet \\
\hline
\end{tabular}%
}
\end{table*}

\begin{acknowledgments}
We wish to acknowledge those who worked to build and maintain the facilities and other infrastructure required for our QLANs, including Mike Wessing of the Griffiss Institute, Michael McGovern of Oneida County, and  David Overrocker and Mike Hartnett of AFRL/RIT. We also thank the contracting professionals in AFRL/RIK who have worked with us for years to build up our laboratories and networks. Thank you for the tireless work you do that makes this science possible.

We also express our gratitude to Mael Flament, Shane Andrewski and Gabriel Portmann at Qunnect Inc. for assisting in the setup and analysis of initial polarization stabilization experiments with the QU-APCs at the Griffiss and Stockbridge QLANs.
\end{acknowledgments}

\section{Author Contributions}
M.F., C.T., A.M.S. and S.P, designed PIC spiral source chips.
N.B., S.D., V.N., R.B. and E.S. installed and connected Griffiss QLAN infrastructure.
M.D., S.F., J.H., V.B., V.N., N.B., C.N., R.B., B.K. and E.S. installed, connected and maintained Stockbridge QLAN infrastructure. 
E.S., N.B., R.B., B.K. and V.N. performed White Rabbit experiments. 
R.B., V.N., S.D., C.T. and J.S. experimentally and theoretically characterized PIC source chips.
V.N., R.B., N.B., C.T., and J.S. designed, performed time-energy CHSH experiments and analyzed data. 
S.S., D.C., D.S., E.S., N.B. and M.L. established and connected cryogenic superconducting network nodes to the QLANs.
E.S., D.H., Z.S., A.M.S., K.S., P.A., M.L., S.S., D.C., L.W., M.F., and D.T. directed and supervised the project. 
All authors contributed to drafting and editing the manuscript.

\section{Data availability statement}
Data sets generated during the current study are available from the corresponding author upon request.

\section{Conflicts of interest}
The authors declare no conflicts of interest.

\section{Disclaimers}
Unless otherwise stated, images are property of the United States Air Force.

\newpage
%\putbib[main] % Uses main.bib
%\end{bibunit}
\clearpage
%\section*{\centering References}
\bibliography{main}% Produces the bibliography via BibTeX.
\onecolumngrid
\newpage
%\begin{bibunit}[plain]
\renewcommand{\thesection}{S\arabic{section}}
\setcounter{section}{0}
\centering
\noindent\textbf{\Large SUPPLEMENTARY INFORMATION}
\vspace{0.5em}
\hrule
\section{Introduction}
This Supplementary note contains additional diagrams and data plots that have been withheld from the Main Text due to length considerations. The details included here provide a complete picture of AFRL's QLAN topologies and capabilities. 

\section{QLAN setups and control planes}

\subsection{Griffiss QLAN}
The Griffiss QLAN ring network consists of four classical telecom interfaces that are available to accept and route signals from quantum nodes located throughout the laboratory space. In assembling these nodes, we focused on sourcing as many commercial-off-the-shelf components as possible to minimize cost and maximize compatibility with existing fielded telecom networks.

\subsubsection{Node setup}
\begin{figure}
    \centering  \includegraphics[width=0.5\linewidth,keepaspectratio]{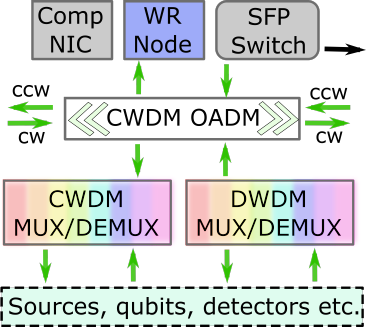}
    \caption{Detailed diagram of the classical LAN node setup for each of the four OADM ring nodes. Signals from WR, SFP switches, and computer NIC cards are routed to and from each node using the CWDM OADMs. CWDM and DWDM multiplexers are in place for general purpose quantum/classical signal multiplexing/de-multiplexing.}
    \label{app fig:griffiss nodes}
\end{figure}

Figure \ref{app fig:griffiss nodes} shows a detailed diagram of the constituent hardware comprising the classical telecom interface of each network node at Griffiss. Coarse-wavelength division optical-add-drop multiplexers (CWDM OADMs) are the interface between each node and the ring. Each OADM has eight C-band wavelength channels, spanning $1470\; \mathrm{nm} - 1610\;\mathrm{nm}$ and spaced by $20\;\mathrm{nm}$,  and an O-band channel with a passband of  $1310\; \mathrm{nm} - 1410\;\mathrm{nm}$. The OADMs allow for bi-directional signal propagation in each of the duplex fiber links, so that the ring has separate clockwise and counter-clockwise signal propagation paths. The ring currently consists of four nodes, but can easily be expanded to include more nodes by adding more OADMs. The number of available wavelength channels at each node can also be expanded by dense wavelength division multiplexers (DWDMs) and O-band CWDMs. Each node can act as a Source node (all-photonic entanglement source), a Memory node (a matter-based qubit that emits an entangled photon) or an analysis node (OTDR, Bell State analyzer, polarization state tomography, polarimeter, etc.) Of course, a node can have more than one of these capabilities at once. 

\subsubsection{Control plane}
The Griffiss QLAN control plane consists of an ethernet-based local area network. Ethernet switches (Netgear GS110MX) connect dedicated computers at each node to each other and to experimental hardware, including White Rabbit nodes, lasers, compensators, analysis hardware, and more. The loop-back, in-laboratory nature of the network allows the control plane to be implemented over ethernet (copper) cables, but SFP switches (FS S3950-4T12S-R) are also available for the extension of the control plane over long-haul fiber links. When the Griffiss QLAN is expanded to include additional deployed fiber, including inter-site connections, a fiber-based classical ethernet LAN will be necessary.

\subsubsection{Detection and analysis capabilities}
The Detection and Analysis manifold at the Griffiss QLAN includes eight superconducting nanowire single photon detectors (SNSPDs, Single Quantum), two time-to-digital converters (TTXs, Swabian Instruments Time Tagger X), fiber-based interferometers for entanglement swapping and analysis of number-state, polarization, or time-bin encoded photons, a polarization tomography module (Oz Optics), WaveShapers for time-energy Bell state analysis (Finisar WaveShaper 1000A and 4000A) and polarimeters (Thorlabs PAX1000 models, Qunnect QU-APC), and an optical time domain reflectometer (OTDR). A sampling of these components is shown in Figure \ref{app fig:griffiss analysis}. An optical switch routes signals from the ring network to the appropriate analysis module for a given experiment.

Additional single photon detectors and time-to-digital converters are located within the laboratory, including e.g. SNSPDs located inside of dilution refrigerators, which can be integrated into the QLAN as needed. This setup maximizes the reconfigurability of the network, so that network nodes can be assigned as Source, Memory, Analysis, or a combination of these functionalities simultaneously, while maintaining the relative autonomy of the control and readout hardware for the independent superconducting, integrated photonics and trapped ion setups. 

The Stockbrdige QLAN is equipped with many identical detection and analysis components, including the same Keysight laser system, polarimeter, time taggers, OTDR, QU-APC, and more. One difference is in the available SNSPDs. At Griffiss we have eight Single Quantum detectors at an operating wavelength of 1550 nm, situated in an Eos cryostat and Atlas driver, and two PhotonSpot SNSPDs at 1550 nm located within a customaized setup inside a dilution refrigerator. At Stockbridge, we have a new Single Quantum Eos R12 rack-mounted cryostat (with a Retina 1200M driver and IGLU compressor) which houses eight 1550 nm detectors and four 1310 nm detectors. Half of the detectors at each wavelength are equipped with an ultra-low jitter upgrade, and the other half have time-gating functionality enabled.  At RRS, we have multiple sets of SNSPDs available that are used within existing photonics and superconducting setups, including multiple cryostat systems from PhotonSpot and Single Quantum, with wavelengths spanning visible to telecom wavelengths, including two 1550 nm PhotonSpot detectors located within a dilution refrigerator.

\begin{figure}
    \centering  \includegraphics[width=0.5\linewidth,keepaspectratio]{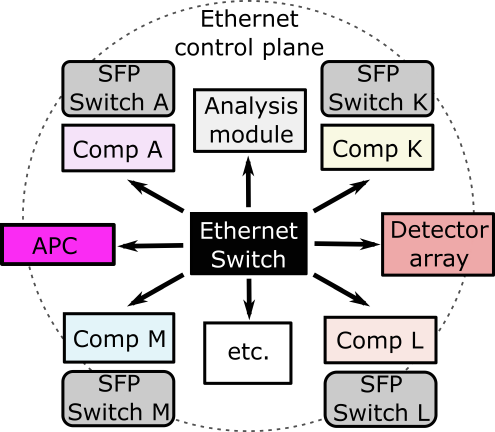}
    \caption{Diagram of the ethernet control plane for the Griffiss QLAN. Four computers (labeled Archer (A), Kreiger (K), Mallory (M) and Lana (L) are connected, as well as experimental hardware. Each network node has an SFP switch to extend the control plane over fiber. The computers are equipped with NIC cards as an optical network interface if/when needed.}
    \label{app fig:griffiss control plane}
\end{figure}

\begin{figure}
    \centering  \includegraphics[width=0.5\linewidth,keepaspectratio]{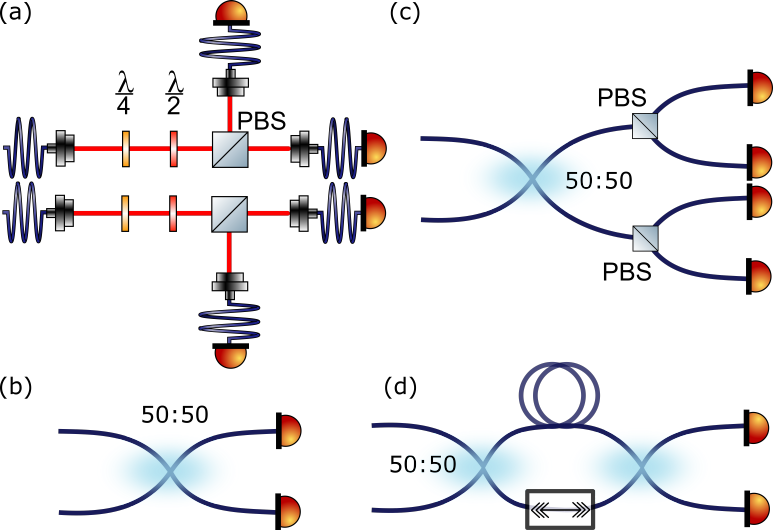}
    \caption{A selection of a few analysis modules at the Griffiss QLAN, to be installed at the Stockbridge QLAN as well. (a) Polarization-state tomography module. $\lambda/4$: quarter waveplate. $\lambda/2$: half waveplate. PBS: polarizing beamsplitter. These are available in the telecom C-band and in a custom bichromatic configuration (795/1324 nm) for the Qunnect entanglement source. (b-d) Fiber-based interferometers to analyze photon qubits in the (b) number state, (c) polarization and (d) time-bin bases. A piezo-stretcher is shown in the unbalanced interferometer in panel (d), but is likely also necessary to stabilize the phase of optical signals propagating through the interferometers shown in (b,c). Not shown: fiber-based polarization controllers.}
    \label{app fig:griffiss analysis}
\end{figure}

\subsection{Stockbridge QLAN}
\subsubsection{Control plane}
\begin{figure*}
    \centering  \includegraphics[width=\linewidth,keepaspectratio]{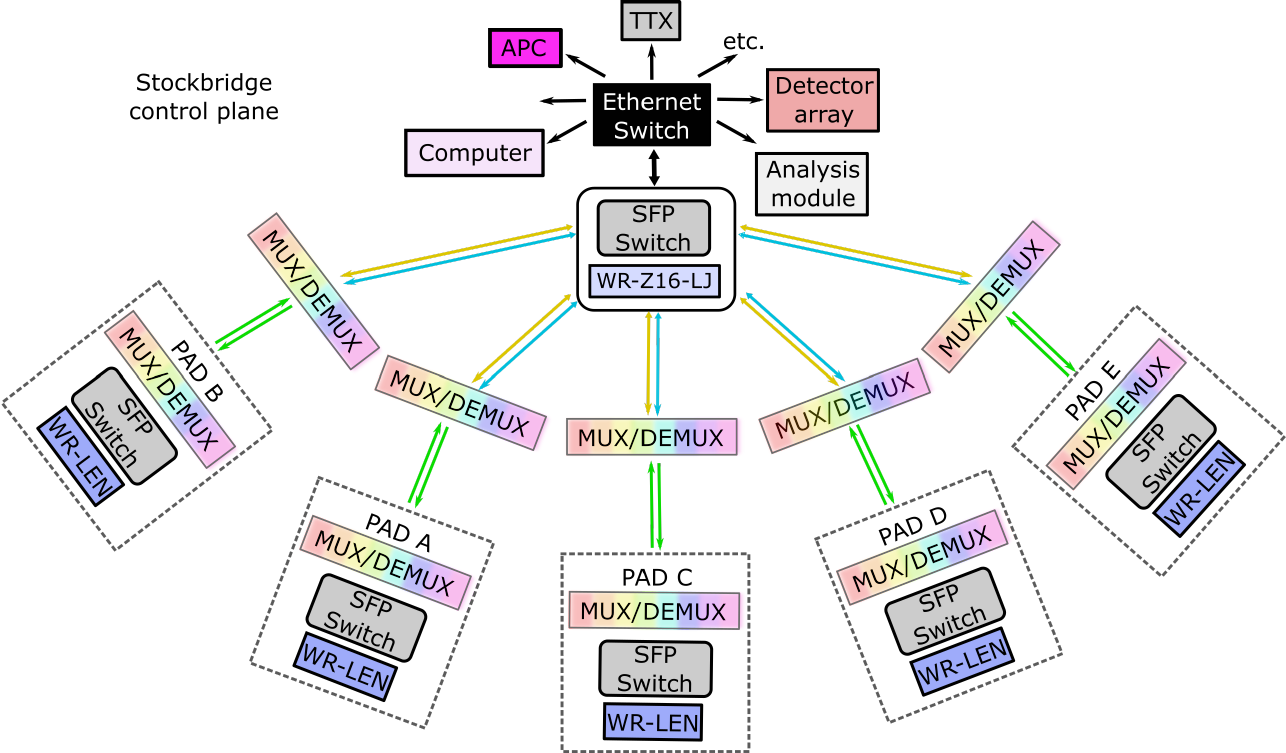}
    \caption{The Stockbridge control plane uses the same hardware components as the Griffiss QLAN. Blue lines: WR links. Yellow lines: ethernet links. Green lines: deployed fiber connecting remote PADs. Here, the ethernet LAN is extended over dedicated classical fiber channels to connect the Quantum PAD network hub to the spoke PADs. CWDM multiplexers/demultiplexers are used to multiplex the ethernet LAN signals with White Rabbit signals for clock distribution. Additional classical signals, such as probe signals for stabilization and test signals for characterization, can also be multiplexed in on unused CWDM or DWDM channels.}
    \label{app fig:stock control plane}
\end{figure*}

The hub-and-spoke topology of the Stockbridge QLAN requires its control plane to be implemented over fiber. As shown in Figure \ref{app fig:stock control plane}, an ethernet switch located in the QPAD is connected to a network hub computer and experimental hardware. The switch is linked to an SFP switch, which extends the control plane to network spoke nodes via dedicated classical fiber links. The control plane signals sit at the $1590$ or $1610\;\mathrm{nm}$ CWDM channels, and can be moved to DWDM channels if  more bandwidth becomes necessary. This classical fiber channel also hosts wavelength-multiplexed White Rabbit (WR) links between the QPAD and the spoke nodes, where WR signals (either $1310/1490\;\mathrm{nm}$,  $1490/1550\;\mathrm{nm}$, or  $1490/1570\;\mathrm{nm}$ wavelength pairs) connect a WR-Z-16-LJ switch in the QPAD to WR-LEN nodes at each node.

\section{White Rabbit topologies at Griffiss QLAN}
\begin{figure*}
    \centering  \includegraphics[width=\linewidth,keepaspectratio]{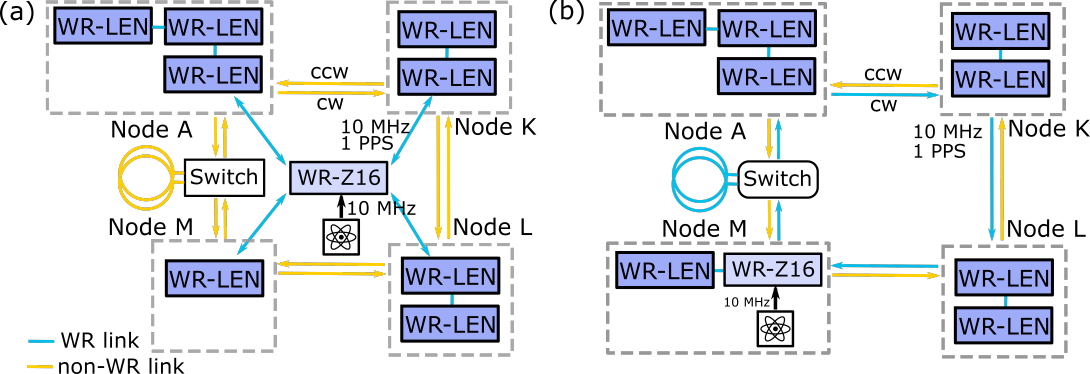}
    \caption{Diagrams of WR configurations that have been implemented at the Griffiss QLAN, including (a) dedicated WR links and (b) shared links with experimental quantum signals. We can toggle back and forth between these configurations depending on the needs of each experiment.}
    \label{app fig:stock bruied pol}
\end{figure*}

\section{\label{app:plots}Additional SOP data}
Here we present an additional set of data plots from an SOP monitoring measurement at the Stockbridge QLAN, of buried fiber only. 

At about 11 hours elapsed, a strong wind gust occurs, and afterwards a ringing behavior can be seen in the fiber SOP. This may indicate that the wind gust hit a small length of fiber mounted on the exterior of a PAD, causing a long series of mechanical oscillations impacting the SOP. Like our other measurements, we calculated linear correlation relationships between the SOP angles and temperature and wind speed, as shown in Tables I and II in the Main Text, and below.

\begin{figure}
    \centering  \includegraphics[width=\linewidth,keepaspectratio]{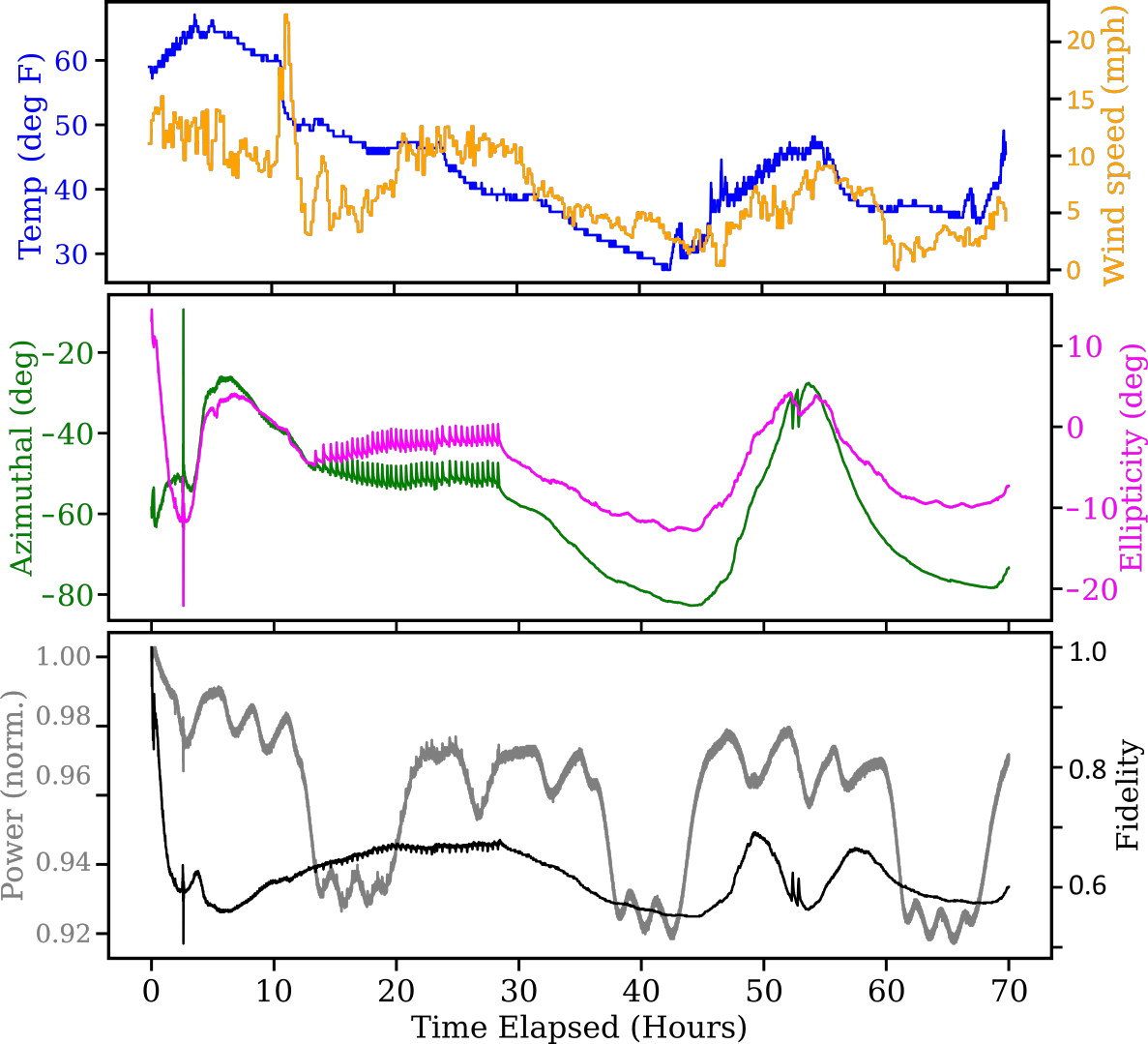}
    \caption{Example polarization drift measurement for buried fiber (no aerial links) connecting PAD A to the Command Center at the Stockbridge QLAN. Temperature and average wind speed are monitored simultaneously.}
    \label{app fig:stock buried pol}
\end{figure}

\section{Correlation results}

% Define colors for shading
\definecolor{headergray}{gray}{0.8}
\definecolor{veryweak}{gray}{0.9}
\definecolor{weak}{RGB}{144,238,144}
\definecolor{moderate}{RGB}{255,255,153}
\definecolor{strong}{RGB}{255,182,193}

\subsection{Tables of Pearson linear fit equations}
\begin{table*}[htbp]
\centering
\caption{TOF Drift Linear Fit Equations corresponding to correlation Table I in the Main Text. Color coding maintained from correlation strength.}
\resizebox{\textwidth}{!}{%
\begin{tabular}{|c|c|c|c|c|c|c|}
\hline
\rowcolor{headergray}
\textbf{Figure} & \textbf{QLAN} & \textbf{Link type} & \textbf{TOF drift vs wind speed} & \textbf{TOF drift vs temp overall} & \multicolumn{2}{c|}{\textbf{TOF drift vs temp split}} \\
\hline
\rowcolor{headergray}
 &  &  & \textbf{Equation} & \textbf{Equation} & \textbf{Below mean} & \textbf{Above mean} \\
\hline
10& G & B & \cellcolor{weak}$1.19x - 310.6$ & \cellcolor{veryweak}$-3.64x - 91.3$ & \cellcolor{veryweak}$-3.10x - 91.3$ & \cellcolor{veryweak}$0.54x - 366.0$ \\
\hline
12& G & B & \cellcolor{weak}$0.008x + 4.76$ & \cellcolor{weak}$0.003x + 4.59$ & \cellcolor{moderate}$0.005x + 4.43$ & \cellcolor{veryweak}$-0.002x + 4.93$ \\
\hline
11(c)& S & B & \cellcolor{weak}$-129.45x + 1230.9$ & \cellcolor{veryweak}$-4.934x + 1183.32$ & \cellcolor{moderate}$-121.224x + 8201.50$ & \cellcolor{weak}$40.393x - 1865.56$ \\
\hline
11(g)& S & A & \cellcolor{veryweak}$-0.02x + 0.2$ & \cellcolor{weak}$-0.014x + 1.225$ & \cellcolor{strong}$0.038x - 2.23$ & \cellcolor{moderate}$-0.057x + 4.38$ \\
\hline
\end{tabular}%
}
\end{table*}

\begin{table*}[htbp]
\centering
\caption{Ellipticity and Azimuthal Drift Linear Fit Equations. Same experimental conditions as Table II in the Main Text. Color coding maintained from correlation strength.}
\resizebox{\textwidth}{!}{%
\begin{tabular}{|c|c|c|c|c|c|c|}
\hline
\rowcolor{headergray}
\textbf{Figure} & \textbf{QLAN} & \textbf{Link type} & \textbf{Ellipticity drift vs wind speed} & \textbf{Ellipticity drift vs temp} & \textbf{Azimuthal drift vs wind speed} & \textbf{Azimuthal drift vs temp} \\
\hline
\rowcolor{headergray}
 &  &  & \textbf{Equation} & \textbf{Equation} & \textbf{Equation} & \textbf{Equation} \\
\hline
4(b)& G & B & \cellcolor{moderate}$-0.162x - 25.56$ & \cellcolor{moderate}$-0.049x - 26.31$ & \cellcolor{moderate}$-0.824x + 66.56$ & \cellcolor{moderate}$-0.226x + 69.79$ \\
\hline
4(e)& G & B & \cellcolor{moderate}$1.839x + 3.36$ & \cellcolor{moderate}$0.491x - 23.24$ & \cellcolor{weak}$5.696x + 44.55$ & \cellcolor{moderate}$2.785x - 119.42$ \\
\hline
12& G & B & \cellcolor{weak}$0.666x + 29.99$ & \cellcolor{weak}$0.166x + 18.72$ & \cellcolor{strong}$4.490x - 87.89$ & \cellcolor{weak}$0.495x - 115.88$ \\
\hline
6& R & A & \cellcolor{veryweak}$-0.097x -30.85$ & \cellcolor{veryweak}$-0.053x -27.10$ & \cellcolor{moderate}$-0.341x-11.94$ & \cellcolor{veryweak}$-0.043x -9.67$ \\
\hline
11(a)& S & B & \cellcolor{moderate}$-0.556x - 12.34$ & \cellcolor{strong}$-0.205x - 0.93$ & \cellcolor{strong}$-1.560x + 68.68$ & \cellcolor{moderate}$-0.284x + 82.27$ \\
\hline
11(e)& S & A & \cellcolor{moderate}$0.528x + 17.26$ & \cellcolor{strong}$-0.296x + 39.73$ & \cellcolor{weak}$1.877x + 67.30$ & \cellcolor{strong}$-1.364x + 34.59$ \\
\hline
7& S & B \& A & \cellcolor{veryweak}$-0.242x - 8.36$ & \cellcolor{veryweak}$-0.207x + 4.49$ & \cellcolor{veryweak}$0.688x - 22.95$ & \cellcolor{weak}$-0.797x - 73.35$ \\
\hline
\ref{app fig:stock buried pol} & S & B  & \cellcolor{strong}$1.249x-11.679$ & \cellcolor{moderate}$0.371x -18.464$ & \cellcolor{strong}$2.211x-74.583$ & \cellcolor{strong}$1.226x-111.462$ \\
\hline
\end{tabular}%
}
\end{table*}

\begin{figure*}
    \centering  \includegraphics[width=\linewidth,keepaspectratio]{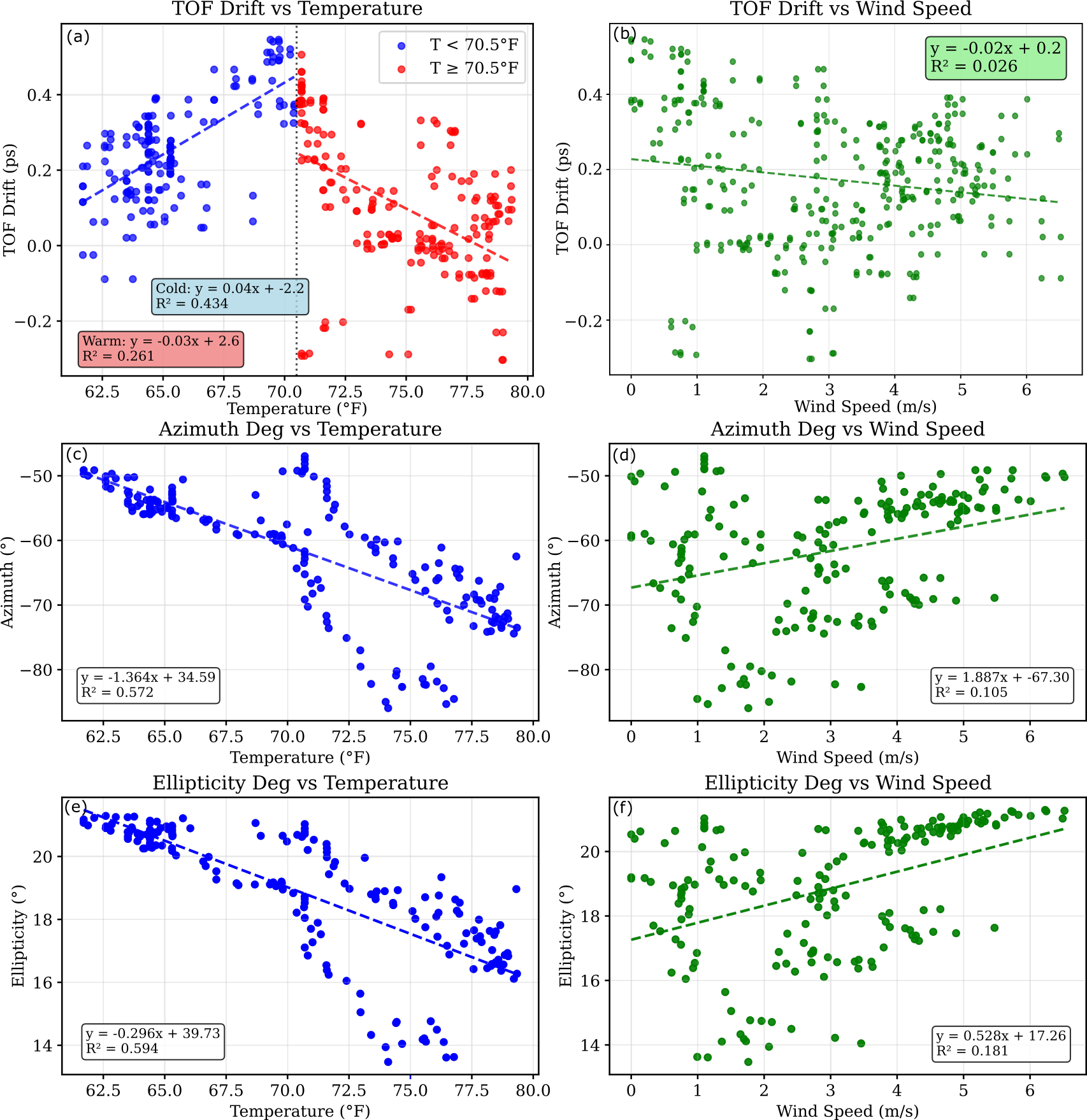 }
    \caption{Linear correlation results for (a,b) TOF drift and (c-f) SOP angle drift vs. environmental parameters for aerial fiber at the Stockbridge QLAN. The TOF drift shows moderate correlations with temperature when asymmetric heating about a mean temperature of $70.5^{\circ}$F is taken into account, and shows minimal correlation with wind speed. The polarization angles exhibit moderate correlations with temperature and weak correlations with wind speed.}
    \label{app fig:stock WUT correlations}
\end{figure*}

Here we show example output from our linear correlation modeling of TOF and SOP drift data. The plots in Figure \ref{app fig:stock WUT correlations} correspond to the experiment summarized Figure 10 in the Main Text; a TOF/SOP drift measurement on aerial fiber at the Stockbridge QLAN.

\section{Quantum-classical signal coexistence}
While not a main thrust of this work, we explored the ability of our QLANs to simultaneously operate quantum and classical signals in a single fiber or free space channel. The impact of the OADM ring hardware on wavelength crosstalk is considered. 

\subsection{Wavelength multiplexing in fiber}
\begin{figure}
    \includegraphics[width = 0.5\linewidth]{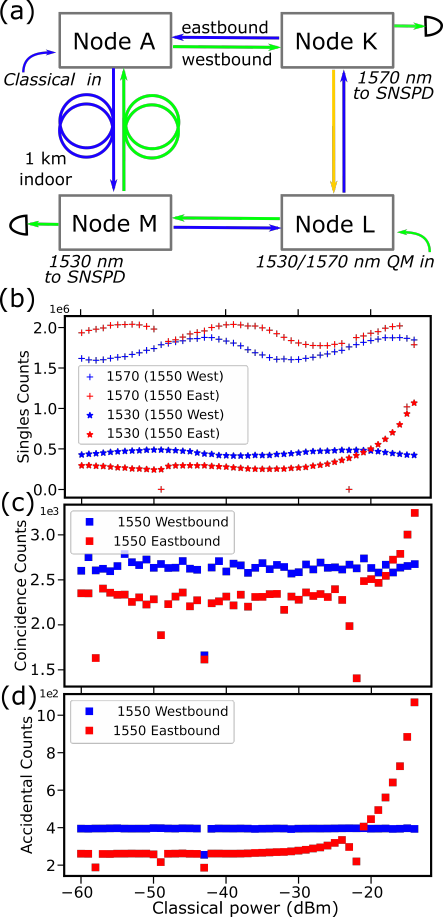}
    \caption{Crosstalk measurements in OADM ring. (a) Experimental setup for counter-propagating quantum and classical signals. In this configuration, the signals travel in separate fibers within a duplex cable. Crosstalk resulting from this configuration can be attributed to the OADMs. (b) Singles counts in the 1530 nm (stars) and 1570 nm (crosses) SNSPD channels vs. the power of a classical 1550 nm signal either co-propagating (Westbound, blue)) or counter-propagating (Eastbound, red). Corresponding coincidence counts (c) and accidental counts (d) are also plotted vs. the 1550 nm signal power.}
    %\vspace{128in}
    \label{Figure OADM crosstalk}
\end{figure}

To learn more about crosstalk in wavelength-multiplexed optical fiber links within the Griffiss QLAN, we send $1530-1570\;\mathrm{nm}$ entangled signal-idler pairs from the PIC source into the ring network, as shown in Figure \ref{Figure OADM crosstalk}. We then inject classically bright tones in the $1490-1610\;\mathrm{nm}$ CWDM channels and characterize channel crosstalk by measuring the changes in singles counts, coincidence counts, and accidental counts at SNSPDs. We input the $1530/1570\;\mathrm{nm}$ quantum signals at the Lana node, traveling counter-clockwise (Westbound), and input the classical signal at the Archer node traveling either counter-clockwise (Eastbound) or clockwise (Westbound). Interestingly, we do not see any crosstalk into the $1530/1570\;\mathrm{nm}$ channels when the classical signal co-propagates in the same fiber, but we see crosstalk when the classical signal counter-propagates, traveling through a separate fiber. We attribute the counter-propagating crosstalk to imperfect directionality and isolation within the CWDM OADMs.

\begin{figure*}
    \includegraphics[width = 1\linewidth]{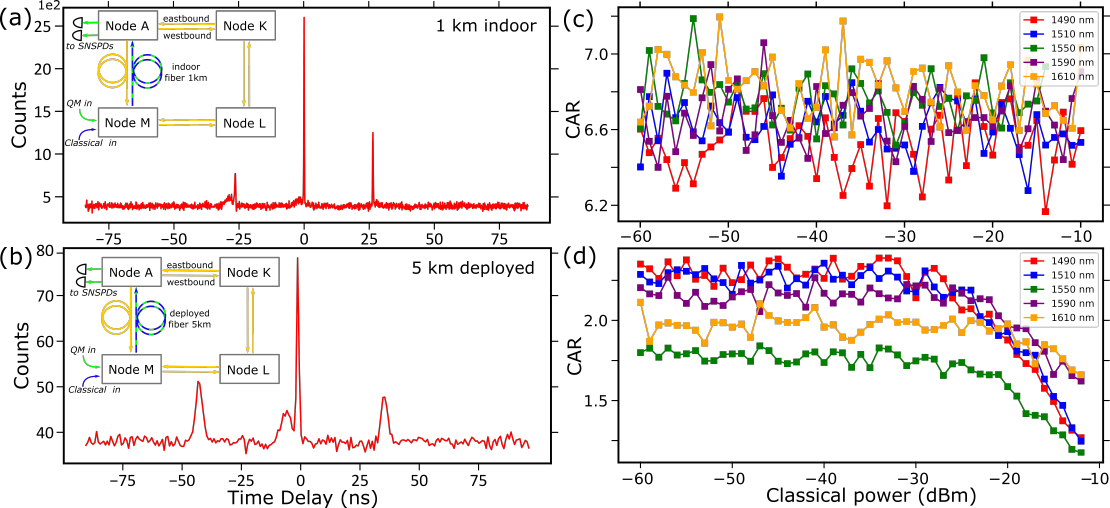}
    \caption{Crosstalk of co-propagating signals in long fiber links, including (a,c) an indoor 1 km fiber spool and (b,d) a deployed 5 km buried fiber. Plots in (a,C) have bin widths of $10\;\mathrm{ps}$, integration times of $5\;\mathrm{s}$, and no averaging. Plots in (b,d) have bin widths of $500\;\mathrm{ps}$, integration times of $5\;\mathrm{s}$, and 50 averages. }
    %\vspace{128in}
    \label{Figure crosstalk coins}
\end{figure*}

In a separate experiment, we send the quantum and classical signals directly to the 5 km deployed link, co-propagating from Mallory to Archer in the OADM ring.  We do this to characterize the wavelength crosstalk over the deployed link itself, eliminating that due to signal counter-propagation OADM ring. Results from this experiment are shown in Figure \ref{Figure crosstalk coins}. The signal-to-noise ratio of a coincidence measurement degrades appreciably for classical powers greater than $-30\;\mathrm{dBm}$, though it started at an already low value of $1.75-2.3$ for each experiment. These results are consistent with previous work showing that the number of noise photons present in the quantum channel due to forward-scattering and back-scattering increases with fiber link length \cite{Burenkov2023}. With the integration of chromatic dispersion compensation, polarization compensation, and lower-loss loopbacks - which will enable smaller coincidence time bins- the SNR should improve to greater than $2.3$ at $5\;\mathrm{km}$.

\subsection{Wavelength multiplexing in free space}
\begin{figure*}
    \includegraphics[width = 1\linewidth]{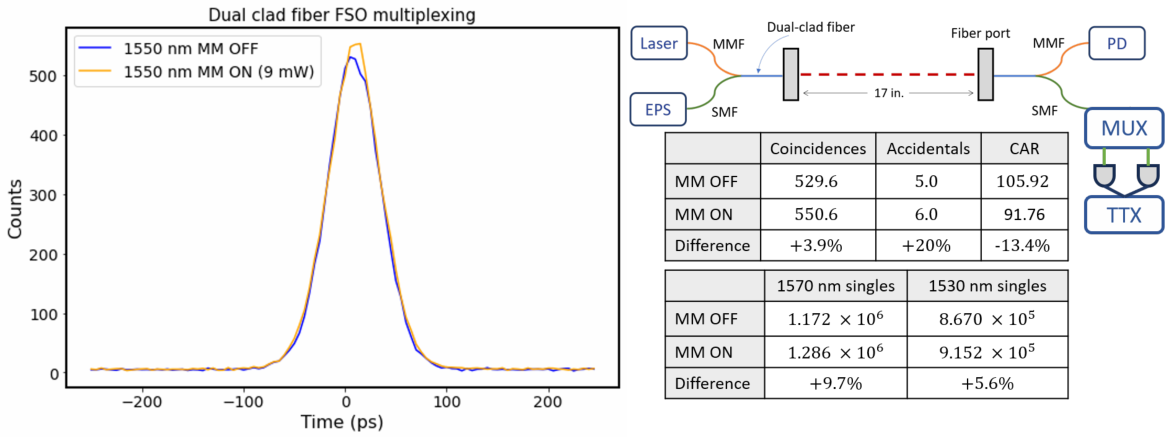}
    \caption{Free space multiplexing experiment performed using double-clad fiber couplers. Singles counts, accidental counts and coincidence counts from quantum signals are sent and received using the single-mode ports of double clad couplers, and measured both without and without high-power co-propagating classical signals transmitted over the same path using multi-mode (MM) ports in the couplers.}
    %\vspace{128in}
    \label{Figure double clad test}
\end{figure*}

Given the availability of dedicated fiber links at our QLANs that we can use to avoid quantum-classical coexistence in fiber, it is likely that we will more often require coexistence in classical network links. As such, we used double-clad fiber couplers \cite{Wroblewski2023} to combine single-mode classical/quantum and multi-mode classical signals for free-space co-propagation indoors at the Griffiss QLAN. Results (Figure \ref{Figure double clad test} show that, at least over short indoor links, the presence of high-power classical multimode signals has little to no effect on the coincidence-to-accidental ratio (CAR) of a coincidence measurement on entangled photon pairs. With 9 mW of classical multi-mode light on at 1550 nm, the number of singles counts in the 1530/1570 nm channels increases by less than 10\%, compared to no classical light, and the number of accidental counts in the coincidence domains increase from only 5 to 6. This has a noticeable effect on the CAR, only because the number of accidentals is so low to begin with. The CAR still remains very high at above 91. Next steps for this effort include repeating over fielded outdoor free space links at the Stockbridge QLAN.

\subsection{White Rabbit crosstalk measurement}
As discussed in the Main Text, the Griffiss QLAN is equipped with WR nodes, and WR links can be instantiated over dedicated fiber or coexisting with experimental signals.

\begin{figure*}
    \includegraphics[width = 0.6\linewidth]{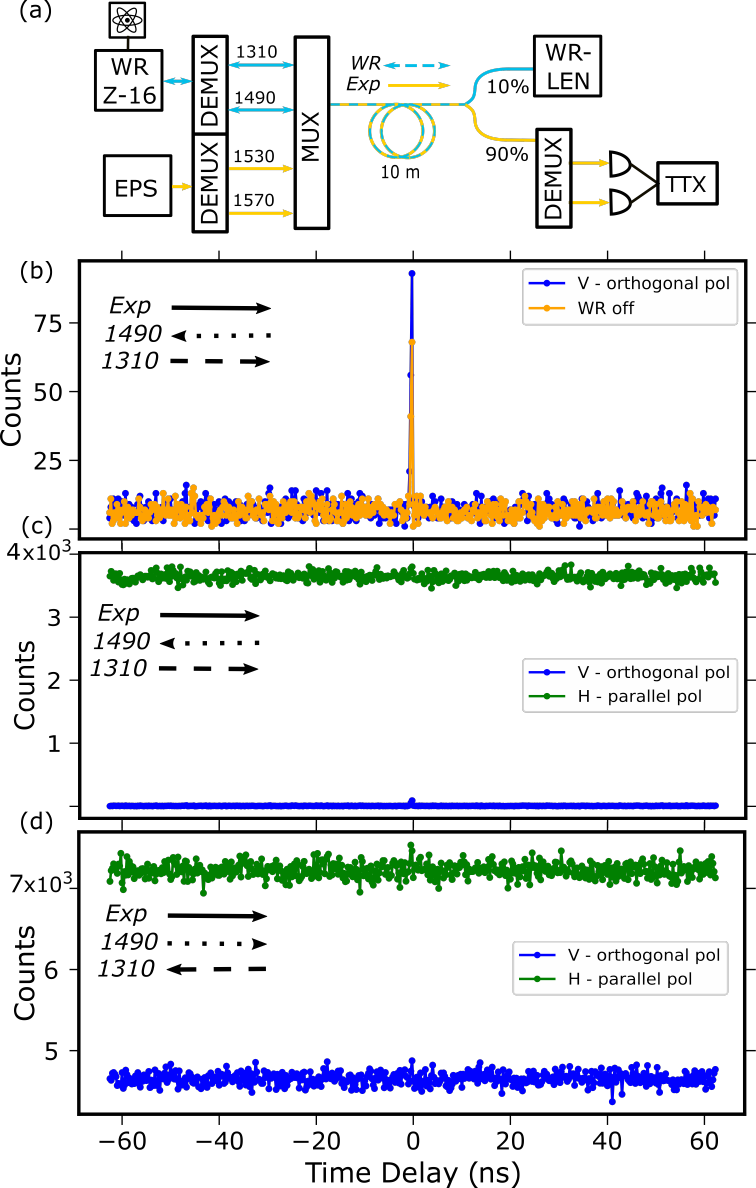}
    \caption{Quantum-classical multiplexing with WR signals. (a) Experimental setup. (b,c) Crosstalk for co-propagating quantum and 1310 nm WR signal and (d) co-propagating quantum and 1490 nm WR signal. (b) shows a comparison of WR off and WR on, cross-polarized, and (c) shows a comparison of cross- and co-polarized signals with WR on.}
    %\vspace{128in}
    \label{Figure WR crosstalk}
\end{figure*}

The Griffiss QLAN OADM ring network topology requires White Rabbit (WR) signals to share a fiber with experimental quantum signals if it is to propagate over a single duplex fiber ring, which immediately raises concerns of wavelength crosstalk and its effects on the quantum signals\cite{Burenkov2023}. To better understand the effects of this crosstalk, we connected the WR-Z16 to a WR-LEN across an experimental link as shown in Figure \ref{Figure WR crosstalk}. The WR nodes use $1310/1490\;\mathrm{nm}$ Bi-di transceivers, and are multiplexed into the same link as a $1530/1570\;\mathrm{nm}$ signal-idler entangled pair, with horizontal (H) polarization, from our PIC source. Coincidence histograms are obtained as a function of the WR signal propagation direction and polarization with respect to the signal-idler pair. When the $1310\;\mathrm{nm}$ WR signal is co-propagating with the signal-idler pair, the amount of crosstalk, as measured by background counts in the coincidence histogram, changes dramatically. If the WR signal is orthogonally polarized with respect to the experimental signal, there is a negligible effect on the coincidence measurement. When co-polarized with the experimental signal, however, the the experimental signal is completely overwhelmed by dark counts from the WR signal, rendering the link unusable for quantum signals. In contrast, when the $1310\;\mathrm{nm}$ signal is counter-propagating with the signal-idler pair, the crosstalk drowns out the quantum signal, irrespective of its polarization. The large amount of crosstalk in Figure \ref{Figure WR crosstalk} can likely be mitigated by attenuating the WR transceiver output powers. Going forward, this experiment can be repeated with longer fiber links to understand how link length affects the crosstalk, as in \cite{Burenkov2023}. 

\section{\label{app:source details}Theoretical description of PIC source}
The biphoton state can be expressed as 

\begin{align} \label{eq:1}
    \ket{\psi} = \left(1-\tfrac{i}{\hbar}\int_{-\infty}^{\infty}d\tau^\prime\hat{H}_{eff}(\tau^\prime)+\dots\right)\ket{\text{vac}} \simeq \ket{\text{vac}} + \ket{II}+\dots,
   % \lab{1}
\end{align}
where we only consider time dependent perturbation to first order such that we only care about biphoton generation. The effective Hamiltonian from Equation \ref{eq:1} is 
\begin{equation}
    \hat{H}_{eff}(\tau) \propto i\hbar\int \text{d}^{3}r\text{d}\omega_{p2}\text{d}\omega_{p1}\text{d}\omega_{1}\text{d}\omega_{2} \alpha(\omega_{p1})\alpha(\omega_{p2})e^{i\Delta k(\omega_{1},\omega_{2},\omega_{p1},\omega_{p2})z}e^{-i\Delta \omega (\omega_{1},\omega_{2},\omega_{p1},\omega_{p2})\tau}\hat{a}_{1}^\dagger(\omega_{1})\hat{a}_{2}^\dagger(\omega_{2}),
    \label{eq:pic src hamiltonian}
\end{equation}

\noindent where $\alpha(\omega)$ is the pump spectral amplitudes and $\Delta\omega = \omega_{1}+\omega_{2}-\omega_{p1}-\omega_{p2}$, $\Delta k = k(\omega_{1})+k(\omega_{2})-k(\omega_{p1})-k(\omega_{p2})$ where $k(\omega) = (\omega/c)n_{\text{eff}}(\omega)$.  Note that in Equation \ref{eq:pic src hamiltonian} we drop the Hermitian conjugate though it is understood to be there; for short interaction times, the $\hat{a}_{1}^\dagger(\omega_{1})\hat{a}_{2}^\dagger(\omega_{2})$ term dominates the evolution.  Assuming a wave-guide of length $L$ propagating in the $z$ direction and using a few integral identities:
\begin{align}
    \frac{1}{L}\int_{-L/2}^{L/2}e^{i\Delta k z}\text{d}^{3}r = \text{sinc}(\tfrac{L\Delta k}{2}), \quad \frac{1}{2\pi}\int_{-\infty}^{\infty} e^{-i\Delta\omega \tau}\text{d}\tau = \delta(\Delta\omega),
   % \lab{3}
\end{align}
\noindent we can plug into the biphoton state term of Eq.~(\ref{eq:1}) to find (assuming the pumps are centered at the same frequency)

\begin{align}
    \ket{II} &= \int\text{d}\omega_{p1}\text{d}\omega_{p2}\text{d}\omega_{1}\text{d}\omega_{2}\Big[\alpha(\omega_{p1})\alpha(\omega_{p2})\delta(\Delta\omega)\text{sinc}(\tfrac{L\Delta k}{2})\Big]\hat{a}_{1}^\dagger(\omega_{1})\hat{a}_{2}^\dagger(\omega_{2})\ket{\text{vac}} \nonumber\\
    &= \int\text{d}\omega_{p1}\text{d}\omega_{1}\text{d}\omega_{2} \Big[\alpha(\omega_{p1})\alpha(\omega_{1}+\omega_{2}-\omega_{p1})\;\text{sinc}\left(\tfrac{L}{2}\left[k(\omega_{1})+k(\omega_{2})-k(\omega_{p1})-k(\omega_{1}+\omega_{2}-\omega_{p1})\right]\right)\times\Big.\nonumber\\
    & \Big.\times\hat{a}_{1}^\dagger(\omega_{1})\hat{a}_{2}^\dagger(\omega_{2})\Big]\;\ket{\text{vac}} \nonumber\\
    &= \int\int\text{d}\omega_{1}\text{d}\omega_{2}\Bigg[\int\text{d}\omega_{p}\alpha(\omega_{p})\alpha(\omega_{1}+\omega_{2}-\omega_{p})\;\text{sinc}\left(\tfrac{L}{2}\left[k(\omega_{1})+k(\omega_{2})-k(\omega_{p})-k(\omega_{1}+\omega_{2}-\omega_{p})\right]\right)\Bigg]\times\nonumber\\
    &\quad\times\hat{a}_{1}^\dagger(\omega_{1})\hat{a}_{2}^\dagger(\omega_{2})\;\ket{\text{vac}} \nonumber\\
    &= \int\int\text{d}\omega_{1}\text{d}\omega_{2} \;\Phi(\omega_{1},\omega_{2})\;\hat{a}_{1}^\dagger(\omega_{1})\hat{a}_{2}^\dagger(\omega_{2})\;\ket{\text{vac}} .
    %\lab{4}
\end{align}

\noindent where $\Phi(\omega_{1},\omega_{2})$ is the biphoton wave function (BWF) given by

\begin{multline}
    \Phi(\omega_{1},\omega_{2}) \propto \int\text{d}\omega_{p}\;\alpha(\omega_{p})\alpha(\omega_{1}+\omega_{2}-\omega_{p})\;\text{sinc}\left(\tfrac{L}{2}\left[k(\omega_{1})+k(\omega_{2})-k(\omega_{p})-k(\omega_{1}+\omega_{2}-\omega_{p})\right]\right).
   % \lab{5}
\end{multline}

\noindent Note that in the limit of a single-frequency pump, tight signal/idler (anti-)correlations lead to 

\begin{equation}
    \Phi^{(\text{nb})}(\omega_{s}) \propto \text{sinc}\left(\tfrac{L}{2}\left[k(\omega_{s})+k(2\omega_{p}-\omega_{s})-2k(\omega_{p})\right]\right).
   % \lab{6}
\end{equation}

\begin{figure}
    \centering  \includegraphics[width=0.6\linewidth,keepaspectratio]{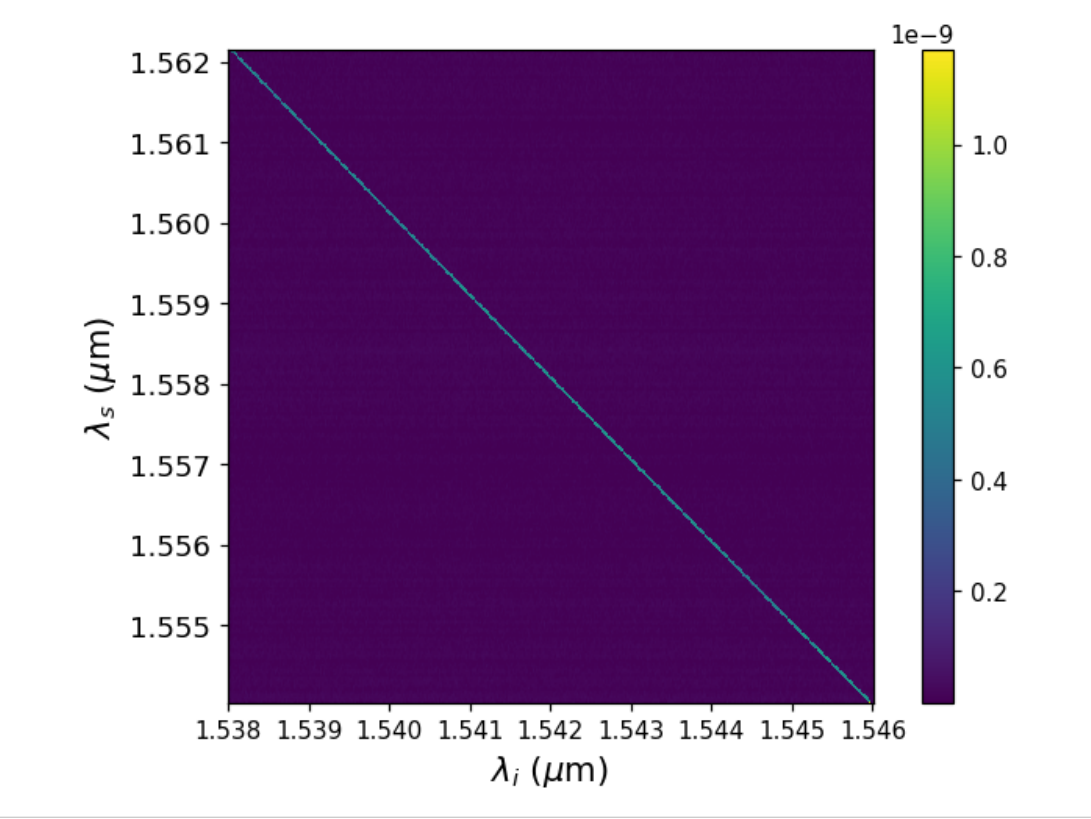}
    \caption{JSI data taken from the spiral source through stimulated FWM.  With a cw pump fixed at 1550nm, a second cw seed laser is swept from 1538-1546nm while an optical spectrum analyzer is used to measure around where the signal is expected to be generated via energy conservation.}
    \label{app fig:0}
\end{figure}

\noindent This is akin to the anti-diagonal line one sees for the joint-spectral amplitude using a cw pump shown in Figure \ref{app fig:0}.

We can also go on to derive the single-photon purity, derived directly from the biphoton wave function.  Without loss of generality between modes, we can derive the idler purity  to find

\begin{equation}
    \gamma = \text{Tr}[\rho_{I}^{2}]=\int\int\text{d}\omega\text{d}\omega^\prime \;|q_{I}(\omega,\omega^\prime)|^{2},
   % \lab{7}
\end{equation}
\noindent where 
\begin{equation}
    q_{I}(\omega,\omega^\prime)=\int\text{d}\omega^{\prime\prime}\;\Phi(\omega^{\prime\prime},\omega) \Phi^{*}(\omega^{\prime\prime},\omega^\prime),
    %\lab{8}
\end{equation}

\noindent which, in the limit of a separable biphoton wave function $\Phi(\omega_{1},\omega_{2})\to A(\omega_{1})\times B(\omega_{2})$ is unity.

\subsection{\label{app:source purity}Spectral purity of PIC source}
For simulation we use Tidy3D, which has a Python package for modeling wave-guides (among other thing) as well as a mode-solver which will allow us to obtain the effective index as a function of frequency $n_{\text{eff}}(\omega)$. The waveguide dimensions are: 500 nm wide at the base, 486 nm wide at the top, and a height of 220 nm. Plots of the joint-spectral intensity (JSI) as well as the marginal bandwidths and Gaussian pump profiles are seen in Figure \ref{app fig:2}. For a CW single-frequency pump, we obtain the expected diagonal of anti-correlations between signal and idler. As the pump bandwidth is broadened, so does the width of the distribution along the anti-diagonal. Further, the distribution becomes increasingly bimodal as the pump bandwidth is broadened and we consider longer wave-guides.  Most notable, however, is the increase in biphoton spectral purity.  For realistic parameters considered here, we can see upwards of 80\% state purity. While this is a marked improvement over the case of a cw pump, it still falls below the upper bound achievable using an MRR-based source.    To better understand the bimodality of the JSI for broader pump bandwidths (or longer wave-guides) consider the following Taylor expansion centered around frequency $\omega_{0}$:

\begin{figure*}
    \includegraphics[width=0.85\linewidth,keepaspectratio]{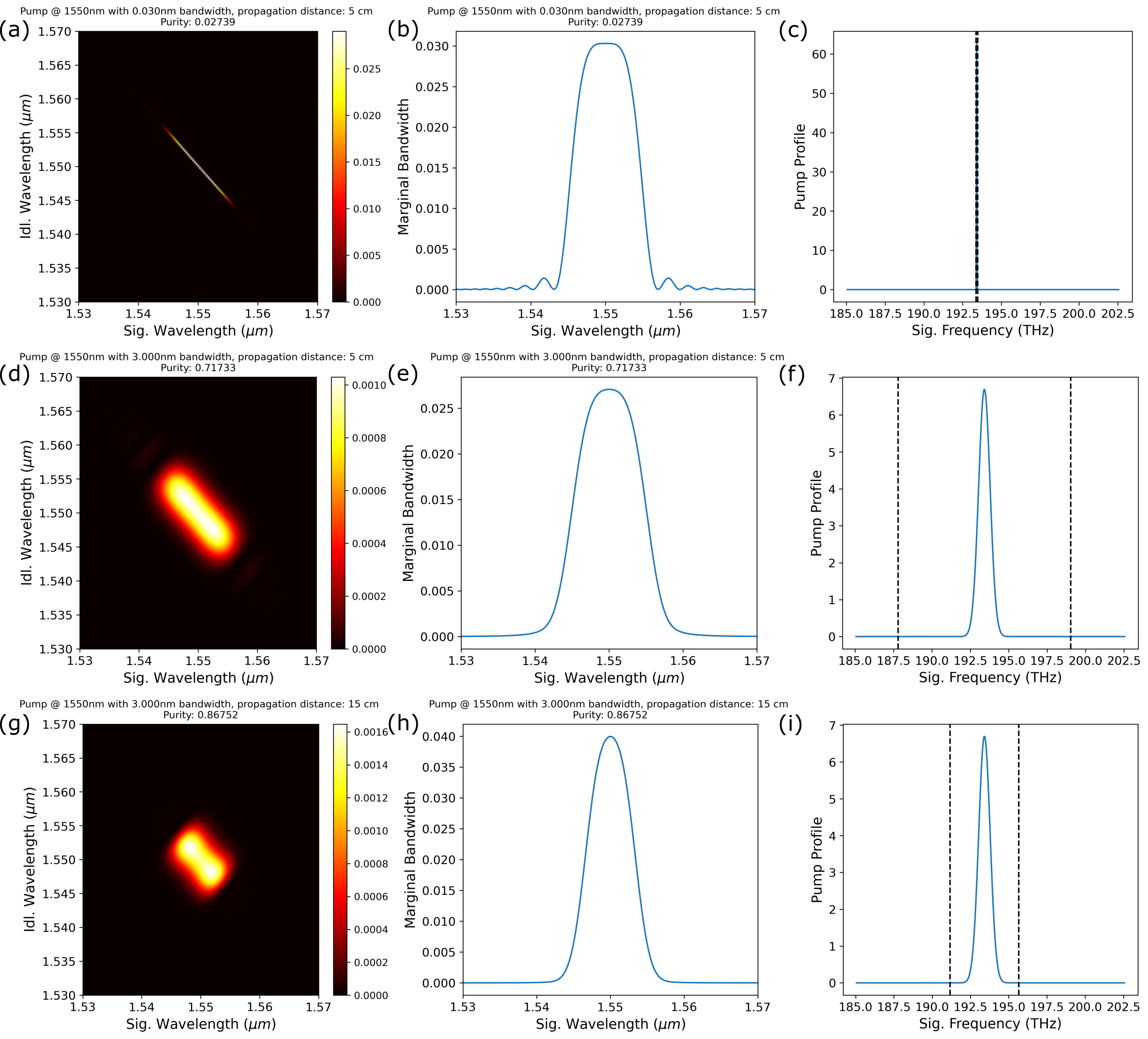}
    \caption{JSI data for (a) cw pump, 5 cm propagation length (b) 3 nm bandwidth Gaussian pump, 5 cm propagation length and (c) 3 nm bandwidth Gaussian pump, 15 cm propagation length.}
    \label{app fig:2}
\end{figure*}

\begin{align}
    k(\omega) &\simeq k_{0} + (\omega-\omega_{0})\frac{\partial k}{\partial\omega}\Bigg|_{\omega=\omega_{0}} + \frac{1}{2}(\omega-\omega_{0})^{2}\frac{\partial^{2} k}{\partial^{2}\omega}\Bigg|_{\omega=\omega_{0}} + \dots \nonumber\\
    &\simeq k_{0} + (\omega-\omega_{0})\left[\frac{1}{c}n_{eff}(\omega)+\frac{\omega_{0}}{c}\frac{\partial n_{eff}}{\partial \omega}\right]\Bigg|  _{\omega=\omega_{0}} + \frac{1}{2}(\omega-\omega_{0})^{2}\left[\frac{1}{c}\frac{\partial n_{eff}}{\partial \omega} + \frac{\omega_{0}}{c}\frac{\partial^{2} n_{eff}}{\partial \omega^{2}}\right]\Bigg|_{\omega=\omega_{0}} + \dots \nonumber\\
    &\simeq k_{0} + (\omega-\omega_{0})\frac{1}{c}\frac{\partial n_{g}}{\partial\omega}\Bigg|_{\omega=\omega_{0}} + \frac{1}{2}(\omega-\omega_{0})^{2}\frac{1}{c}\frac{\partial^{2}n_{g}}{\partial\omega^{2}}\Bigg|_{\omega=\omega_{0}} + \dots \nonumber\\
    &\simeq k_{0} +  (\omega-\omega_{0})\;f^{(1)}(\omega_{0}) +  \frac{1}{2}(\omega-\omega_{0})^{2}\;f^{(2)}(\omega_{0}) + \dots\;,
   %\lab{7}
\end{align}

\noindent where we used $n_{g}(\omega)=c\partial_{\omega}k(\omega)$ and $\partial_{\omega}n_{g}(\omega) = c\partial^{2}_{\omega}k(\omega)$. Note the functions $f^{(1)}(\omega_{0})\;\text{and}\;f^{(2)}(\omega_{0})$ are scaling factors that can be largely ignored without loss of generality.  Under this approximation we have 

\begin{align}
    \Delta k(\omega_{1},\omega_{2},\omega_{p}) &= k(\omega_{1}) + k(\omega_{2}) - k(\omega_{p}) - k(\omega_{1}+\omega_{2}-\omega_{p}) \nonumber\\
    &\simeq k_{0} + (\omega_{1}-\omega_{0})f^{(1)}(\omega_{0}) + \frac{1}{2} (\omega_{1}-\omega_{0})^{2}f^{(2)}(\omega_{0}) + \nonumber\\
    &+ k_{0} + (\omega_{2}-\omega_{0})f^{(1)}(\omega_{0}) + \frac{1}{2} (\omega_{2}-\omega_{0})^{2}f^{(2)}(\omega_{0}) - \nonumber\\
    &- k_{0} + (\omega_{p}-\omega_{0})f^{(1)}(\omega_{0}) + \frac{1}{2} (\omega_{p}-\omega_{0})^{2}f^{(2)}(\omega_{0}) - \nonumber\\
    &- k_{0} + (\omega_{1}+\omega_{2}-\omega_{p}-\omega_{0})f^{(1)}(\omega_{0}) + \frac{1}{2} (\omega_{1}+\omega_{2}-\omega_{p}-\omega_{0})^{2}f^{(2)}(\omega_{0}) \nonumber\\
    &\;\;\vdots\nonumber\\
    &\simeq \left(\omega_{p}(\omega_{1}+\omega_{2})-\omega_{p}^{2}-\omega_{1}\omega_{2}\right)f^{(2)}(\omega_{0}) = \Delta \tilde{k}.
   % \lab{8}
\end{align}

\begin{figure*}
    \centering  \includegraphics[width=1\linewidth,keepaspectratio]{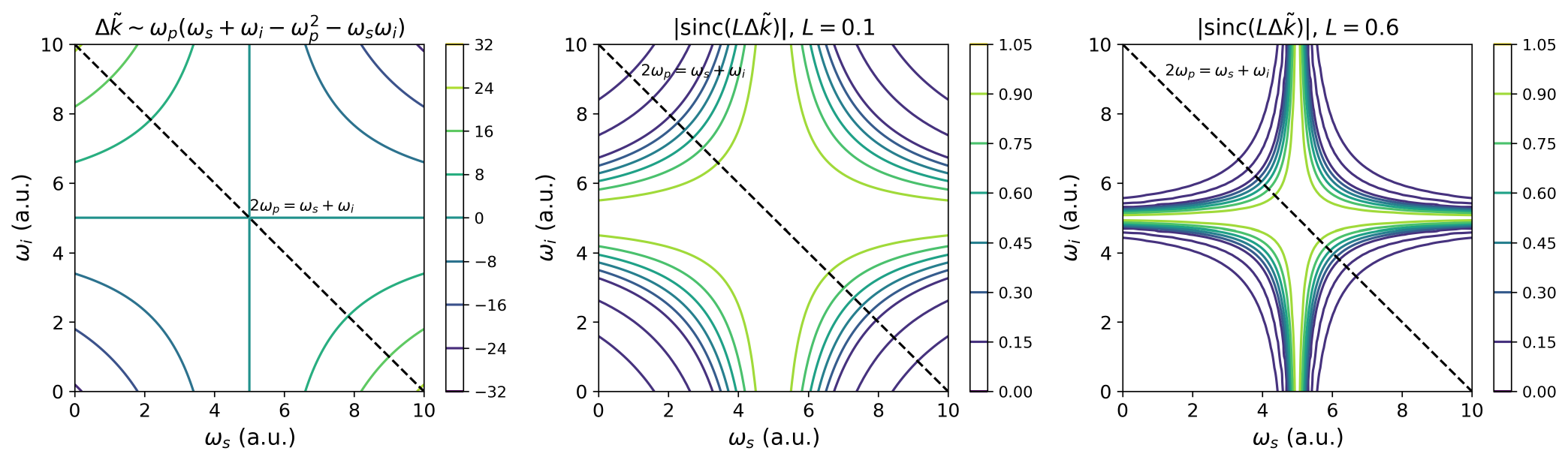}
    \caption{Contour plot of $\Delta\tilde{k}$ with $\omega_{p}=5$ (left). Plots of $\text{sinc}L\Delta\tilde{k}$ for $L=0.1$ (middle) and $L=0.6$ (right).}
    %\vspace{128in}
    \label{app fig:sinc_contours}
\end{figure*}

Dropping the scaling factor $f^{(2)}(\omega_{0})$, we can investigate when $\text{sinc}(L\Delta \tilde{k}/2)$ is maximal, i.e., when $\Delta \tilde{k} \sim 0$ (and the features of the sinc contours in general).  In Figure \ref{app fig:sinc_contours}, we plot $\Delta\tilde{k}$ as well as $\text{sinc}L\Delta\Tilde{k}$ for $L=0.1,0.6$, each with $\omega_{p}=5$. In each we include the anti-diagonal line corresponding to $2\omega_{p}=\omega_{s}+\omega_{i}$.  Note that as we increase $\omega_{p}$, the entire distribution will simply shift up towards the top-right corner of the plots. From these figures, as we integrate over broader pump bandwidths, more of the sinc features will appear, causing the distribution to bow inwards.  Note also that as the length is increased, the width of the distribution along the anti-diagonal line decreases, which is also confirmed transitioning from Figure \ref{app fig:2}(b-c).

With this, it is worth pointing out that with realistic values of wave-guide lengths, we can expect at most around ~80\% bi-photon purity.  The marginal bandwidths will still be quite broad.  On the other hand, for micro-ring resonator (MRR) based sources, the marginal bandwidths are limited by the resonance linewidths.  Further, it has been shown that with a single-bus MRR, one can get upwards of 93\% purity. In what follows, we will dive more into the topic of \textit{Si}-based MRRs for biphoton production. 

%\subsection{\label{MRR sources} Micro-ring resonator PIC sources}
%\todo[inline]{RICH: maybe drop the MRR-based source stuff for this since we don't have any real results using them?  Can save it for a future version.}
%For a MRR PIC source, the bi-photon wavefunction can be written as 
%\begin{equation}
%    \phi(\omega_{s},\omega_{i}) = F_{p}(\omega_s+\omega_i)\times\mathbf{l}_{s}(\omega_{s})\times\mathbf{l}_{i}(\omega_{i}),
    \label{eq: MRR wavefunction}
%\end{equation}
%\noindent where $\mathbf{l}_{j}(\omega_{j})$ is the complex Lorentzian factor describing the $j^{th}$ resonance and where $F_{p}(\omega)$ is a convolution of the pump spectral amplitudes and the pump resonance line-widths
%\begin{equation}
%    F(\omega) = \int\text{d}\omega_{p}\;\alpha(\omega_{p})\alpha(\omega-\omega_{p})\mathbf{l}_{p}(\omega_{p})\mathbf{l}_{p}(\omega-\omega_{p}).
  %  \lab{eq:}
%\end{equation}
%\noindent For a broad enough pulse pump, $F$ is nearly constant over the domain spanned by the signal/idler frequencies, and as a result, the bi-photon wavefunction is nearly pure. Note that the Lorentzian factors in Equation \ref{eq: MRR wavefunction} effectively serve as signal/idler filter functions.

%We are working with a chip that features arrays of  15$\mu$m radius MRRs with gap sizes varying from 1000 nm to 100 nm in (mostly) steps of 10 nm. Larger gaps couple TM modes that have a much different FSR from TE mode coupling, resulting in resonance placements that are not phase-matched for FWM, so many of the rings can be disregarded outright (see Figure \ref{app fig:TE_vs_TM_coupling}).

%\begin{figure}
%	\centering
%	\includegraphics[width=\linewidth,keepaspectratio]{Figures/app fig MRR TM TE.pdf}
%	\caption{Phase-matching condition $2\omega_{p}=\omega_{s}+\omega_{i}$ for  TM-mode coupling and  TE-mode coupling. }
%\label{app fig:TE_vs_TM_coupling}
%\end{figure}

\section{\label{app: CHSH}Time-Energy CHSH measurement details}

\subsection{Phase modulation with WaveShapers}
\begin{figure}
    \centering \includegraphics[width=0.95\linewidth ]{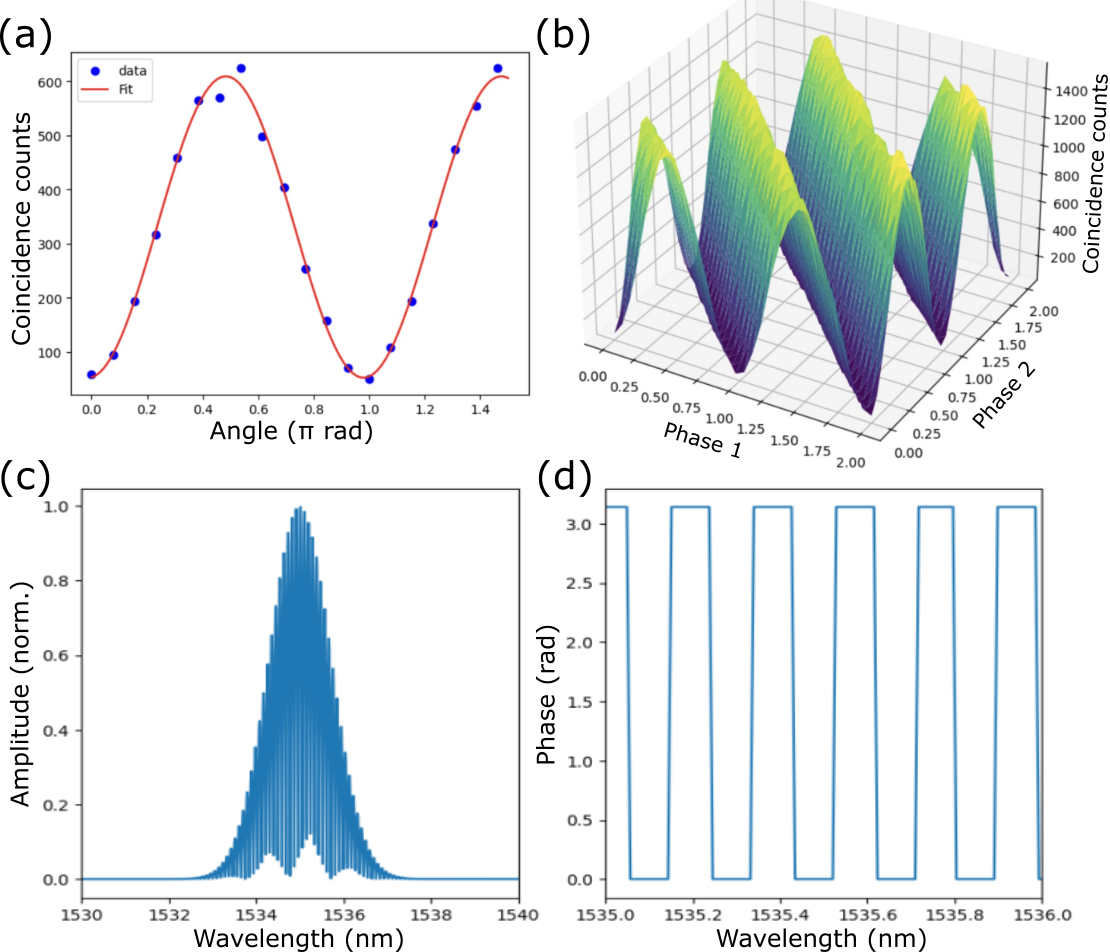}
    \caption{(a)An example plot showing the sinusoidal fringe pattern of the number of coincidence counts at the output of a Franson interferometer versus the relative phase offset between WaveShapers A and B. The 3D plot in (b) shows this data for phase offset sweeps in both WaveShapers A and B. (c) Amplitude modulation function applied to the $1530\;\mathrm{nm}$ CWDM channel, showing a Gaussian envelope with a 2 nm FWHM and fast $\cos^2(\lambda)$ oscillations with a period of $\sim45\;\mathrm{ps}$,  and  (d) phase oscillations between $0$ and $\pi$ radians with a period of $0.2\; \mathrm{nm}$.}
    %\vspace{128in}
    \label{app fig:finisar phase mod}
\end{figure}

\begin{figure}
    \centering \includegraphics[width=1\linewidth ]{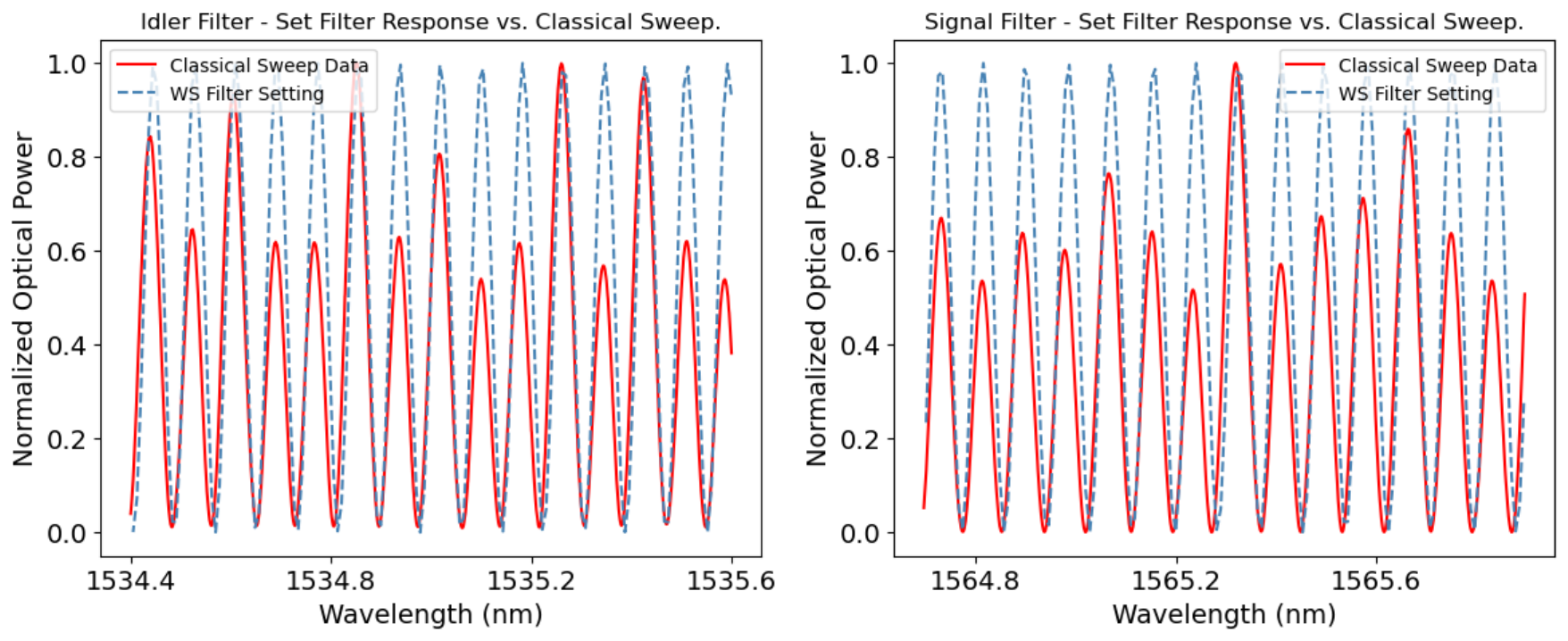}
    \caption{Set WS profile (blue dashed) compared to the actual filter response from a swept laser (red solid), for both the signal and idler channels. Note the coarse resolution of the WS.}
    %\vspace{128in}
    \label{app fig:filter_response}
\end{figure}

To apply the cosine filter, we use the WaveShapers to apply both amplitude modulation, which resembles the absolute value of the cosine function, and a phase modulation, where the phase toggles between 0 and $\pi$ (see Fig.~(\ref{app fig:filter_response}) for an analysis on the filter fidelity). These two modulation curves, combined, give the effect of a general cosine modulation, as would a phase shifter in a bulk optics Franson interferometer.

In Figure \ref{fig:franson_interference}, we plot the calculated joint-spectral intensity (JSI) following Equation 7 in the Main Text. 

\begin{figure}
    \centering \includegraphics[width=0.8\linewidth ]{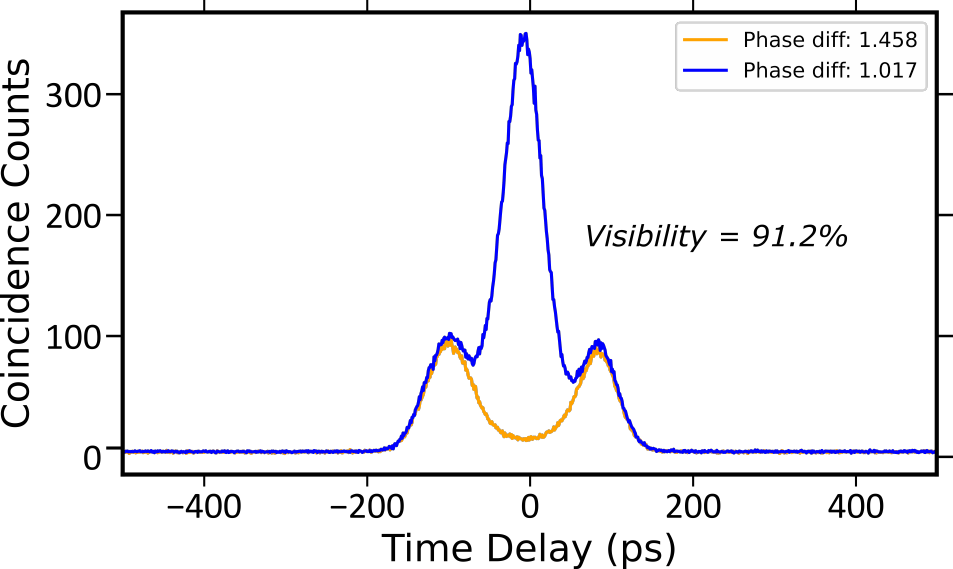}
    \caption{Example plot showing the achievable visibility of WaveShaper phase modulation over indoor (10 m) fiber length. Visibilities of $>90\%$ are routinely achieved.}
    %\vspace{128in}
    \label{app fig:indoor visibility}
\end{figure}

\begin{figure*}
\centering
    \includegraphics[width = 1\linewidth]{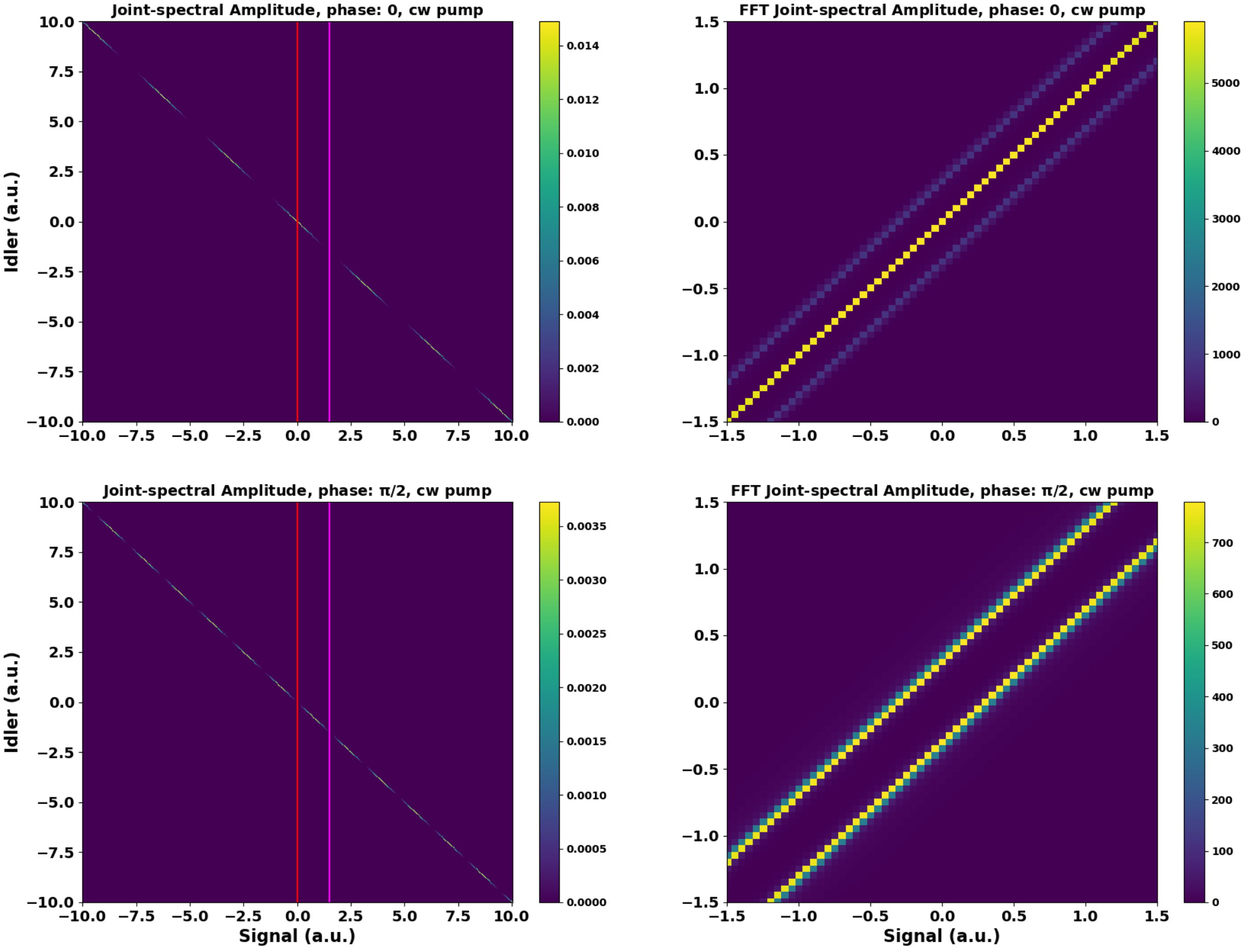}
    \caption{(left column) Joint-spectral intensity $\text{JSI}(\omega_s,\omega_i)=|\psi(\omega_{s},\omega_i)\phi(\omega_s,\omega_i)|^{2}$ for relative phases between simulated Franson interferometers of $\theta=0,\pi/2$. (right column) The corresponding Fourier transforms showing the path indistinguishability between SS,LL paths and the destructive interference between terms for $\varphi=\pi/2$.  Note the axis scaling is in arbitrary units and the color map scalings vary between plots.}
    %\vspace{128in}
    \label{fig:franson_interference}
\end{figure*}

The period of the cosine amplitude modulation curve is finely tuned to maximize the modulation visibility. We used a cosine function time period of $42\; \mathrm{ps}$, as it provides the best resolution for the peaks. A higher time period would require a higher spatial resolution from the LCOS device to accurately represent the rapid changes in the waveform, translating to a need for more pixels on the LCOS to achieve the same modulation precision at higher frequencies. If we go over a period of $42\; \mathrm{ps}$, we encounter issues with insufficient pixel density and loss of resolution, which causes coincidence peaks to bleed into one another. On the other hand, a lower cosine period may also cause separate peaks to merge into one another. Overall, using a spatial light modulator with more pixels would increase measurement resolution.

The variable B in the modulation function represents the relative phase between the two cosine profiles on WaveShapers A and B. Changing this relative phase will change the number of coincidences obtained for the short-short (SS) and long-long (LL) cases. The phase values for maximum and minimum coincidences can be obtained via a sinusoidal fit of the phase modulation curve. We perform a triple Gaussian fit of each coincidence histogram, extract the center and FWHM of the central peak, integrate the curve within the FWHM to obtain the total number of coincidence counts, then plot that number versus phase. Using this method, the maximum visibility achieved with a one-input-four-output WaveShaper with a band-pass amplitude filter is $82.6\%$. 

A Gaussian amplitude filter can also be used. Switching to a WaveShaper configuration using one-input-one-output modules, instead of one-input-four-output modules, and implementing a Gaussian amplitude filter with an FWHM of 1 nm, boosts the maximum observed visibility to $92.3\%$. 
%We keep our central wavelength to be 1535 nm and 1565 nm for the signal and idler photons respectively. This satisfies the energy conservation condition as well. 

\subsection{CHSH implementation}
Here we explore the analogy between sinusoidal phase modulation and the polarization degree of freedom of a photon. To perform a CHSH inequality test for two photons entangled in the polarization degree of freedom, two polarizers are placed in the paths of the photons and rotated to specific angles to obtain correlations.

For a time-energy Bell state, the phase applied in
the longer arm of the asymmetric MZI acts in the same manner as a polarizer. All we need to do in a CHSH inequality test is to change the 'B’ parameter to get our curve of coincidence vs phase. We can label the usual cosine modulation as a polarizer oriented in such a way that only the vertically-polarized (V) photons pass. Similarly, we can label the sine modulation to be the polarizer oriented in such a way that only the horizontally-polarized (H) photons pass through.

For the case where the signal polarizer remains steady at $V$ and the idler polarizer is rotated, we set $B = 0$ for the idler Finisar and change $B$ on the signal Finisar from $0$ to $3\pi/2$. For the case where the signal polarizer remains at $A$ and the idler polarizer is rotated, we keep $B = \pi/8$ for the idler Finisar and change B on the signal Finisar from $0$ to $3\pi/2$. For the case where the signal polarizer remains at ’D’ and the idler polarizer gets rotated, we keep $B =- \pi/8$ for the idler Finisar and change B on the signal Finisar from $0$ to $3\pi/2$. For the case where the signal polarizer remains at 'H’ and the idler polarizer is rotated, we keep $B = \pi/2$ for the idler Finisar and change B on the signal Finisar from $0$ to $3\pi/2$. 

A rotation by angle $\delta$ modifies the polarization state of a photon as
\begin{equation}
    \ket{V_{\delta}} = \cos(\delta)\ket{V} + \sin(\delta)\ket{H}
    \ket{V_{\delta}} = -\sin(\delta)\ket{V} + \cos(\delta)\ket{H} 
\end{equation}
The probability of detecting a coincidence when both photons are vertically polarized:
\begin{multline}
    P_{VV}(\alpha, \beta) = |\bra{V_{\alpha}}\braket{V_{\beta}}{\psi_{Bell}}|^2 
    = \left\vert [\bra{V} \cos(\alpha) ] [\bra{H} \sin{\alpha}] \right\vert^2 \times [\bra{V} \cos(\beta) ] [\bra{H} \sin{\beta}] \times  \frac{1}{\sqrt{2}} (\ket{V}\ket{V} + \ket{H} \ket{H})
\end{multline}

\subsection{Spectral dispersion compensation with WaveShapers}

\begin{figure*}
    \includegraphics[width = 1\linewidth]{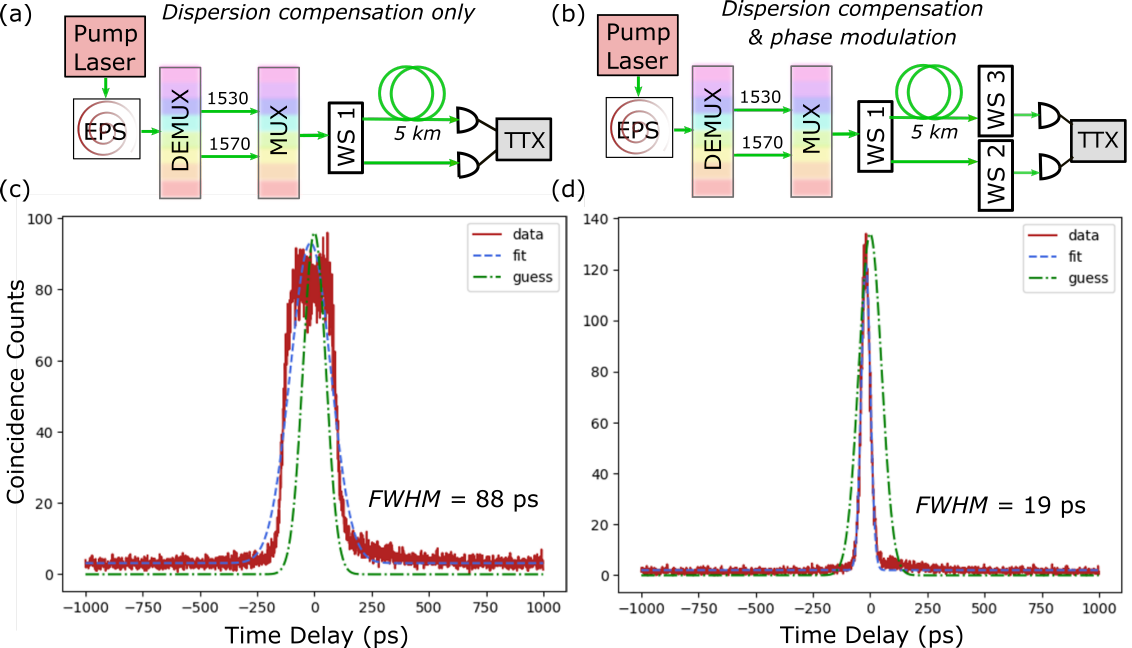}
    \caption{Experimental diagrams and sample coincidence histograms for (a,b) and (c,d)}
    %\vspace{128in}
    \label{Figure finisar disp comp}
\end{figure*}

Here, a one-input-four-output WaveShaper (Finisar Waveshaper 4000A) applies a pre-calibrated phase offset to the signal to compensate the dispersion and restore the signal-idler coincidence peak to a narrow Gaussian lineshape. Adding a phase term of the form $e^{i \phi(\lambda)}$ to the biphoton signal introduces an wavelength-dependent phase shift that pre-compensates the phase shifts arising from spectral dispersion in fiber links.

GVD as a function of wavelength follows a quadratic relationship, with the parabola centered at the optical signal's central wavelength $\lambda_0$. As such, we apply a  pre-compensation phase shift of 
\begin{equation}
    \phi(\lambda) = -\pi c L\frac{D}{\lambda_{0}^{2}} (\lambda - \lambda_0)^2,
    \label{eqn:dispcomp}
\end{equation}

\begin{figure}
\centering
    \includegraphics[width = 0.7\linewidth]{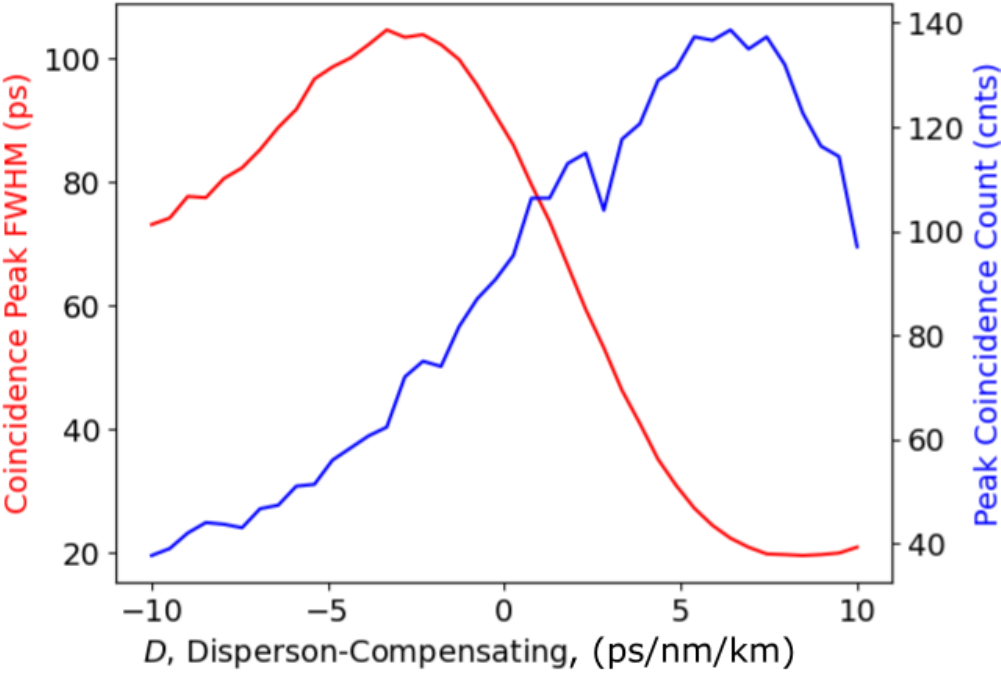}
    \caption{Coincidence FWHM and binned 20ps integrated counts around the center coincidence peak as a function of $D$, the dispersion-compensation parameter (ps/nm/km).  We find $D\sim8.8$p/nm/km yields the narrowest coincidence FWHM.  Note as the width becomes smaller, the number of total coincidences accumulating in the center bin increases, confirmed by the blue curve.}
    %\vspace{128in}
    \label{fig:dispcompfig}
\end{figure}

\noindent for a given dispersion parameter $D$ and propagation distance $L$. The dispersion-compensating phase $\phi(\lambda)$ is pre-calibrated by tuning the value of $D$ over a given length of fiber and measuring the width of the coincidence peak. Using this method, we extracted a value of $D \simeq 8.8\;\mathrm{ps/(nm \cdot km)}$, as shown in Fig.~(\ref{fig:dispcompfig}). Example dispersion compensation results are plotted in Figure \ref{Figure finisar disp comp}(c,d). Chromatic dispersion over a $5\;\mathrm{km}$ deployed link spreads the coincidence peak to a width of $88\;\mathrm{ps}$. Using a WaveShaper pre-compensation phase, the FWHM is decreased to $19\;\mathrm{ps}$.

\subsection{CHSH over 5 km deployed fiber at Griffiss QLAN}
As mentioned in the Main Text, we performed a CHSH inequality measurement over the 5 km deployed fiber link  at the Griffiss QLAN. Here, we used the Finisar WaveShapers for spectral dispersion compensation, in place of a standalone tunable dispersion compensation module (TDCM). In this experiment the achievable visiblity was 58.8\%, far lower than that achieved elsewhere in this work. The low visibility limited the achievable CHSH value to $1.76$, falling short of a violation. We attribute this poor result to (a) inferior spectral dispersion compensation, which can be improved by using the TDCM, (b) higher dark counts in our single photon detectors caused by room lights, and (c) decreased efficiency of the detectors. At the time, the cryostat housing our detectors was operating above optimal base temperature, at about 4 Kelvin. As such, the performance of the detectors was degraded. With improved dispersion compensation, better fiber insulation from background light, and detectors operating to specification, we should be able to achieve a CHSH violation over deployed fiber at the Griffiss QLAN in the near future.

\begin{figure}
\centering
    \includegraphics[width = 0.8\linewidth]{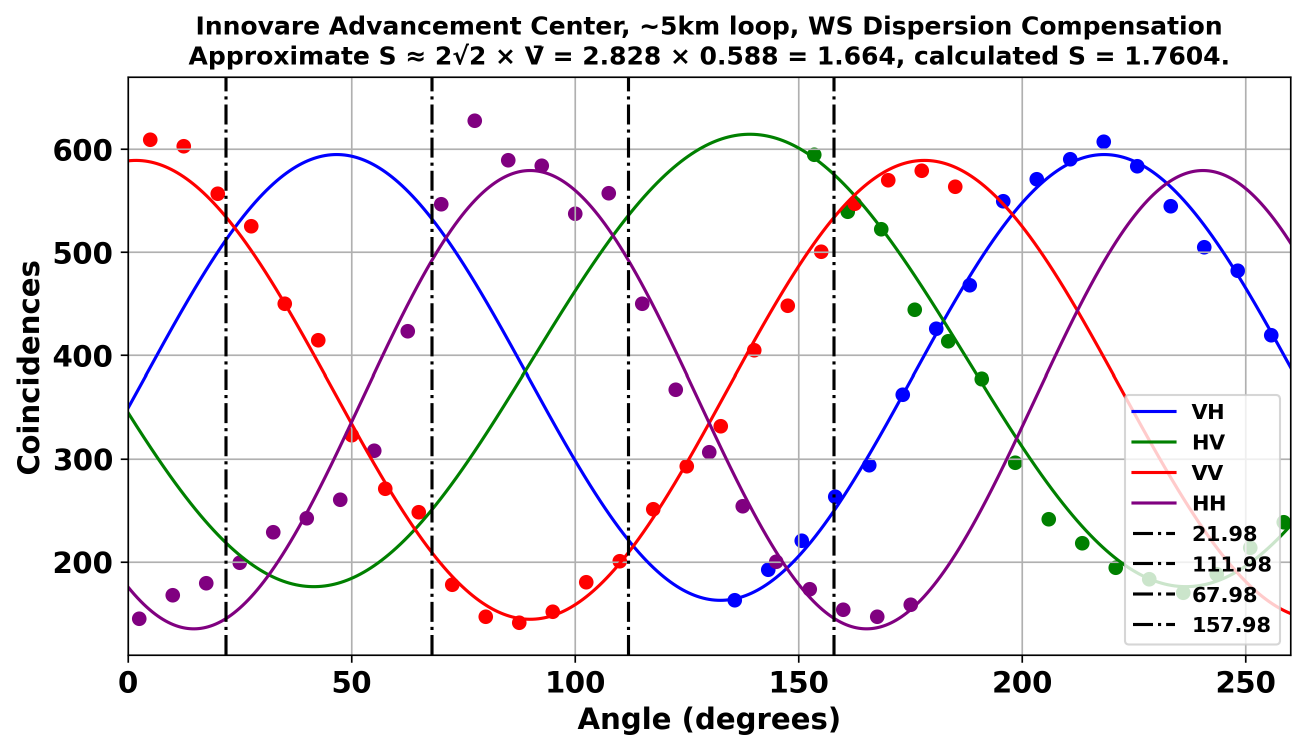}
    \caption{CHSH measurement results over $5\;\mathrm{km}$ deployed fiber at the Innovare Advancement Center. Dispersion was pre-compensated for using Finisar WaveShapers applying a quadratic wavelength-dependent phase, as per Eq.~(\ref{eqn:dispcomp}).}
    %\vspace{128in}
    \label{fig:5km_CHSH_IAC}
\end{figure}

\subsection{\label{app:chsh_uncertainty}Uncertainty in CHSH measurements}

\noindent Given a correlation function of the form $E=(N_{++}+N_{--}-N_{-+}-N_{+-})/(N_{++}+N_{--}+N_{-+}+N_{+-})=N_{\text{corr.}}/N_{\text{tot.}}$, we wish to calculate the measurement variance $\sigma^{2}_{E}$ to subsequently determine the Bell violation uncertainty and corresponding significance factor.  We can assume pair generation through spontaneous FWM shares statistics with the CW pump driving the interaction, i.e. Poissonian statistics. Note the signal and idler modes themselves remain super-Poissonian in photon number statistics.  Thus $\sigma_{N_{k}}^{2}=N_{k},\;k\in\{++,--,+-,-+\}$.  If we call $A=N_{\text{corr.}}\;\text{and}\;B=N_{\text{tot.}}$, making $E=A/B$, then we have by error propagation

\begin{equation}
    \sigma_{E}^{2} = \sum_{k}\left(\frac{\partial E}{\partial N_{k}}\right)^{2}\sigma_{N_{k}}^{2} = \sum_{k}\left[\frac{B\tfrac{\partial A}{\partial N_{k}}-A\tfrac{\partial B}{\partial N_{k}}}{B^{2}}\right]^{2}\sigma_{N_{k}}^{2}\approx \frac{1-E^{2}}{N_{\text{tot.}}},
    \label{eqn:var_exact}
\end{equation}

\noindent where we have used $\partial_{N_{k}}B=1\;\forall k$ and 

\begin{equation}
    \frac{\partial A}{\partial N_k} =
    \begin{cases}
        +1, & \text{if } k \in \{++, --\}, \\
        -1, & \text{if } k \in \{+-, -+\}.
    \end{cases}
\end{equation}

\noindent Alternatively, we can instead approximate $\sigma_{A}^{2} = \sigma_{B}^{2} = N_{\text{tot.}}$, treating both $A$ and $B$ as fluctuating with equal variance set by the total counts. While this does not strictly follow from Poisson statistics, it serves as a conservative simplification.  Again through error propagation we can write

\begin{equation}
    \sigma_{E}^{2} = \left(\frac{\partial E}{\partial A}\right)^{2}\sigma_{A}^{2} + \left(\frac{\partial E}{\partial B}\right)^{2}\sigma_{B}^{2} = \left(\frac{1}{B}\right)^{2}\sigma_{A}^{2} + \left(\frac{A}{B^{2}}\right)^{2}\sigma_{B}^{2}\xrightarrow{\sigma_{A}^{2}=N_{\text{corr.}},\;\sigma^{2}_{B}=N_{\text{tot.}}}\approx\frac{N_{\text{tot.}}}{B^{2}}\Big(1 + \left(\tfrac{A}{B}\right)^{2}\Big) = \frac{1+E^{2}}{N_{\text{tot.}}}.
    \label{eqn:E_var}
\end{equation}

\noindent  This overestimation may better account for real-world experimental uncertainties such as detector noise, background counts, or imperfect state preparation, and serves as a pseudo-upper bound on the expected uncertainty. Finally, for the uncertainty in $S$ and the corresponding significance factor $\sigma$, we have

\begin{equation}
    \sigma_{S} = \sqrt{\sum_{i}\sigma_{E_{i}}^{2}},\;\;\;\sigma=\frac{S_{\text{meas.}}-2}{\sigma_{S}}.
    \label{eqnn:S_uncertainty}
\end{equation}

\section{\label{app:OTDR}Fiber OTDR reports}
A Jonard OTDR-1000 optical time-domain reflectometer is used to measure loss events in the QLAN fiber links. Reports from OTDR measurements at Griffiss, Stockbridge and RRS show that our QLAN fiber links are low-loss across the three locations. Loss comes primarily from connectors within deployed links.

\newpage

\begin{figure*}
    \centering  \includegraphics[width=0.8\linewidth,keepaspectratio]{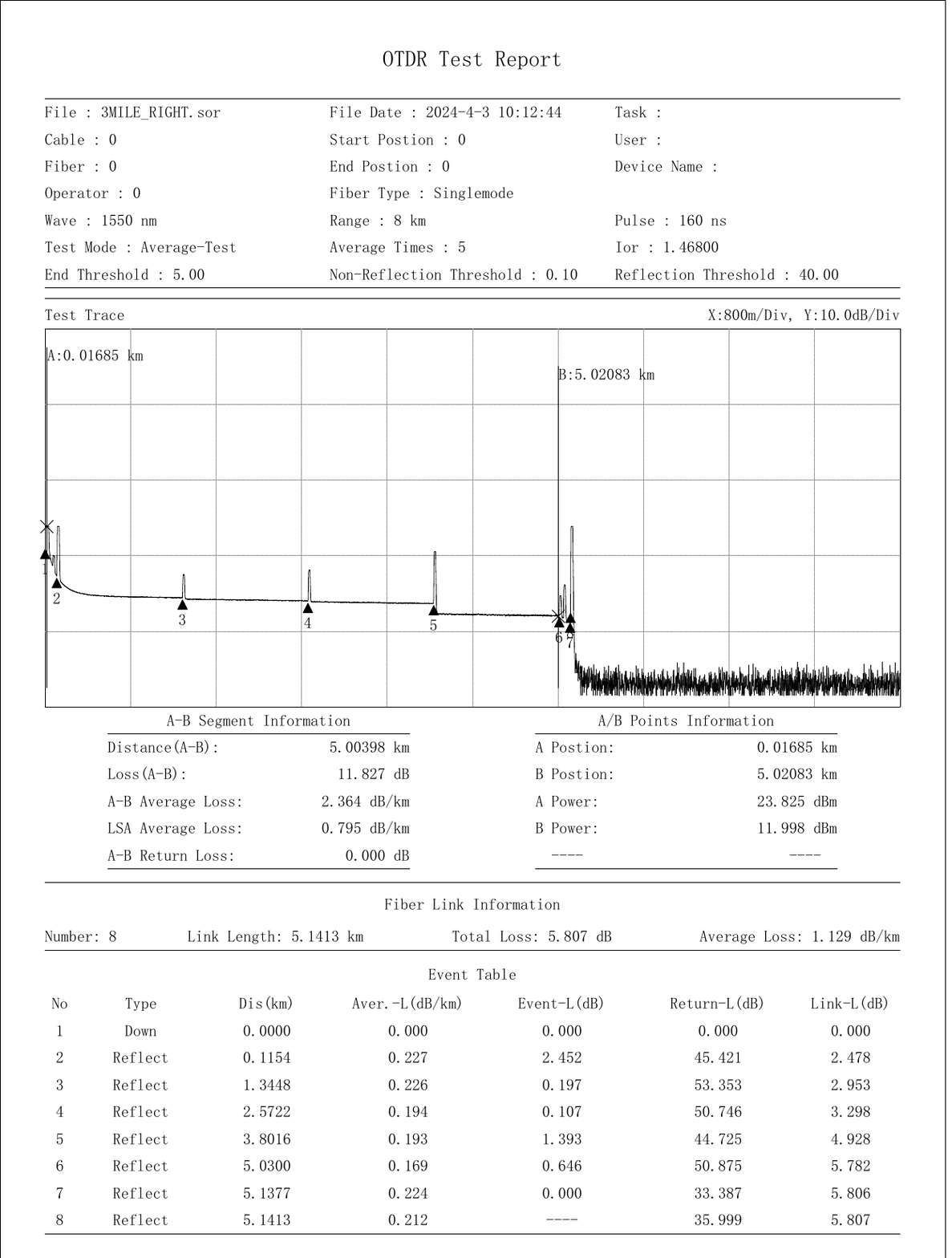}
    \caption{OTDR measurements of Griffiss QLAN deployed buried fibers. While the fiber links are low-loss, loss from loopback connectors accumulates to 7-8 dB total.}
    \label{app fig:Griffiss OTDR}
\end{figure*}

\begin{figure*}
    \centering  \includegraphics[width=0.85\linewidth,keepaspectratio]{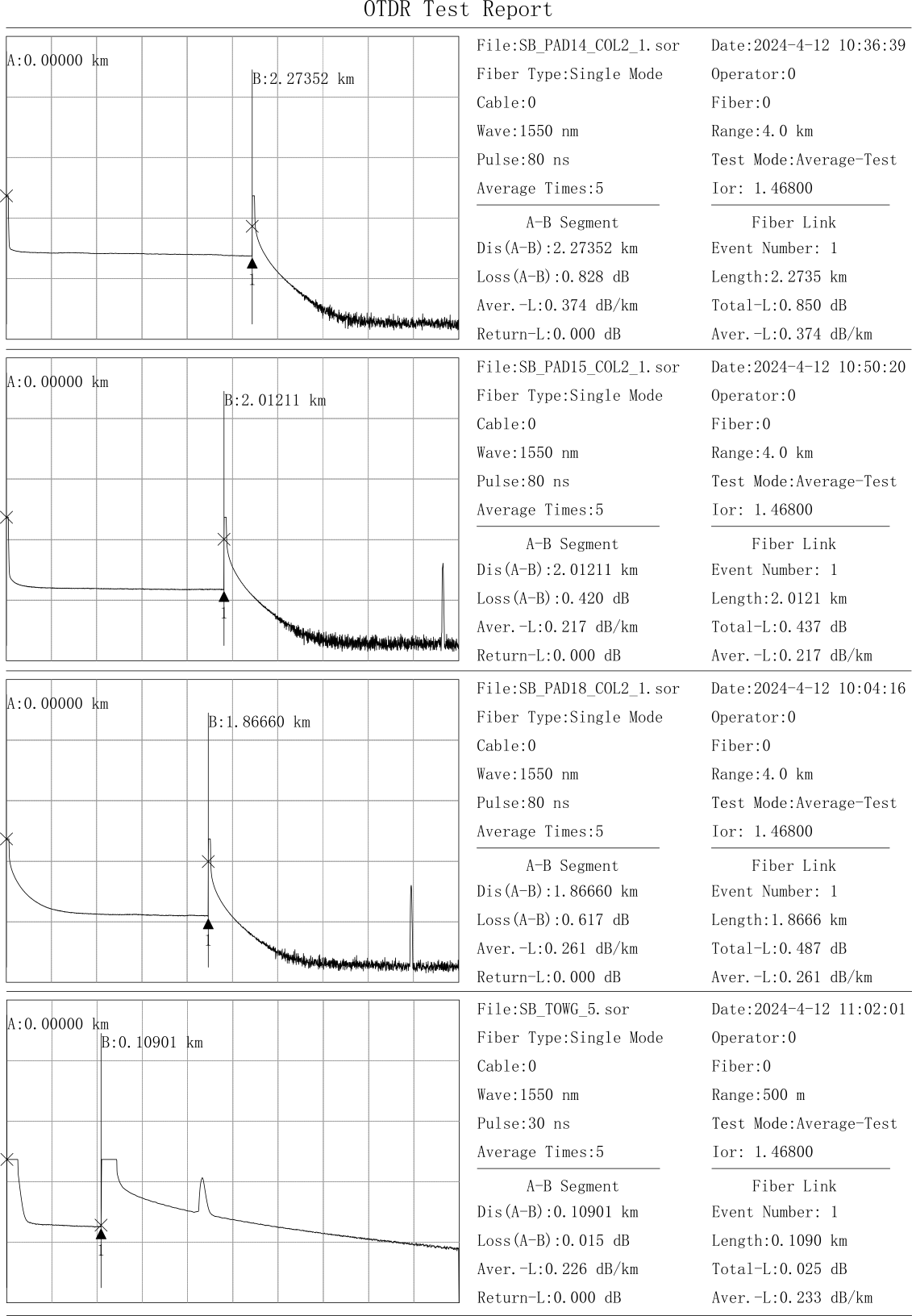}
    \caption{Example OTDR measurements of buried and aerial fiber at the Stockbridge QLAN. Fibers connecting the command center to Pads A, B, and C are low loss, with average loss values of $\sim 0.19 \;\mathrm{dB/km}$}.
    \label{app fig:Stockbridge OTDR}
\end{figure*}

\begin{figure*}
    \centering  \includegraphics[width=0.85\linewidth,keepaspectratio]{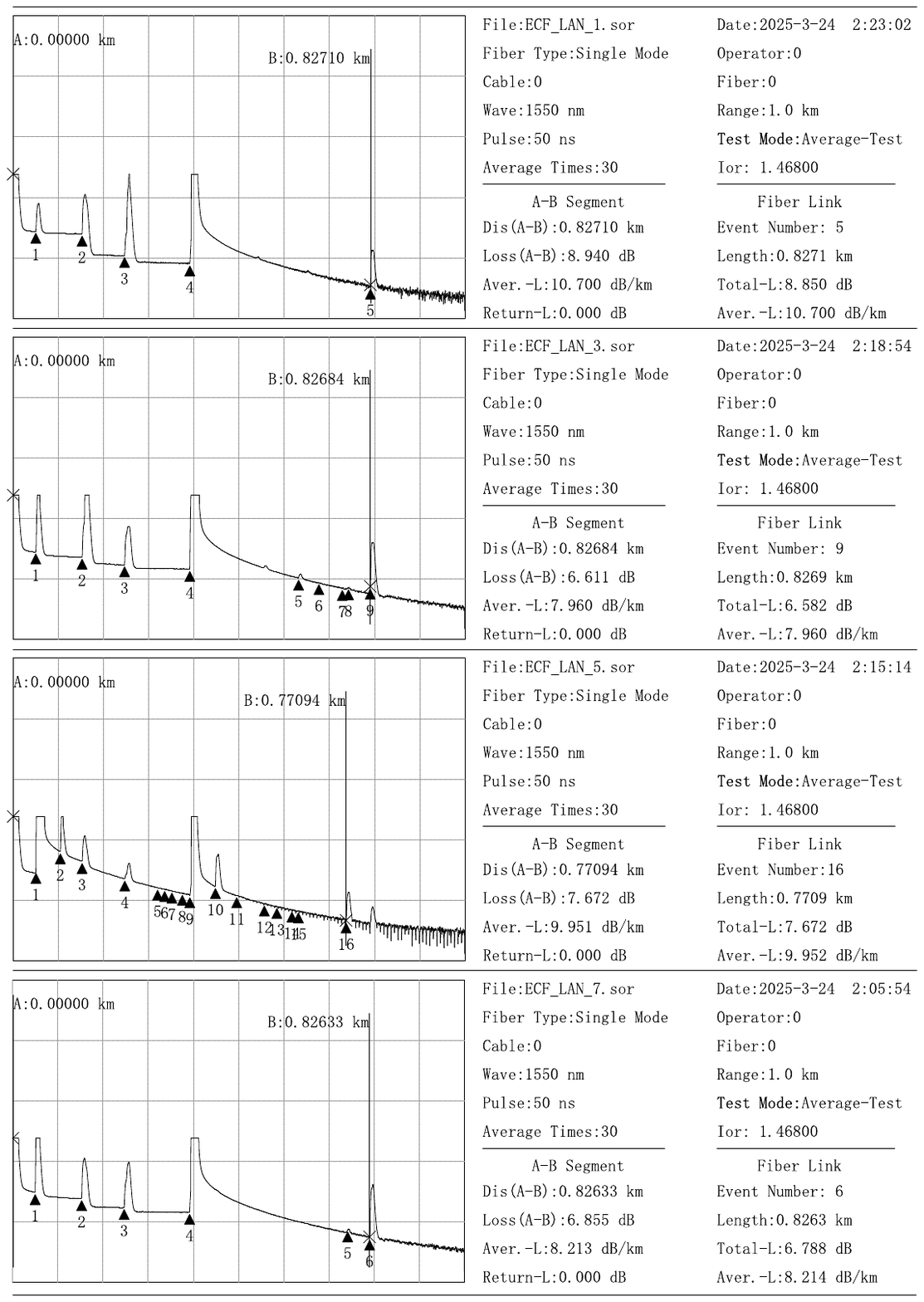}
    \caption{Example OTDR measurements of aerial fiber connected to the network hub at the RRS QLAN.Link loss values fall between $6-10\;\mathrm{dB}$.}
    \label{app fig:RRS OTDR}
\end{figure*}

%\nocite{*}
\clearpage
\section*{Supplementary References}

\addcontentsline{toc}{section}{Supplementary References}

{[S1] I.~A.~Burenkov \textit{et al.}, ``Synchronization and coexistence in quantum networks,'' \textit{Optics Express} \textbf{31}(7), 11431 (2023).}

{[S2] A.~C.~Wroblewski, ``Space-and-wave-division de-multiplexing of a quantum key distribution and classical channels into a single receiving optical fiber,'' US Patent, February 2023.}
%\bibliography{library}% Produces the bibliography via BibTeX.
%\newpage
%\section*{\centering References}
%\bibliography{supplement}% Produces the bibliography via BibTeX.

%\putbib[supplement] % Uses supplement.bib
%\end{bibunit}
\end{document}